\documentclass[%
  aps,
  pre,
  superscriptaddress, letterpaper,
  longbibliography,
  onecolumn,
  showpacs,
  showkeys,
  amsfonts, amssymb, amsmath]{revtex4-1}

\usepackage{inputenc}
\usepackage[caption=false]{subfig}
\usepackage{hyperref} % Required for customising links and the PDF
\hypersetup{pdfpagemode={UseOutlines},
bookmarksopen=true,
bookmarksopenlevel=0,
hypertexnames=false,
colorlinks=true, % Set to false to disable coloring links
citecolor=blue, % The color of citations
linkcolor=red, % The color of references to document elements (sections, figures, etc)
urlcolor=black, % The color of hyperlinks (URLs)
pdfstartview={FitV},
unicode,
breaklinks=true,
}
\usepackage{graphicx,amssymb,amsmath}
\usepackage[english]{babel}
\usepackage{grffile}
\usepackage{color}
\usepackage{ulem}
\DeclareMathOperator{\sinc}{sinc}
\def \etal {\textit{et al.}}

\begin{document}

\title{Identifying four wave resonant interactions in a surface gravity wave turbulence experiment.}
\author{Antoine Campagne, Roumaissa Hassaini, Ivan Redor, Thomas Valran, Samuel Viboud, Jo\"{e}l Sommeria, Nicolas Mordant}
\email[]{antoine.campagne@univ-grenoble-alpes.fr}
\affiliation{Laboratoire des Ecoulements G\'eophysiques et Industriels, Universit\'e Grenoble Alpes, CNRS, Grenoble-INP,  F-38000 Grenoble, France}

\begin{abstract}
The nonlinear dynamics of waves at the sea surface is believed to be ruled by the Weak Turbulence framework. In order to investigate the nonlinear coupling among gravity surface waves, we developed an experiment in the Coriolis facility which is a 13-m diameter circular tank. An isotropic and statistically stationary wave turbulence of average steepness of 10\% is maintained by two wedge wave makers. The space and time resolved wave elevation is measured using a stereoscopic technique. Wave-wave interactions are analyzed through third and fourth order correlations. We investigate specifically the role of bound waves generated by non resonant 3-wave coupling. Specifically, we implement a space-time filter to separate the dynamics of free waves (i.e. following the dispersion relation) from the bound waves. We observe that the free wave dynamics causes weak resonant 4-wave correlations. A weak level of correlation is actually the basis of the Weak Turbulence Theory. Thus our observations support the use of the Weak Turbulence to model gravity wave turbulence as is currently been done in the operational models of wave forecasting. Although in the theory bound waves are not supposed to contribute to the energy cascade, our observation raises the question of the impact of bound waves on dissipation and thus on energy transfers as well.

\end{abstract}

\maketitle

%%%%%%%%%%
\section{Introduction}
%%%%%%%%%%
\label{sec_intro}

Hasselman made the first developments of the Weak Turbulence Theory (WTT) in the 1960's to describe the statistical properties of the deformation of the sea surface due to surface gravity waves~\cite{hasselmann_non-linear_1962}. 
The main hypothesis underlying the WTT is that of weak nonlinearity. The nonlinearity enables energy transfers among waves. Its weakness implies that the dynamics remains dominated by quasi-linear waves and only resonant wave interactions can have a long term cumulative effect to transfer significant amounts of energy. 
These resonances imply $N$-tuples of waves with $N\geq3$ where 3 is the lowest order in the case of a quadratic non linearity. For $N=3$, resonant waves of frequency $\omega$ and wavevector $\mathbf k$ should follow:
\begin{eqnarray}
\omega_1&=\omega_2+\omega_3,
\label{eq:res_cond3_om}\\
%\end{eqnarray}
%\begin{eqnarray}
{\bf k}_1&={\bf k}_2+{\bf k}_3,
\label{eq:res_cond3_k}
\end{eqnarray}
where $\mathbf k_i$ and $\omega_i$ are related by the linear dispersion relation (LDR) $|{\bf k}_i|=k_{LDR}(\omega_i)$ or $\omega_i=\omega_{LDR}(|{\mathbf k}_i|)$. Using the additional hypothesis of very large systems, it is then possible to derive an equation for the slow evolution of 
the wave spectrum~\cite{zakharov_kolmogorov_1992} for dispersive waves.
The nonlinear transfer terms is one of the core ingredients of operational models of prediction of the evolution of the sea state such as Wavewatch III~\cite{tolman18user}. 
Weak turbulence concerns potentially any kind of dispersive waves and was actually applied to many systems in addition to surface waves: inertial waves in rotating fluids~\cite{galtier_weak_2003}, 
Kelvin waves in quantum vortices~\cite{lvov_energy_2004}, vibrating plates~\cite{during_weak_2006}, Bose Einstein condensates~\cite{nazarenko_wave_2006}, magnetized plasmas~\cite{zakharov_hamiltonian_1985}, 
optics in nonlinear media~\cite{bortolozzo_optical_2009},... 

Here we focus on the case of turbulence of surface gravity waves propagating in deep water and in a 2D horizontal space (i.e. a 3D flow).
In that case, four-wave interaction is the elemental process of energy transfers in the framework of Weak Turbulence Theory (WTT)~\cite{phillips_dynamics_1960, zakharov_kolmogorov_1992,nazarenko_wave_2011}.
%The applicability of the theory to gravity wave turbulence relies on the hypotheses of a large domain compared to the largest energetic scales and of weak non-linearities. 
The weak nonlinearity hypothesis implies that the linear time $T^L=2\pi/\omega$ is much smaller than the typical non-linear time $T^{NL}$ related to the non-linear wave interactions. Since the non-linearity is quadratic, resonant interactions should involve 3-wave interactions at the lowest order which impose the following resonant conditions (\ref{eq:res_cond3_om}) and (\ref{eq:res_cond3_k}). For surface gravity waves, the LDR is
\begin{equation}
\omega^2=gk\tanh(kh),
\label{shal}
\end{equation}
where $k=|{\bf k}|$, $h$ is the water depth (here $h=0.9$~m) and $g$ is the acceleration of gravity. {At high frequencies (getting close to 10~Hz) capillarity should be taken into account as well}. For $kh<1$ (i.e. wavelength shorter than 5 m) the equation (\ref{shal}) reduces to 
\begin{equation}
\omega^2=gk.
\label{LDR}
\end{equation}
However, this set of resonance equations (\ref{eq:res_cond3_om},\ref{eq:res_cond3_k},\ref{LDR}) does not have solutions due to the negative curvature of the linear dispersion relation $\omega_{LDR}(k)=\sqrt{gk}$. Hence higher order terms i.e. four-wave resonant interactions have to be considered~\cite{phillips_dynamics_1960,Hammack:1993vz} with the following resonant conditions (together with (\ref{LDR})):
\begin{eqnarray}
\omega_1+\omega_2&=\omega_3+\omega_4,
\label{eq:res_cond4_om}\\
%\end{eqnarray}
%\begin{eqnarray}
{\bf k}_1+{\bf k}_2&={\bf k}_3+{\bf k}_4.
\label{eq:res_cond4_k}
\end{eqnarray}
This set of equations can be reduced to

\begin{equation}
k_{LDR}(-\omega_2+\omega_3+\omega_4){\bf e}_1+
k_{LDR}(\omega_2){\bf e}_2 =
k_{LDR}(\omega_3){\bf e}_3+
k_{LDR}(\omega_4){\bf e}_4.	
\label{eq:4_waves_resonances}
\end{equation}
where ${\bf e}_i=\cos\theta_i{\bf e}_x+\sin\theta_i{\bf e}_y$ {and $k_{LDR}(\omega)=\omega^2/g$}.
It should be noted that because of nonlinear widening of the dispersion relation, a slight detuning of the resonance conditions enables approximate resonances that are known to play a major role in the dynamics of weak turbulence~\cite{annenkov_role_2006,shemer_evolution_2007,Kartashova:2008bv}.

In the situation of an out of equilibrium and stationary forced system, solutions of the wave spectrum can be analytically computed. 
It leads to the so-called Kolmogorov-Zakharov (KZ) power spectrum. 
For surface gravity waves propagating on a 2D surface, the predicted spatial spectrum of the surface elevation in the isotropic case is given by $E^\eta(k)\propto g^{-1/2}P^{1/3}k^{-5/2}$ where $g$ is the acceleration of gravity and $P$ the averaged {energy flux}~\cite{nazarenko_wave_2011}. 
The use of the linear dispersion relation of gravity waves makes straightforward the deduction of its temporal spectrum $E^\eta(\omega)\propto g P^{1/3}\omega^{-4}$.

In situ measurement of ocean surface waves~\cite{hwang_airborne_2000, romero_airborne_2010, leckler_analysis_2015} as well as direct numerical simulations of gravity wave turbulence~\cite{lvov_discreteness_2006, yokoyama_statistics_2004, onorato_freely_2002, zakharov_mesoscopic_2005}
have shown a good agreement with the predicted $\omega^{-4}$ or $k^{-5/2}$ decay of the spectrum. On the contrary, experimental studies of weak gravity wave turbulence~\cite{campagne_impact_2018, aubourg_three-wave_2017, deike_role_2015, 
%miles_surface-wave_1967, %henderson_diane_m._role_2013,
nazarenko_wave_2016, denissenko_gravity_2007,lukaschuk_gravity_2009} 
report a strong variability of the surface elevation spectrum which stand quite steeper than the theoretical prediction. 

Several reasons are invoked to explain this disagreement between laboratory experiments and theoretical predictions: (i) the effect of dissipation which is not weak enough, (ii) the strength of  nonlinearity which is too strong, (iii) the finite size of the experimental basins which are too small.
Concerning dissipation (i): For technical reasons, experimental facilities have sizes of the order of a couple times 10 meters and the forced wavelengths are of the order of a few 
meters (i.e. a forcing frequency usually only slightly below 1~Hz). As analyzed in \cite{campagne_impact_2018}, at these scales, the ratio between the bulk viscous time scale $T^d\propto\nu^{-1}k^{-2}=\nu^{-1}g^{2}\omega^{-4}$ and the period of the waves $T^L$ is not very large {($10^{5}$ for a 1~m wavelength gravity wave)} and {above all} decreases dramatically as $\omega^{-3}$. Thus the scale separation between the period of the wave and the dissipative time scale may not be large enough to accommodate 4-wave coupling that occurs over very long times. Moreover, this estimation for the dissipative time is probably overestimated since it does not take into account the surface pollution nor the friction at the walls of the tank which can increase significantly the dissipation~\cite{campagne_impact_2018, miles_surface-wave_1967,dorn_boundary_1966, henderson_single-mode_1990}. The development of a true conservative inertial range may be difficult to achieve in experiments. We then expect a non negligible dissipation which steepens the spectrum as the cascade is leaking energy at all scales (see for instance~\cite{campagne_impact_2018,Humbert,R23,Deikedis2}). Concerning (ii), the spectral exponents of frequency spectra come closer to the theoretical predictions when the {average} steepness $\epsilon$ of the waves (i.e. the strength of the nonlinearity) is increased to very high values ($\epsilon\geq 0.3$). Such high steepness are {very rarely} observed in the ocean (see fig.~2 of \cite{Socquet} that show a scatter plot of observed values of the steepness) in which values of the steepness above 0.1 are very rare. 1D spatial measurements in experiments have shown that spectral exponents in wavenumber in the very strong regimes are not compatible with the weak turbulence prediction and suggest rather a regime dominated by strongly nonlinear structures~\cite{NazLuk}. Note that in field measurements, the -4 theoretical exponent of the frequency spectrum is observed at the largest scales (called usually the equilibrium spectrum) but that the spectrum becomes steeper at small scales with an exponent close to -5 which is interpreted as a saturated spectrum~\cite{Phillips:2006jt,leckler_analysis_2015} with whitecapping or overturning waves (measurement in the spatial domain show the same feature \cite{romero_airborne_2010}). These observations suggest that experimental data have a too high level of nonlinearity and thus do not remain in the weak turbulence framework.  
The third reason (iii) is the effect of finite size. The resulting discretization in the space of wavevectors makes the number of 
solutions of resonant conditions~(\ref{LDR}-\ref{eq:res_cond4_k}) considerably lower than in the infinite case~\cite{Kartashova:2008bv} which leads to a steepening of the spectrum. 
We notice that numerical simulations, which are computed on a finite number of modes, share the former remark with experiments.
Note that all three reasons may be acting simultaneously. In order to overcome the effect of dissipation (i) and finite size (iii), the nonlinearity has to be increased so that to reduce the nonlinear time scale so that it is much smaller than the dissipative time and so that spectral broadening can overcome the spacing between discrete modes by allowing quasi-resonances (i.e. a slight detuning is allowed in the resonance conditions). The risk is then to be too strongly nonlinear (ii).

The analysis developed in the present article focusses on point (ii): do the experiments remain in a regime of weak turbulence in which the main process of energy transfer is 4-wave resonant coupling, even though the level of nonlinearity is quite large ? In order to answer this question, we use 3 and 4-wave correlation estimators (bicoherence and tricoherence tools) to directly probe the nonlinear coupling among waves. In the weak turbulence framework, 4-wave correlations or rather 4-wave cumulants must remain weak to support the validity of the multiscale expansion at the core of the theory~\cite{NewellR}. Our analysis is aimed at checking this hypothesis. 
Aubourg~\etal~\cite{aubourg_three-wave_2017} investigated experimentally four-wave interactions in a surface gravity wave turbulence with fourth-order correlations of the vertical velocity. 
At their degree of {statistical convergence of the averaging}, no clear trace of a significant correlation appears {most likely due to an insufficient amount of data}. The authors proposed the dissipation as the cause of the absence of four-wave correlation. 
The dissipation time scale $T^d$ is of the same order of the non linear time $T^{NL}$ which should reduce the efficiency of the energy transfers. Nonetheless, the lack of {statistical} convergence do not permit to draw any definite conclusion. 
Here we report experiments with a much larger dataset so that to converge the fourth order statistics. Another technical difficulty in experiments is that our observation variable, the surface elevation, is not the one used in the development of the theory. The reason is that {quadratic nonlinearities among 3 Fourier modes} yield the development of so-called bound waves which can be quite strong if $\epsilon$ is large. {Bound waves (that should rather be called bound modes as they are not waves) are Fourier modes that do not follow the dispersion relation of the waves.} The impact of the bound waves may hide the contribution of free waves. In the theory, the bound waves are eliminated by an adequate change of variable (see below) that retains only the free waves but that cannot be performed in experiments due to limitations of the measurements. To circumvent these limitations we take advantage of our stereoscopic measurement technique that provides us with a time and space resolved measurement of the surface elevation. We can thus perform a specific filtering in $(\omega,\mathbf k)$ space so that to remove the contribution of bound waves and retain only the free waves to compute the correlations.

The structure of the article is the following: in part~\ref{sec_ecperimental_setup}, we describe briefly the experimental setup, in part~\ref{sec_structure} we present the surface elevation spectrum that shows in particular the importance of bound waves. We also introduce the filtering in this part. Then in part~
\ref{sec_hamiltonian} we recall a few elements of the theory concerning the elimination of the bound waves. The statistical analysis of the high order correlation starts in part~\ref{sec_C3} with 3-wave interaction due to bound waves. 4-wave correlations are investigated in section~\ref{sec_correlation} and \ref{sec_filtC4}.

%%%%%%%%%%%%%%
\section{Experimental setup}
%%%%%%%%%%%%%%
\label{sec_ecperimental_setup}
\begin{figure}[!htb]
\includegraphics[width=7cm]{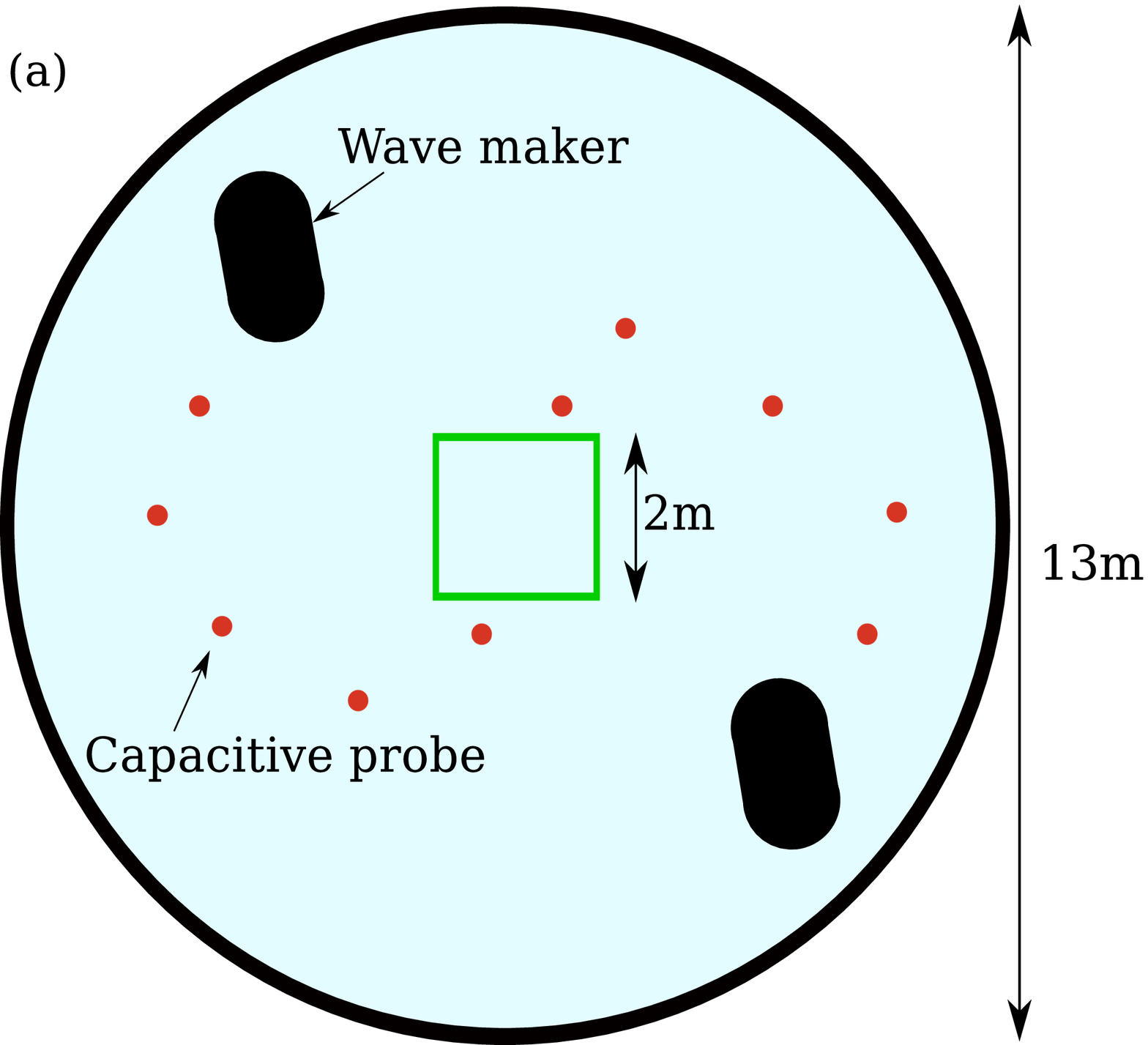}
\includegraphics[width=7cm]{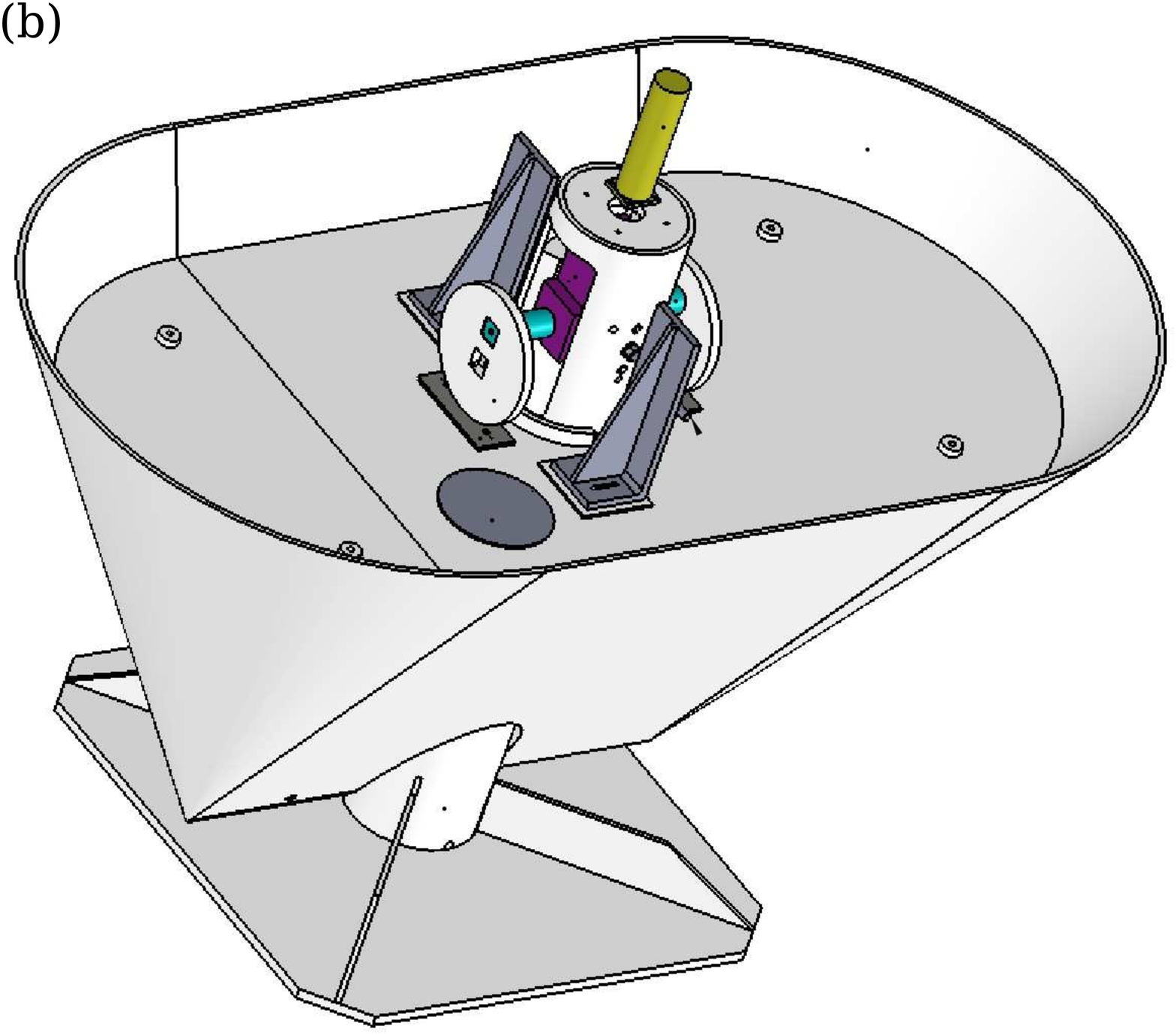}
\caption{(a) Global schematics of the setup in the Coriolis facility. A 13m diameter circular tank is filled with $0.9$~m of water. Surface gravity waves are generated by two wedge wavemakers.. Surface elevation is recorded by 10 capacitive wave gauges (red points) and by a stereoscopic reconstruction on  a $2\times2$~m$^2$ area at the center of the tank (green square). (b) Perpective view of a wedge wavemaker.  It is oscillating vertically at a  randomly modulated frequency $\omega_f/2\pi = 0.59 \pm 0.15$~Hz with an amplitude of 2~cm.}
\label{setup}
\end{figure}

The setup is the same as in~\cite{campagne_impact_2018} and similar to the one described previously in~\cite{aubourg_etude_2016,aubourg_three-wave_2017} (fig.~\ref{setup}) so we only briefly recall the main features. The wave tank is circular with a diameter equal to 13~m and the water depth is $h=0.9$~m (thus waves with wavelength below 1~m can be considered as in the deep water regime). Waves are generated by two wedge wavemakers of size $2\times 1 \times 1$~m$^3$ oscillating vertically with an amplitude of 2~cm at a randomly modulated frequency $\omega_f/2\pi=0.59\pm0.15$~Hz (fig.~\ref{setup}(b)). It corresponds to wavelengths around 2~m). We wait about 15 minutes after starting the forcing to make sure to have reached a statistically stationary state. The water surface is maintained as clean a possible by pumping the surface through a skimmer located near the wall and by flowing the pumped water through a large active carbon filter. The water is then reinjected near the wall at a location diametrically opposite to the skimmer. Surface tension is known to be very sensitive to surface contamination. Thus we measure surface tension by sampling the water regularly during the experiment. After several days of filtration the measured surface tension was $0.074\pm0.005$~N/m which was the value measured from fresh tap water at the same temperature and very close to the value for pure water. This process ensures that {as little extra dissipation as possible} due to surface contamination is present, so that to have the weakest possible dissipation.

\begin{figure}[!htb]
\includegraphics[width=16cm]{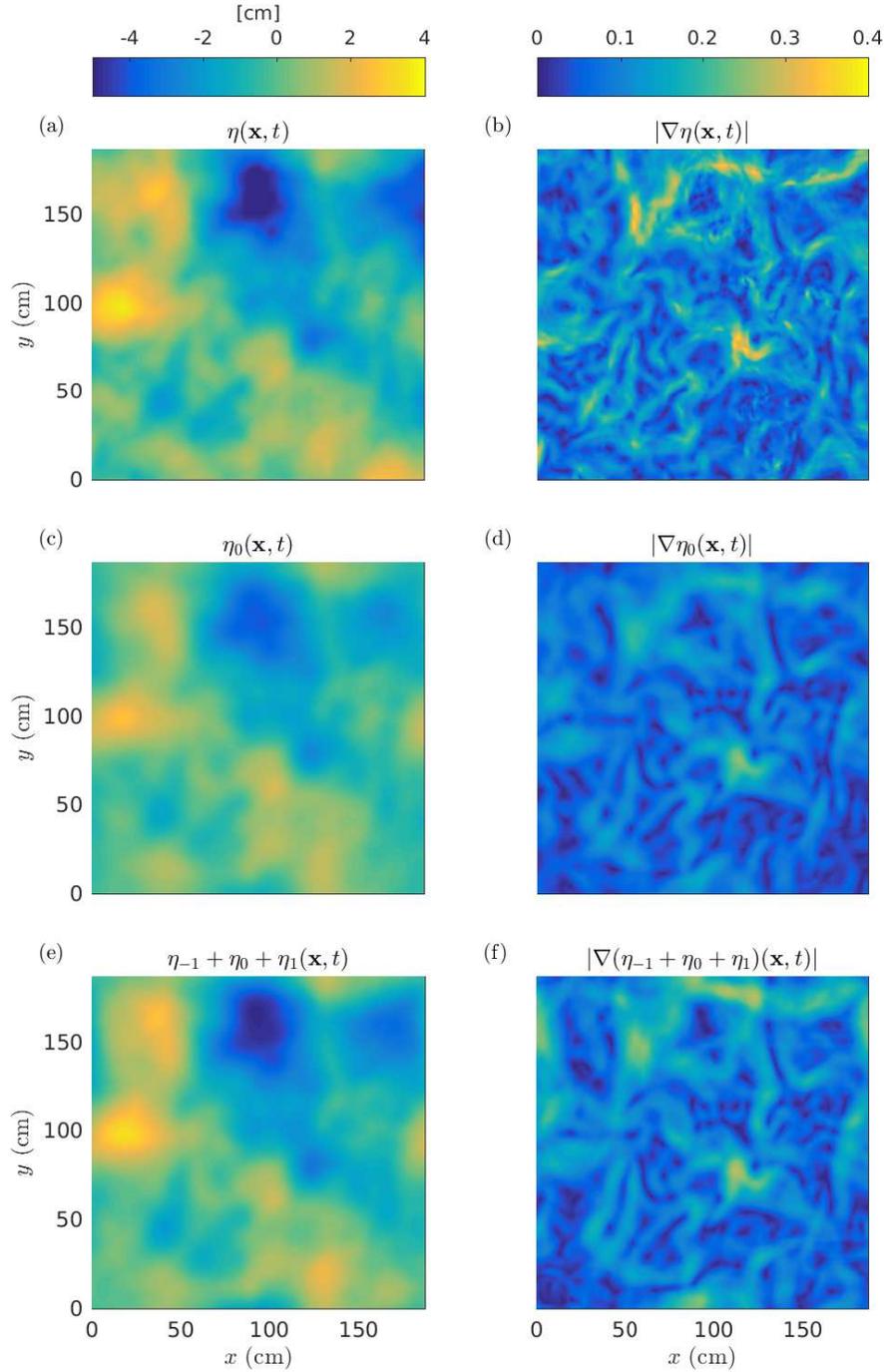}
\caption{Instantaneous surface elevation $\eta$ (a-c-d) and norm of its spatial gradient ${|\boldsymbol \nabla}\eta|$ (b-d-f) for raw data $\eta$ (a-b) and filtered data $\eta_0$ (c-d) {(free waves)} and $\eta_{-1}+\eta_{0}+\eta_{1}$ (e-f) {(see text for definitions). One sees that fine scale details are associated to bound waves.}}
\label{fig_fieldz}
\end{figure}

The surface elevation is recorded using two different methods. The first one is performed by 10 capacitive wave gauges (their positions are shown in fig.~\ref{setup}(a)).
The second one is a stereoscopic reconstruction, called Stereo-PIV in the following. Buoyant particles (700~$\mu$m) are seeded on the surface in order to form a random pattern which is recorded simultaneously by three cameras (5 Mpixel, PCO Edge). 
A cross-correlation between the three images taken at the same time is computed to reconstruct surface elevation $\eta({\bf x}, t)$ in a rectangular area of size $2\times2$~m$^2$ with a resolution of $\Delta x=1$~cm at 
the center of the tank using a stereoscopic algorithm~\cite{aubourg_etude_2016, aubourg_three-wave_2017}. Analysis of the performance of the stereoPIV algorithm shows that the accuracy of the surface reconstruction is of the order of 0.5~mm \cite{aubourg_etude_2016, 2019arXiv190205819A}. 9~hours of image acquisition at a frequency of 40~frames/s have been collected which correponds to more than 50 terabytes of raw images. Note that our method is quite different to that used in the field and in the lab by \cite{Benetazzo:2006dp,Zavadsky:2017fm,leckler_analysis_2015} due to the fact that the surface is materially marked by particles and the angle of view between the cameras is very large.

\begin{figure}[!tb]
\centerline{
\includegraphics[width=18cm]{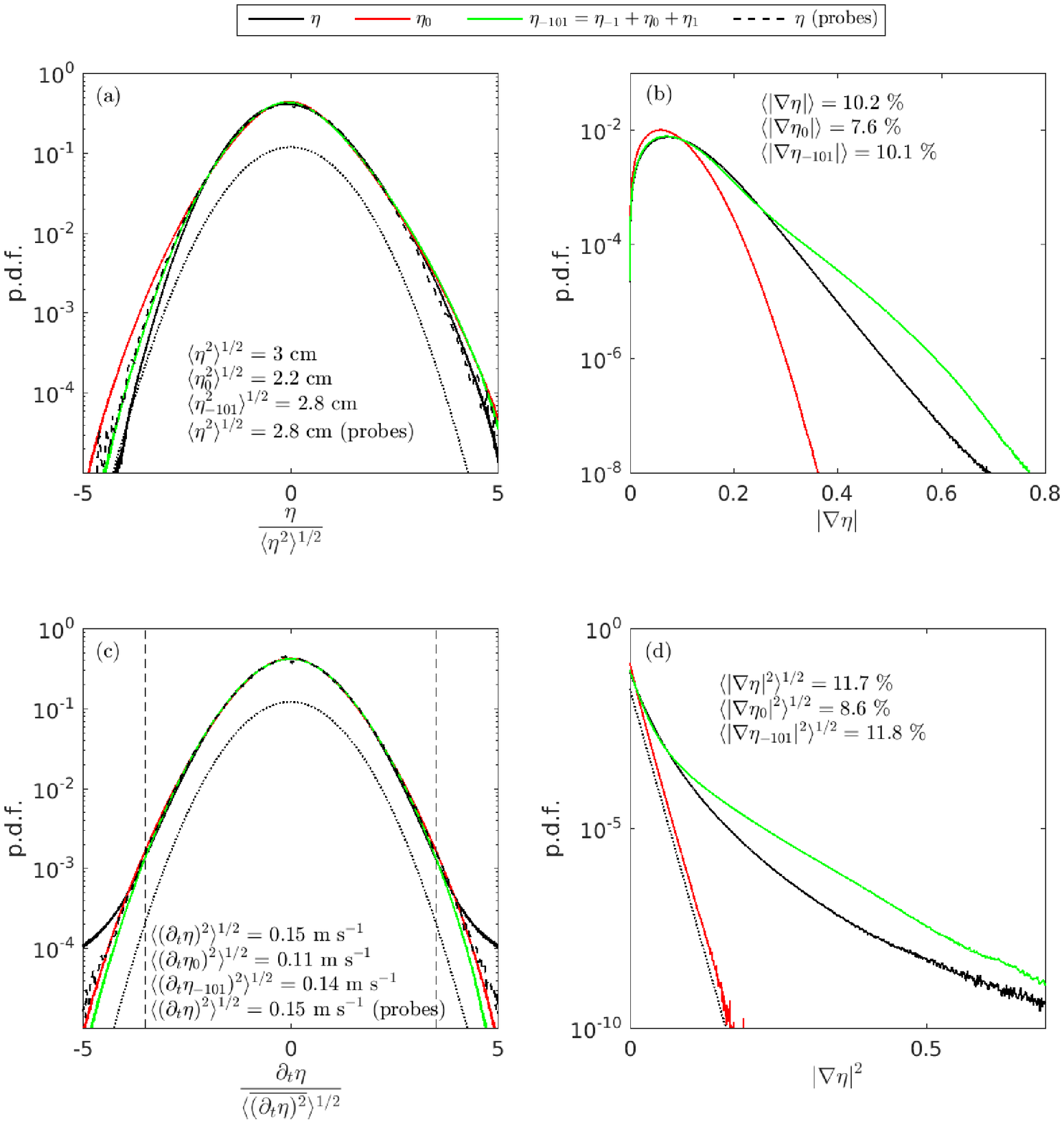}}
\caption{Probability density function of the norm  of the surface elevation $\eta$ (a), of local slope $|{\boldsymbol \nabla}\eta|$ (b), of the time derivative of surface elevation $\partial_t \eta$ (c) and of the squared local slope $|{\boldsymbol \nabla}\eta|^2$ (d). Raw data $\eta$, filtered data $|{\boldsymbol \nabla}\eta_0|$ and $|{\boldsymbol \nabla}(\eta_{-1}+\eta_0+\eta_1)|$ and data from probes are represented by black, red, green and black dashed lines respectively. Dotted lines represent a gaussian distribution for figures (a) and (c) and a Rayleigh distribution for figure~(d). {Departure from Gaussianity can clearly be attributed to contribution of bound waves}.}
\label{fig_slope}
\end{figure}
Instantaneous surface elevation $\eta({\bf x}, t)$ and its PDF are shown in the Fig.~\ref{fig_fieldz}(a) and ~\ref{fig_slope}(a) respectively. 
The wave elevation is typically of the order of few centimeters.
% with an energetic length scale 
%of the order of one meter. This is smaller than the wavelength associated to the central forcing 
%frequency $2\pi/k_{LDR}(\omega_f)\simeq 4$~m. 
A good consistency  between probes (dashed line) and stereoscopic (black) measurements is observed in fig.~\ref{fig_slope}(a). This confirms the accuracy of the stereoPIV reconstruction. 
A slight asymmetry with larger tails for positive elevations is observed. This asymmetry is
usually related to the presence of Stokes or bound waves~\cite{stokes1880theory} and {is included in the so-called Tayfun distribution~\cite{Tayfun} (which reproduces correctly our observed distribution as reported previously in~\cite{aubourg_three-wave_2017})}.
Furthermore, the PDF of the time derivative of elevation $\partial_t\eta$ (Fig.~\ref{fig_slope}(c)) is symetric with a good agreement between probes and sterescopic measurements up to $3.5v_{rms}$ with $v_{rms}=\langle\overline{(\partial_t\eta)^2}\rangle^{1/2}$. Here, the angular brackets stand for the spatial average and the overline for the temporal average. The over representation for larger values for stereoscopic measurement is due to a lower signal-to-noise and to the 
presence of rare false elevation reconstruction associated to poor image correlations. We stress that events such that 
$|\partial_t\eta|>3.5v_{rms}$ represents less than $0.2\%$ of the whole statistics.

The instantaneous local slope $|{\boldsymbol \nabla}\eta({\bf x}, t)|$ is shown in fig.~\ref{fig_fieldz}(b) (top).
The average slope $\varepsilon=\langle\overline{|{\boldsymbol \nabla}\eta({\bf x}, t)|}\rangle$ is evaluated 
to be $10~\%$.
We observe that the local slope shows small scale structures (about few centimeters) with a local values much 
larger than $\varepsilon$. This is confirmed by the probability density function (PDF) of the local slope 
(see Fig.~\ref{fig_slope}(a)). The PDF of $|{\bf \nabla}\eta |^2$ (Fig.~\ref{fig_slope}(d) blue curve) displays a
very wide tail corresponding to extreme events of the slope (that can reach values of order 1 i.e. about ten 
times larger than the average). The presence of such high local slopes highlights the fact that strongly 
non-linear events occur in a bath of wave turbulence of finite non-linearity. 
The presence of such singularities may appear as an important violation of the hypothesis of weak non-linearity 
of the WTT. However, the possible role of such singularities on the dynamics cannot be discussed at this stage since their structure and their coupling with linear waves are unknown. The next section is then dedicated to describe the spatiotemporal structure of the flow.

%%%%%%%%%%%%%%%
\section{Structure of the flow}
%%%%%%%%%%%%%%%
\label{sec_structure}

\begin{figure}[!htb]
\includegraphics[width=12cm]{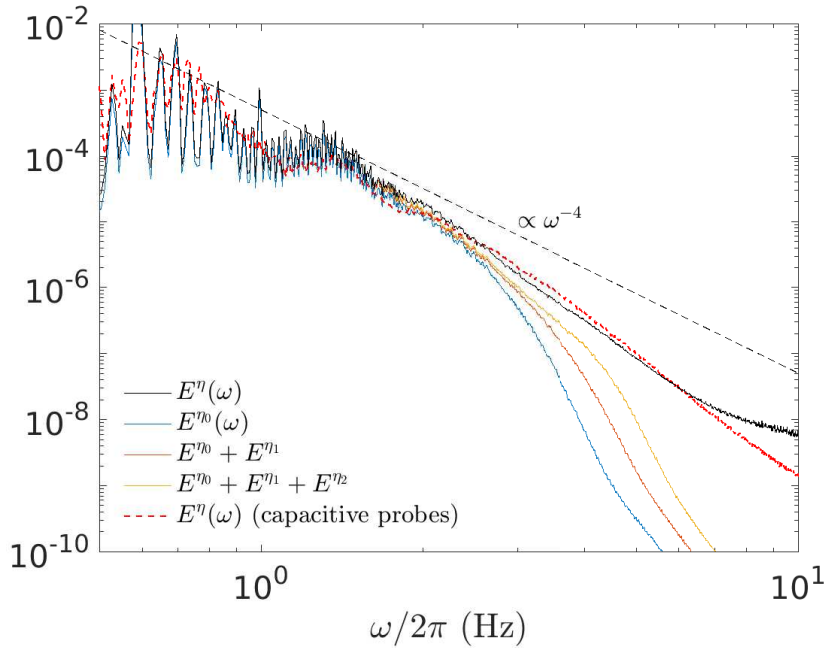}
\caption{Frequency power spectrum of the surface elevation from stereoscopic measurements and capacitive probes. {The black curve is the spectrum of the full signal measured by the stereoscopic PIV. The other continuous curves correspond to filtered data as shown in the legend. The red dashed line is the spectrum measured by the capacitive probes. The black dashed straight line is an eye guide for the $\omega^{-4}$ theoretical behavior. The spectra measured by capacitive probes and by the stereoscopic contributions are in agreement up to about 7~Hz. At higher frequencies, the stereoscopic spectrum deviates from that of the local probes due to measurement noise. The difference between the spectrum of $\eta_0$ (free waves) and that of the full signal corresponds to the energy of the bound waves.}}
\label{fig_spectra_w}
\end{figure}

In this section, we report general features of various spectra of the surface elevation that show the presence of significant bound waves.

\subsubsection{Frequency spectrum}
%%%%
We first consider the energy distribution of wave elevation as a function of angular frequency $\omega$.
Our estimator of the temporal energy spectrum is:
\begin{eqnarray}
E^\eta(\omega)=\frac{2}{T}\langle |\widetilde{\eta}({\bf x}, \omega)|^2\rangle,
\end{eqnarray}
where
\begin{eqnarray}
\widetilde{\eta}({\bf x}, \omega)=\frac{1}{\sqrt{2\pi}}\int_0^T \eta({\bf x}, t)e^{-i\omega t}h_T(t)dt
\label{eq_temporal_FFT}
\end{eqnarray}
is the temporal Fourier transform of the wave elevation $\eta({\bf x}, t)$ over a time window of finite duration $T$ (with a Hanning window $h_T(t)$ of length $T$). Note that we use integrals in the formula but the fields being discrete in time, the sum is actually a discrete sum in practice. It will be the same for space variations below. {This transposes in Fourier space by the fact that frequencies and wavevectors are discrete as well. We keep the notations with integrals to avoid a very heavy writing of the equations but the discreteness is implicit.} The normalization is such that $\langle\overline{\eta^2}\rangle=\int_0^\infty E^\eta(\omega)d\omega$. {The angular brackets refers here to the space average together with the use of the standard Welch method with a time window of $T=125$~s. In doing this we assume that the system is homogeneous in space and time. Assuming ergodicity, we perform average in space and time to improve the statistical convergence of the estimators.} Figure~\ref{fig_spectra_w} shows the temporal spectrum $E^\eta(\omega)$ evaluated either from stereoscopic reconstruction or from capacitive probes. A good correspondance of the spectra is observed for the two measurement methods up to 7~Hz. The measurement dynamics of the stereo surface reconstruction is thus about 6 orders of magnitude which is quite good for an image based measurement. At higher frequencies, the stereoscopic measurement is dominated by noise. The spectral exponent is about $-6$ which is significantly steeper than the theoretical prediction $-4$. This value is consistent with other experimental studies at the same average wave steepness $\varepsilon$~\cite{deike_role_2015}. This has been related to the important effect of the dissipation at $\gtrsim 1$~Hz scale which strongly decreases the energy content at these scales~\cite{campagne_impact_2018}.

\subsubsection{Wavenumber-frequency spectrum}
%%%%
\begin{figure}[!htb]
\vspace{-1cm}
\centering
\includegraphics[width=12cm]{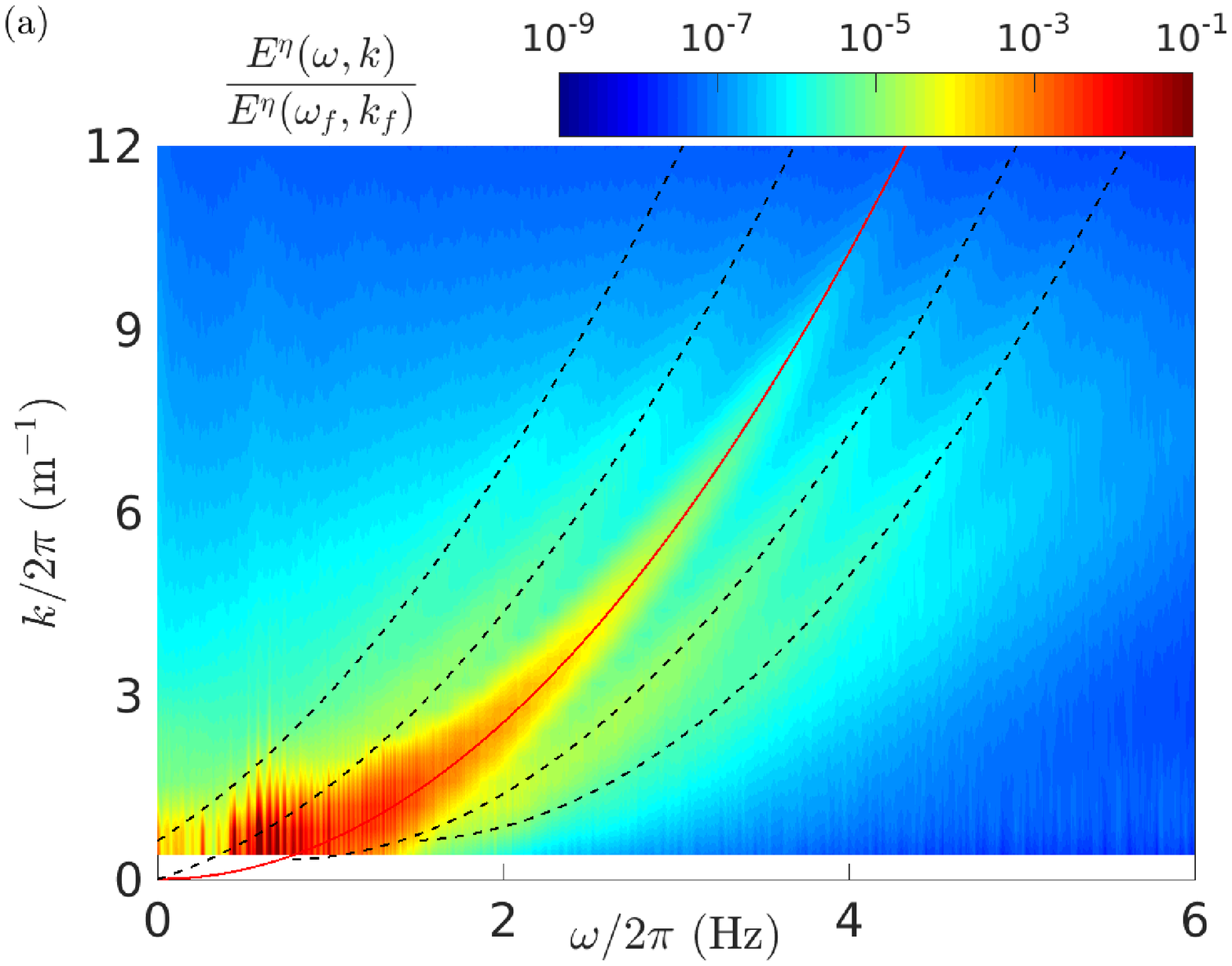}

\hspace{-0.8cm}\includegraphics[width=12cm]{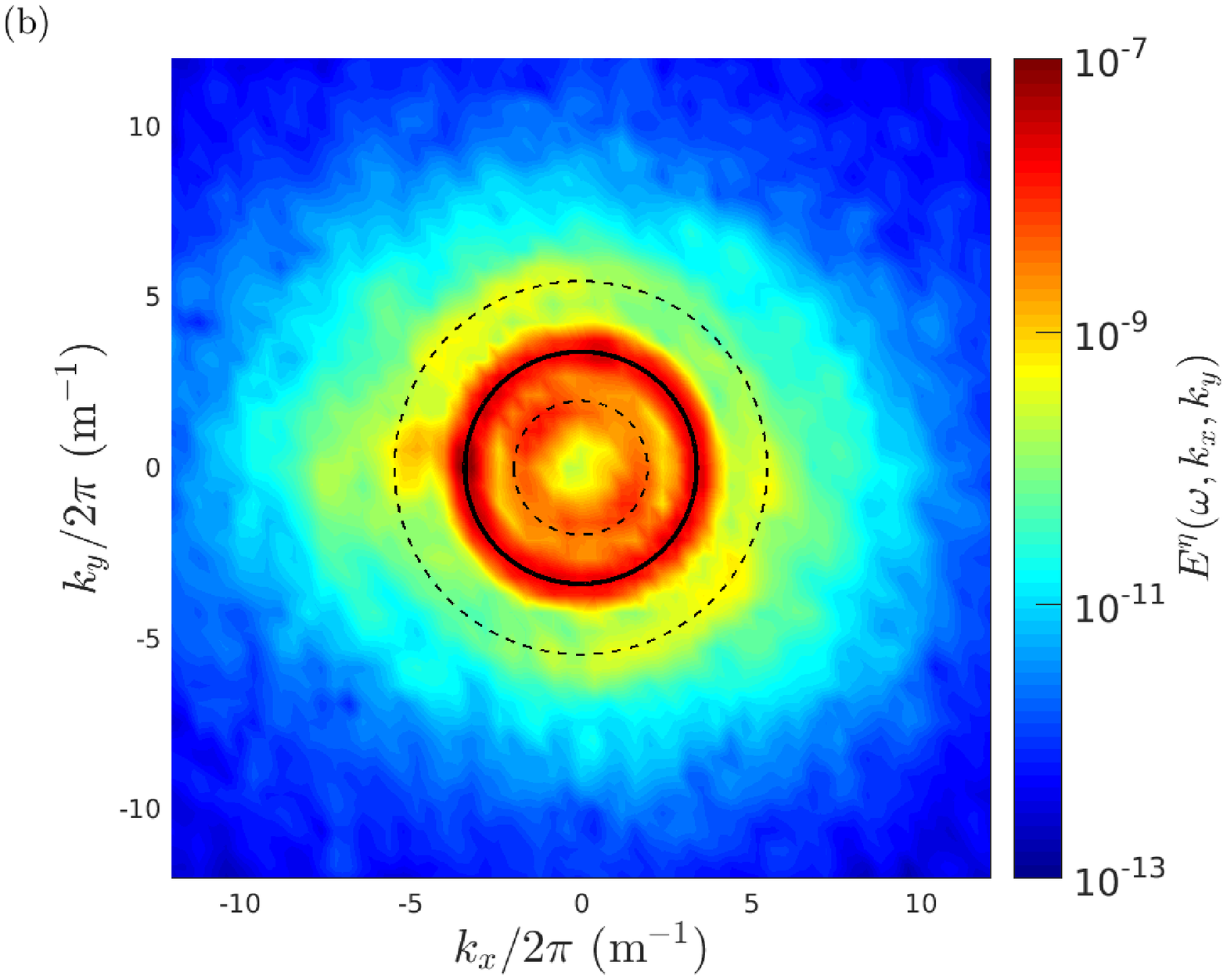}
\caption{(a) Normalized angular average of the spatio-temporal spectrum of the surface elevation. The red line corresponds to the linear dispersion relation of gravity waves $k_{LDR}(\omega)$ and dotted lines correspond to $k_n(\omega)$, with $n\in[-2,-1,1,2]$ (eq. (\ref{BW})). (b) Spatio-temporal spectrum of the surface elevation for $\omega/2\pi = 2.3$~Hz. Solid and dashed lines represent the expected wavenumber for free and bound waves (with $n=\pm1$) respectively. {In both figures most of the energy lies on the linear dispersion relation, but a significant part of energy is spread over a large area with a special concentration on the specific bound waves $k_n(\omega)$.}}
\label{fig_spectra}
\end{figure}

To go further in the characterization of the energy content of the flow in terms of waves, we now turn to a spatio-temporal analysis. Such an analysis is required to discriminate free and bound waves. Our estimator of the spatio-temporal energy spectrum is
\begin{eqnarray}
E^\eta({\bf k}, \omega)=\frac{2}{T L^2}\langle |\widetilde{\eta}({\bf k}, \omega)|^2\rangle,
\end{eqnarray}
where
\begin{equation}
\widetilde{\eta}({\bf k}, \omega)=
(2\pi)^{-3/2}\int_0^T \int_0^{L}\int_0^{L}\eta({\bf x}, \omega)e^{-i(\omega t-{\bf k}\cdot{\bf x})}h_T(t)h_L({\bf x})dtdxdy
\label{eq_spatiotemporal_FFT}
\end{equation}
is the spatio-temporal Fourier transform of the wave elevation. $\eta({\bf x}, t)$, $L$ is the spatial size of the measured field. $h_T(t)$ and $h_L(\bf x)$ are respectively a 1D and a 2D Hanning window. {In this definition of the spatio-temporal energy spectrum, the average is done only over time by the use the standard Welch method to improve the statistical convergence of the power spectrum with a time window of duration 125~s. No average is performed in space because of the Fourier transform in space.} The normalization is such that $\langle\overline{\eta^2}\rangle=\int_0^\infty \int_{-\infty}^\infty\int_{-\infty}^\infty E^\eta({\bf k}, \omega)d\omega dk_xd k_y$.

Figure~\ref{fig_spectra} (a) shows the normalized spatio-temporal spectrum $E^\eta(k, \omega)/E^\eta(k_{LDR}(\omega_f), \omega_f)$ summed over the direction angle: $E^\eta(k, \omega)= \int_0^{2\pi} E^\eta({\bf k}, \omega)kd\theta$ with ${\bf k}=k\cos\theta{\bf e}_x+k\sin\theta{\bf e}_y$. We observe that most of the energy is concentrated around the linear dispersion of gravity waves (solid red line) up to 4~Hz. This is what is expected for weak gravity waves turbulence. However energy is also present out of the LDR. In particular secondary lines of energy can be distinguished on both sides of the LDR that are highlighted by the black dashed lines. These lines are defined as
 \begin{eqnarray}
k_n(\omega) = k_{LDR}(|\omega|-n\omega_0) +nk_{LDR}(\omega_0),
\label{BW}
\end{eqnarray}
with $\omega_0/2\pi=0.7$~Hz and $n\in\mathbb Z$ ($n=0$ corresponds to free waves). $\omega_0$ is within the range of forced frequencies and also corresponds roughly to the peak of energy at large scales i.e. the most energetic waves in the tank.
These lines corresponds to so-called bound waves. These not freely propagating waves result from a triadic interaction between freely propagating waves. In our case, the observed lines can be obtained by assuming that the very energetic linear gravity wave at $(k_{LDR}(\omega_0), \omega_0 )$ is interacting with all free waves on the linear dispersion relation propagating in the same direction. This corresponds to $n=\pm1$. Higher values correspond to triadic interactions of free waves with the harmonics of the forcing peak $(|n|k_{LDR}(\omega_0), |n|\omega_0 )$. Note that a continuum of weak bound waves is also present almost everywhere in addition to these most energetic lines, including (but not only) at frequencies $\omega/2\pi$ larger than $5$~Hz or wavenumbers $k/2\pi$ larger than 10~m$^{-1}$.

Figure~\ref{fig_spectra} (b) displays a cut of $E^\eta({\bf k}, \omega)$ for $\omega/2\pi = 2.3$~Hz. The energy distribution is seen to be quite isotropic. The contribution of bound waves $n=\pm1$ is clearly visible. The contribution of the bound wave $n=1$ is isotropic as well while that of $n=-1$ seems more anisotropic.

\subsubsection{Space-time filtered field}
%%%%
In order to evaluate the contribution of each spatio-temporal structure in the temporal spectrum $E^{\eta}(\omega)$, we introduce a filter of the spatio-temporal signal in the spirit of~\cite{peureux_observation_2017}. The filter is designed to retain either only the free waves that are concentrated in the vicinity of the linear dispersion relation or the most energetic bound waves discussed above appearing as dashed lines in fig.~\ref{fig_spectra}.
We define the filtered spatio-temporal Fourier transform
\begin{eqnarray}
\widetilde{\eta}_n({\bf k}, \omega) = \widetilde{\eta}({\bf k},\omega) f_n({\bf k}, \omega) .
\end{eqnarray}
The matrix $f_n({\bf k}, \omega_0)$ filters the signal around the bound wave of order $n$: 
{
\begin{equation}
\begin{array}{lr}
 f_n({\bf k},\omega) = 1 -\delta k, \quad & \text{if} \quad \delta k\in[0, 1],\\
 f_n({\bf k}, \omega) = 0,& \text{if} \quad \delta k\notin[0, 1],
\end{array}
\end{equation}
}
with 
{
\begin{eqnarray}
\delta k = \frac{\left |k_n(\omega) - |{\bf k}| \right|}{1.5\Delta_k},
\end{eqnarray}
}
and $\Delta_k=0.5\times2\pi${~m$^{-1}$} is the spatial spectral resolution of our measurement.
%(restricted by the size of our images). 
We emphasize that $\eta_0(\omega)$ is the temporal Fourier transform filtered around the linear dispersion relation of gravity waves and thus it corresponds to free waves only. Note that due to the limited resolution of the Fourier transform in space (related to finite size of the images), the filter has a finite width around the dispersion relation. Thus our filter retains the possibility of approximate resonances due to nonlinear spectral widening around the linear dispersion relation.

Fig.~\ref{fig_fieldz} shows a snapshot of the water elevation and the local slope of the filtered $\eta_0({\bf x},t)$ as well as $\eta_0({\bf x},t)+\eta_1({\bf x},t)+\eta_{-1}({\bf x},t)$ where $\eta_i({\bf x},t)$ is the spatio-temporal inverse Fourier transform of $\widetilde{\eta}_i({\bf k}, \omega)$. We see that $\eta_0$ captures the coarse features of the total surface elevation $\eta$ with $\langle\overline{\eta_0^2}\rangle^{1/2}=2.2$~cm being only slightly lower than {the variance of the full signal} $\langle\overline{\eta^2}\rangle^{1/2}=2.9$~cm. 

The PDF of $\eta_0$ and of its time derivative are displayed in fig.~\ref{fig_slope}(a) and (c) (red curve). The filtered field has a more symmetric distribution than for the total field $\eta$ with a time derivative statistics comparable with the probe measurements. This is consistent with the fact the the asymmetry is related to the presence of bound waves. We now include the bound waves, i.e. we consider the filtered field $\eta_{-101}=\eta_{-1}+\eta_0+\eta_1$ (green curve). The PDF of wave elevation recovers its asymmetry and is comparable to the capacitive probes measurements ($\langle\overline{\eta_{-101}^2}\rangle^{1/2}=2.8$~cm$\simeq\langle\overline{\eta^2}\rangle^{1/2}=2.9$~cm). We note also that no significative change occurs in the statistics of the time derivative with $\langle\overline{(\partial_t\eta_{-101})^2}\rangle^{1/2}=14$~cm~s$^{-1}\simeq\langle\overline{(\partial_t\eta)^2}\rangle^{1/2}=16$~cm~s$^{-1}$.  

The PDF of the local slope of $\eta_0$ is displayed in fig.~\ref{fig_slope}(b) \& (d) (red curves). It shows a much narrower tail than that of the total surface elevation, with an averaged slope which is reduced from 0.1 to 0.076 (i.e. 25~\% smaller than the average slope of the full field). The PDF of $|\nabla\eta_0|^2$ (bottom) is decaying exponentially while the PDF of $|\nabla(\eta_0+\eta_1+\eta_{-1})|^2$ is very close to that of the full field and is decaying with much wider tails. Thus these wide tails can clearly be associated to the bound waves. An exponential PDF of $|\nabla\eta_0|^2$ is consistent with a Rayleigh distribution, i.e. Gaussian statistics of the gradient. This strongly suggests that the part of the wave field that corresponds to free waves could be consistent with Weak Turbulence, for which the wave field is very close to Gaussian statistics.

We then define the filtered frequency spectrum 
%\begin{eqnarray}
%E^{\eta_n}(\omega)=\langle |\widetilde{\eta}_n({\bf x}, \omega)|^2\rangle,
%\end{eqnarray}
{
\begin{eqnarray}
E^{\eta_n}(\omega)=\frac{2}{TL^2}\iint \langle |\widetilde{\eta}_n({\bf k}, \omega)|^2\rangle dk_x dk_y.
\end{eqnarray}
}
%with 
%$\widetilde{\eta_n}({\bf x}, \omega) = \int_0^\infty\int_0^\infty\widetilde{\eta}_n({\bf k}, \omega)dk_xdk_y$.

We observe in fig.~\ref{fig_spectra_w} that the temporal spectrum $E^{\eta}(\omega)$ is well reproduced by that of freely propagating gravity waves $E^{\eta_0}(\omega)$ up to 2-3~Hz. The energy associated to higher frequencies is the combination of freely propagating linear gravity waves and bound waves and is actually dominated by the contribution of the bound waves at frequencies greater than 4~Hz.

\section{Hamiltonian formulation of the Weak Turbulence Theory}
%%%%%%%%%%%%%%%%%%%
\label{sec_hamiltonian}

We recall here some elements of the theoretical background of the WTT applied to surface gravity waves (following~\cite{zakharov_statistical_1999}). 
We consider an inviscid, homogeneous and incompressible fluid in which potential gravity waves propagates on the free surface. The Hamiltonian formulation in spatial Fourier space that describes this system is
\begin{eqnarray}
i\frac{\partial a({\bf k})}{\partial t} = \frac{\delta\mathcal{H}}{\delta a^*({\bf k})},
\label{eq_hamiltonian_form}
\end{eqnarray}
with $a({\bf k})$  being the canonical complex variable defined as
\begin{eqnarray}
\widetilde{\eta}({\bf k}, t) =\sqrt{\frac{k}{2\omega_{LDR}(|{\bf k}|)}}\left[a({\bf k}, t)+a^*({\bf k}, t)\right], 
\end{eqnarray}
and $\mathcal{H}$ the non linear Hamiltonian. The Hamiltonian is classically presented in a power expansion of $a({\bf k})$ as
\begin{eqnarray}
\mathcal{H} &=& \mathcal{H}_{lin} + \mathcal{H}_3 + \mathcal{H}_4 + \cdots,
\end{eqnarray}
with $\mathcal{H}_{lin} = \int \omega_{LDR}(|{\bf k}|)|a({\bf k})|^2d{\bf k}$ being the part of the Hamiltonian associated to the linear dynamics. The non linear part is splitted into $l$-wave interactions contributions defined by
\begin{eqnarray}
\mathcal{H}_{l} &=& \sum_{n\leq l}\frac{1}{n!(l-n)!}\int d{\bf k}_1\cdots d{\bf k}_{l}\left[V^{n, l}\delta_n^l c_n^l(a)\right]
\end{eqnarray}
with $c_{n}^l(a)=a^*({\bf k}_1)\dots a^*({\bf k}_n)a({\bf k}_{n+1})\dots a({\bf k}_{l})$ and $\delta_n^l=\delta({\bf k}_1+\dots+{\bf k}_n-{\bf k}_{n+1}-\cdots-{\bf k}_{l})$ is the Dirac delta function. The coefficients $V^{n, l}$ are expressed in the reference~\cite{zakharov_statistical_1999, krasitskii_reduced_1994} and are not recalled here. 
This formulation highlights the fact that energy transfers are strongly linked to the correlations $c_n^l(a)$ for l-tuples satisfying the spatial resonance ($\delta_n^l$=1). By definition of $\delta_n^l$, the contributions labeled with $(n, l)$ are related to interactions of the type $n\textrm{ waves}\longleftrightarrow l-n\textrm{ waves}$.

In the limit of weak non-linearity, one expects that 
%$c_n^3\ll c_n^4\ll c_n^5\ll\cdots$. 
$\mathcal{H}_{lin} \gg \mathcal{H}_3 \gg \mathcal{H}_4 \gg \cdots$. The negative curvature of the linear dispersion of surface gravity waves makes impossible for a triplet $({\bf k}_1,{\bf k}_2,{\bf k}_3)$ of free waves satisfying the spatial resonance $\pm{\bf k}_1\pm{\bf k}_2\pm{\bf k}_3=0$ to verify also the time resonance $\pm\omega_1\pm\omega_2\pm\omega_3=0$. The consequence, for statistically stationary turbulence, will be a scrambling with time of $\mathcal{H}_{3}$ which will vanish on average and will not contribute to a net energy transfer. This term expresses the non linear coupling between free waves and bound waves. 

Exclusion of the cubic term in the Hamiltonian can be made by a canonical nonlinear transfomation $a({\bf k}, t)\longrightarrow b({\bf k}, t)$ of general structure:
\begin{equation}
a(\mathbf k)=b(\mathbf k)+ b^{(2)}(\mathbf k)+b^{(3)}(\mathbf k)...
\end{equation}
where the terms $b^{(n)}(\mathbf k)$ are of order $n$ in $b(\mathbf k)$ (see Krasitskii~\cite{krasitskii_reduced_1994} for details). In the limit of vanishing non linearity $a(\mathbf k)\approx b(\mathbf k)$.
The terms of order $n>1$ are chosen so that to cancel the cubic contribution in the Hamiltonian, i.e. such that eq.~(\ref{eq_hamiltonian_form}) has the following structure
\begin{eqnarray}
i\frac{\partial b({\bf k})}{\partial t} = \frac{\delta\widetilde{\mathcal{H}}}{\delta b^*({\bf k})}
\label{eq_hamiltonian_form2}
\end{eqnarray}
with the non linear Hamiltonian expressed as 
\begin{eqnarray}
\widetilde{\mathcal{H}}=\widetilde{\mathcal{H}}_{lin}+\widetilde{\mathcal{H}}_4+\widetilde{\mathcal{H}}_5\cdots
\end{eqnarray}
with $\widetilde{\mathcal{H}}_{lin} = \int \omega_{LDR}({\bf k})|b({\bf k})|^2d{\bf k}$. The first non linear contribution $\widetilde{\mathcal{H}}_4$ which appears is related to 4-wave interactions

\begin{equation}
\widetilde{\mathcal{H}}_{4}({\bf k}_1) = \frac{1}{4}\int T({\bf k}_1,{\bf k}_2,{\bf k}_{3},{\bf k}_{4}) b^*({\bf k}_1)b^*({\bf k}_2)b({\bf k}_{3})b({\bf k}_{4})
\delta({\bf k}_1+{\bf k}_2-{\bf k}_{3}-{\bf k}_{4}) d{\bf k}_1\cdots d{\bf k}_{4}.
\label{eq_hamiltonian4wave}
\end{equation}
Here, only the $2\longleftrightarrow 2$ interactions contributions of $\widetilde{\mathcal{H}}_{4}$ are kept since other combinations ($0\longleftrightarrow 4$, $1\longleftrightarrow 3\cdots$) can not satisfy the temporal resonance for the same reasons as for 3-wave interactions. The coefficients $T$ are expressed in ~\cite{zakharov_statistical_1999, krasitskii_reduced_1994} and are not recalled here. These coefficients have the interesting property to cancel out~\cite{webb_non-linear_1978} for unidimensional 4-wave interactions, i.e. for ${\bf k}_1\propto{\bf k}_2\propto{\bf k}_3\propto{\bf k}_4$.
In this formulation, the contribution of bound waves through triple correlations is no more explicitely visible but is actually hidden in the canonical transformation $a({\bf k}, t)\longrightarrow b({\bf k}, t)$ as well as in the terms $\widetilde{\mathcal{H}}_4$, $\widetilde{\mathcal{H}}_5\cdots$. 

From a practical point of view, the characterization of the net energy transfers from experimental measurements is very challenging. The evaluation of $\widetilde{\mathcal{H}}_{4}$ requires to compute a multiple integral over the whole range of wave vectors ${\bf k}_i$ which is barely accessible in experiments. Moreover, the variable $b({\bf k})$ is not a direct observable of the flow. Its evaluation necessitate to compute the canonical transformation which involves a multiple integral over the whole range of wave vectors ${\bf k}_i$ (and expressed in an implicit way in the literature). It requires a very accurate knowledge of the parameters. Furthermore the experimental resolution in $\mathbf k$ is strongly reduced by the size of the images and measurement noise can not be avoided. Unfortunately, it appears totally unrealistic to us to perform such a canonical change of variable from experimental data to remove the contribution of bound waves.

In the present article, we keep the observation variable $\eta$ and describe the energy transfers through high order correlations. 
In the next part~\ref{sec_C3}, we first describe the bound waves contributions to the non linear Hamiltonian through the third order correlation in the temporal frequency space. The correlated spatio-temporal structures are then clearly identified thanks to the third order correlation both in temporal and spatial frequency spaces.
In the subsequent sections, we analyze the 4-wave interactions through the fourth order correlation in the temporal frequency space. The identified bound waves are also expected to contribute to the fourth order correlations and are not related to the net energy transfers discussed by the WTT. We then propose, in part~\ref{sec_filtC4}, a spatio-temporally filtered fourth order correlation as a surrogate of the canonical change of variable $a(\mathbf k)\longrightarrow b(\mathbf k)$ in order to separate the contributions of free and bound waves to the total fourth order correlation.
% 
% As for Zakharov's canonical variable in the previous discussion, we expect a dominance of the third order correlations for water elevation $\eta$ which reflects the link between free and bound waves.
% 
% As no canonical transformation is performed, contributions of bound waves in the correlations is expected to be important both in third order and fourth order correlations. 
% 
%  This correlation in spatial Fourier space lies on a 8 dimensions space $({\bf k}_1, {\bf k}_2, {\bf k}_3, {\bf k}_4)$ which makes it hardly tractable. Since the linear dispersion relation is bijective, we propose in a first step to evaluate four temporal frequency correlations instead. As for Zakharov's canonical variable in the previous discussion, we expect a dominance of the third order correlations for water elevation $\eta$ which reflects the link between free and bound waves. In the same way, those bound waves have a contribution to the fourth order correlations which is not related to the net energy transfers discussed by the WTT. We then propose, in the second part, a spatio-temporally filtered fourth order correlation as a surrogate of the canonical change of variable $a(\mathbf k)\longrightarrow b(\mathbf k)$ in order to separate the contributions of free and bound waves to the total fourth order correlation.

\section{Third order correlation}
\label{sec_C3}
We now investigate the three waves resonances i.e. such that resonant conditions~(\ref{eq:res_cond3_om}) and ~(\ref{eq:res_cond3_k}) are satisfied. 
As previously discussed, we recall that those conditions cannot be satisfied for three free waves (with the additional condition $|{\bf k}_i|=k_{LDR}(\omega_i)$ {for all $i$}). {However the nonlinear dynamics makes possible to have quadratic coupling between Fourier modes that are not free waves. In the correlation estimators that we introduce below, we will probe Fourier components that fulfil the resonance conditions ~(\ref{eq:res_cond3_om}) and ~(\ref{eq:res_cond3_k}) but at least one of the free Fourier modes is not a free wave, i.e. $|{\bf k}_i|\neq k_{LDR}(\omega_i)$ for at least one value of $i$. This is the process that generates the bound waves.}
Since bound waves bring also their contributions to the four waves resonances {that will be discussed in the next section, the description of the coupling between 3 modes} is necessary for the purpose of a physical interpretation of the energy transfers in the next section.
\subsection{Third order frequency correlation}
We investigate the 3-wave resonant interactions using the 3rd-order correlation defined as
% We now investigate the contribution of bound wave to the non linear Hamiltonian using the 3rd-order correlation defined as
\begin{equation}
c_3(\omega_1, \omega_2, \omega_3) =
\langle \widetilde{\eta}^*({\bf x},\omega_1)\widetilde{\eta}({\bf x},\omega_2)\widetilde{\eta}({\bf x},\omega_3)\rangle.
\label{eq:c3}
\end{equation}
{Here again, the brackets $\langle \cdot \rangle$ correspond in practice to an average both over space and over successive temporal windows.}

It must be noted first that for stationary signals, such correlations are non zero only for resonant frequencies such that $\omega_1=\omega_2+\omega_3$. It will be similar for any order correlations of such Fourier amplitudes {thus also for the 4-wave correlations studied in the section \ref{sec_correlation}}. 

Assuming spatial homogeneity and due to the spatial averaged performed in our definition of the estimator, the third order correlation can be rewritten as (see Appendix \ref{app_cor}): 
\begin{equation}
\begin{array}{ll}
c_3(\omega_1, \omega_2, \omega_3) 
% \frac{1}{(2\pi)^{8}}\int_{{\bf k}_1}\int_{{\bf k}_2}\int_{{\bf k}_3}\int_{{\bf k}_4}\left\langle  \widetilde{\eta}^*({\bf k}_1,\omega_1) \widetilde{\eta}^*({\bf k}_2,\omega_2) \widetilde{\eta}({\bf k}_3,\omega_3)\widetilde{\eta}({\bf k}_4,\omega_4) e^{i({\bf k_1}+{\bf k_2}-{\bf k_3}-{\bf k_4})\cdot{\bf x}}\right\rangle d{\bf k}_1d{\bf k}_2d{\bf k}_3d{\bf k}_4\\
=\frac{1}{(2\pi)^{3/2}} \iiint_{{{\bf k}_1}, {\bf k}_2,{\bf k}_3}\left\langle  \widetilde{\eta}^*({\bf k}_1,\omega_1) \widetilde{\eta}({\bf k}_2,\omega_2) \widetilde{\eta}({\bf k}_3,\omega_3) \right\rangle \delta({\bf k_1}-{\bf k_2}-{\bf k_3})d{\bf k}_1d{\bf k}_2d{\bf k}_3,
\label{eq_c3_developped}
\end{array}
\end{equation}
%where $\delta^2({\bf k})=\delta(k_x)\delta(k_y)$ with $\delta(k_\alpha)$ is the Dirac delta function which is coming from the equality $\int e^{ik_\alpha \alpha}d\alpha=2\pi\delta(k_\alpha)$. 
This formulation highlights the fact that the correlation associated to the 3-tuple $(\omega_1, \omega_2, \omega_3)$ probes {indirectly}  all spatial resonances such that ${\bf k}_1={\bf k}_2+{\bf k}_3$. {Actually $c_3(\omega_1, \omega_2, \omega_3)$ contains the cumulative effects of all resonances ${\bf k}_1={\bf k}_2+{\bf k}_3$ compatible with the 3-tuple $(\omega_1, \omega_2, \omega_3)$, and not only an individual triad}. Thus, by nature, the correlation (\ref{eq:c3}) is non zero only for Fourier components that fulfill both resonant conditions on frequencies and wavevectors.
We emphasize that no relation is imposed between $\omega$ and $\mathbf k$ so that the correlation incorporates the effect of all resonant 3-tuples $({\bf k}_1,{\bf k}_2,{\bf k}_3)$ including those with $|{\bf k}_i|\neq k_{LDR}(\omega_i)$. 
Thus it includes the contribution of bound waves into the correlation $c_3$ but possibly other structures. 
In the case of our finite size domain, the Dirac $\delta$-function would be replaced by the following product of cardinal sine function $L^2\sinc(L{\bf e}_x\cdot({\bf k}_{1}-{\bf k}_{2}-{\bf k}_{3}))\sinc(L{\bf e}_y\cdot({\bf k}_{1}-{\bf k}_{2}-{\bf k}_{3}))$ which relaxes to some extent the resonant condition ${\bf k}_1={\bf k}_2+{\bf k}_3$. Hence quasi-resonances may also contribute to the correlation $c_3$. By definition of the cardinal sine function, the largest contributions are for $|({\bf k_1}-{\bf k_2}-{\bf k_3})\cdot{\bf e}_i|\lesssim\pi/L=\Delta_k$ ($i=(x, y)$) which corresponds to our spatial spectral resolution. Hence eq.~(\ref{eq_c3_developped}) remains a good approximation of the measured correlation.

\begin{figure}[!htb]
\centerline{
\includegraphics[width=20cm]{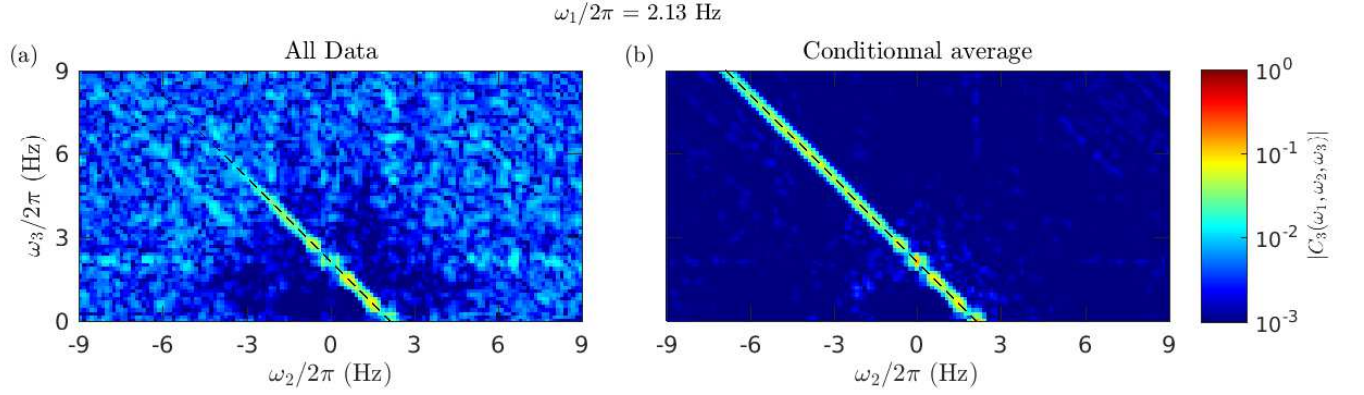}}
\caption{Normalized third order correlation of the temporal Fourier transform of the wave elevation $C_3(\omega_1, \omega_2, \omega_3)$ for $\omega_1/2\pi =2.13$~Hz.
Left: averaged on all data. Right: conditionnally averaged on data for which $|\partial \eta/\partial t|\leq 3.5 v_{rms}$ ($99.8\%$ of total data). Dotted lines indicate the resonant line $\omega_3 = \omega_1-\omega_2$. It shows that spurious events in the stereoscopic reconstruction are responsible for the high background level in (a).}
\label{fig_C3_2freqs}
\end{figure}

We define the normalized third order correlation (bicoherence)
\begin{equation}
C_3(\omega_1, \omega_2, \omega_3) =
\frac{|c_3(\omega_1, \omega_2, \omega_3)|}{\langle|\widetilde{\eta}(\omega_1)|^2\rangle^{1/2}\langle|\widetilde{\eta}(\omega_2)\widetilde{\eta}(\omega_3)|^2\rangle^{1/2}},
\label{eq_C3}
\end{equation}
which is by construction in the range $[0, 1]$. Such tools are quite common in signal processing to identify and quantify non linear coupling between Fourier modes and have been already used to investigate nonlinearity in water waves (see for instance~\cite{elgar:2006hj}). 
Figure~\ref{fig_C3_2freqs}(a) represents the third order correlation $C_3$ in the $(\omega_2, \omega_3)$ plane for $\omega_1/2\pi=2.13$~Hz. {The resonant line $\omega_3=\omega_1-\omega_2$ with a correlation of the order $10^{-1}$ emerges from the fluctuating background for frequencies $\omega_3/2\pi$ up to 6~Hz. For stationary signals, one expects that only the resonant line must be visible. The origin of the fluctuating background lies in spurious measurement points.} Indeed, fig.~\ref{fig_spectra_w} shows a departure at {high frequency} of the temporal spectrum from the one measured from probes measurements. This departure is due to a poorer signal-to-noise ratio of the stereoscopic reconstruction compared to the probes measurements. The signature of the noise {is also clearly visible in the distribution} of the time derivative of wave elevation as seen in fig.~\ref{fig_slope}(c) {(black curve)}. Events associated to $|\partial_t \eta| > 3.5 v_{rms}$ likely corresponds to unphysical signal. {To cure this issue,} we then apply a conditional average on data satisfying the condition $|\partial_t \eta| \leq 3.5 v_{rms}$ which removes only $0.2\%$ of the total volume of data. In practice, for each time window, positions $(x,y)$ for which the condition is not respected are detected and removed from the average. The conditionally averaged third order correlation $C_3$ is represented in fig.~\ref{fig_C3_2freqs}~(b). The resonant line is now two orders of magnitude larger than the {remaining background level} and is extended over the whole range of frequencies. {The background level is now due to the level of statistical convergence}. This illustrate the extreme sensitivity of these correlations to the quality of data. This conditional average will then be systematically adopted in the following.

{As the correlation is non zero only on the resonance line we can now focus on the bicoherence which extracts the correlation only for resonant frequencies:}
\begin{equation}
B(\omega_2, \omega_3)=C_3(\omega_2 +\omega_3, \omega_2, \omega_3).
\label{eq_B}
\end{equation}
\begin{figure}[!htb]
\centerline{
\includegraphics[width=7cm]{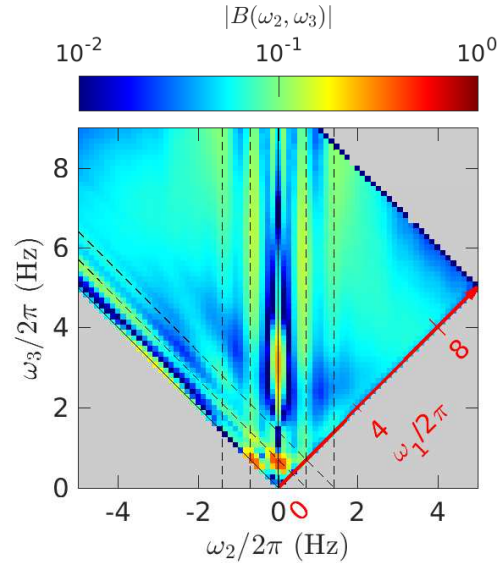}}
\caption{(Top Left) {Example of the} bicoherence of the temporal Fourier transform of the wave elevation $B(\omega_2, \omega_3)$ defined by eqs~(\ref{eq_C3}) and~(\ref{eq_B}). This correlation intrinsically have the following symmetries 
$B(\omega_2, \omega_3)=B(\omega_3, \omega_2)=B(-\omega_2,-\omega_3)$. For sake of simplicity, the non-redundant part of the correlation is represented here i.e. $\omega_3>\omega_2$ and $\omega_1=\omega_2+\omega_3>0$. 
Dashed lines indicate $\omega_1, \omega_2= \pm n \omega_0$ with $n=1, 2$. {One can see several components in this picture: discrete lines  highlighted by the dashed lines and a diffuse continuum. All these contributions are due to bound waves as free wave have no 3-wave resonance possible.}}
\label{fig_C3}
\end{figure}
The bicoherence owns the following obvious symmetries
\begin{eqnarray}
B(\omega_2, \omega_3)&=B(\omega_3, \omega_2)&=B(-\omega_2,-\omega_3).
\end{eqnarray}
Fig.~\ref{fig_C3} represents the minimalistic (i.e. non redundant) part of the bicoherence in the plane $(\omega_2, \omega_3)$ which corresponds to the region defined as $\omega_3>\omega_2$ and $\omega_1=\omega_2+\omega_3>0$. We observe a diffuse region for $\omega_3/2\pi>3$~Hz with a correlation slightly lower than $10^{-1}$. The bicoherence is higher for identified frequencies $\omega_1, \omega_2, \omega_3 = \pm n \omega_0$ with $n=1, 2$ with a correlation larger than $10^{-1}$. {These lines correspond to the bound waves $k_n(\omega)$ identified before.}This signature gives a first clue of the correlation between free and bound waves. However, at this stage, we are not able to discriminate the contribution of free waves, bound waves, singularities or correlated noise to the observed level of coherence.  Indeed, as previously discussed, any Fourier modes verifying the resonance conditions would contribute to $C_3$.

\subsection{Third order spatio-temporal correlation}
\begin{figure}[!htb]
\centerline{
\includegraphics[width=11cm]{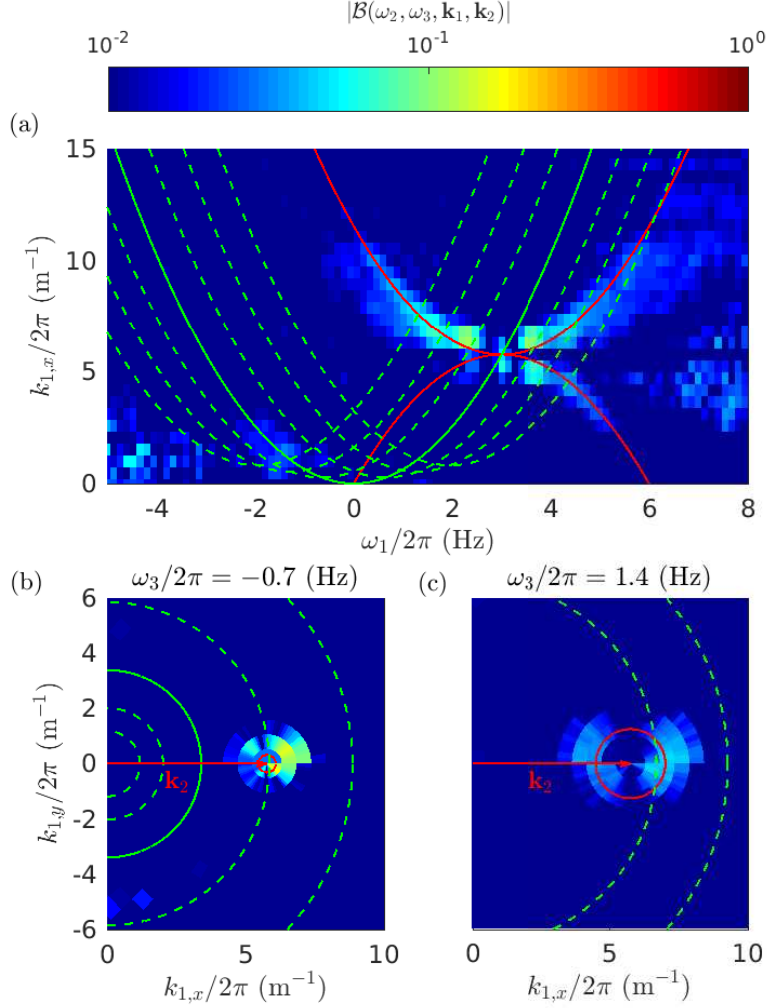}}
\caption{Cut of the bicoherence $\mathcal{B}$ for $\omega_2/2\pi=3$~Hz and ${\bf k}_2 = k_{LDR}(\omega_2){\bf e}_x$ in the planes defined by (a): $k_{1, y}=k_{3, y}=0$ , (b): $\omega_3=\omega_1-\omega_2=-0.7~2\pi$~rad.s$^{-1}$ and (c): $\omega_3=1.4~2\pi$~rad.s$^{-1}$ (c). {The red line corresponds to the linear dispersion relation ${\bf k}_1 = \mathbf k_{LDR}(\omega_1)$ shifted by $\omega_2$ in frequency and $\bf k_2$ in wave vector, i.e. ${\bf k}_1 = {\bf k}_2+\mathbf k_{LDR}(\omega_1-\omega_2)$. This equation corresponds to a surface in the 3D space $(\mathbf k_1,\omega_1)$ which cuts in the various subfigures gives the red lines. The green continuous lines are cuts of the linear dispersion relation. The dashed green lines are cuts of the bound waves dispersion relations $\mathbf k_n(\omega)$ for $n=-3...3$. Wave resonances lie at the intersection of red and green lines. In (b) and (c) the red arrows represent $\mathbf k_2$. One clearly sees that high levels of bicoherence are observed near the intersections of the free wave dispersion relation (red) and the bound waves (green), highlighting the contribution of the latter. No such intersection (3-wave resonances) exists for free waves only except for trivial solutions $\mathbf k_1=\mathbf k_2$ and $\omega_1=\omega_2$. Thus the observed non zero bicoherence is clearly attributed to wave resonances involving at least one bound wave.}}
\label{fig_C3_kw}
\end{figure}
The natural next step to go further is then to look at the spatio-temporal third order correlation
\begin{equation}
C_3^{k\omega}(\omega_1, \omega_2, \omega_3, {\bf k}_1, {\bf k}_2, {\bf k}_3) =
\frac{\langle \widetilde{\eta}^*({\bf k}_1,\omega_1)\widetilde{\eta}({\bf k}_2,\omega_2)\widetilde{\eta}({\bf k}_3,\omega_3)\rangle}{\langle|\widetilde{\eta}({\bf k}_1,\omega_1)|^2\rangle^{1/2}\langle|\widetilde{\eta}({\bf k}_2,\omega_2)\widetilde{\eta}({\bf k}_3,\omega_3)|^2\rangle^{1/2}}
\label{eq:c3kw}
\end{equation}
with $\langle \rangle$ corresponding to the use of the usual Welch method (in time).
This correlation has the advantage to separate the correlations between well distinguished spatio-temporal structures in the space $(\omega, {\bf k})$. 
We focus on temporal and spatial resonant conditions through the bicoherence
\begin{eqnarray}
\mathcal{B}(\omega_2, \omega_3, {\bf k}_2, {\bf k}_3)&=C_3^{k\omega}(\omega_2 + \omega_3, \omega_2, \omega_3, {\bf k}_2+{\bf k}_3, {\bf k}_2, {\bf k}_3).
\end{eqnarray}

Figure~\ref{fig_C3_kw} represents the bicoherence for $\omega_2/2\pi=3$~Hz and ${\bf k}_2 = k_{LDR}(\omega_2){\bf e}_x$ in the planes $k_{1, y}=k_{3, y}=0$ (a), $\omega_3=\omega_1-\omega_2=-0.7\times2\pi$~rad.s$^{-1}$ (b) and $\omega_3=1.4\times2\pi$~rad.s$^{-1}$ (c). The red curve is given by $k_{2, x}\pm k_{LDR}(\omega_3)$ and the green curves are given by $k_n (\omega_2+\omega_3)$, with $n=\pm 1,2$. The points at the crossing between red and green curves indicate graphically solutions of {the resonance} equations~(\ref{eq:res_cond3_om}) and~(\ref{eq:res_cond3_k}). {The intersection of the red line and the continuous green line corresponds to resonant free waves. As expected no resonance exists beyond the trivial one for which $\omega_1=\omega_2$ and $\omega_3=0$.} We observe significant correlation far above the background correlation at the vicinity of {the intersections of dashed green lines and the red line that correspond to resonances with bound waves}.
We turn back to the total third order bicoherence $B$ (figure.~\ref{fig_C3}). We are now able to claim that the observed coherence is dominated by correlations of the type bound-free-free waves. We notice however that correlations larger that the background noise for $\mathcal{B}$ are observed outside the predicted solutions which means that other non-identified structure also contribute to $B$.

\subsection{Filtered third order time correlation}
%%%%%%%%%%%%%%%%%%%%%%%%%
\begin{figure}[!htb]
\centerline{
\includegraphics[width=12cm]{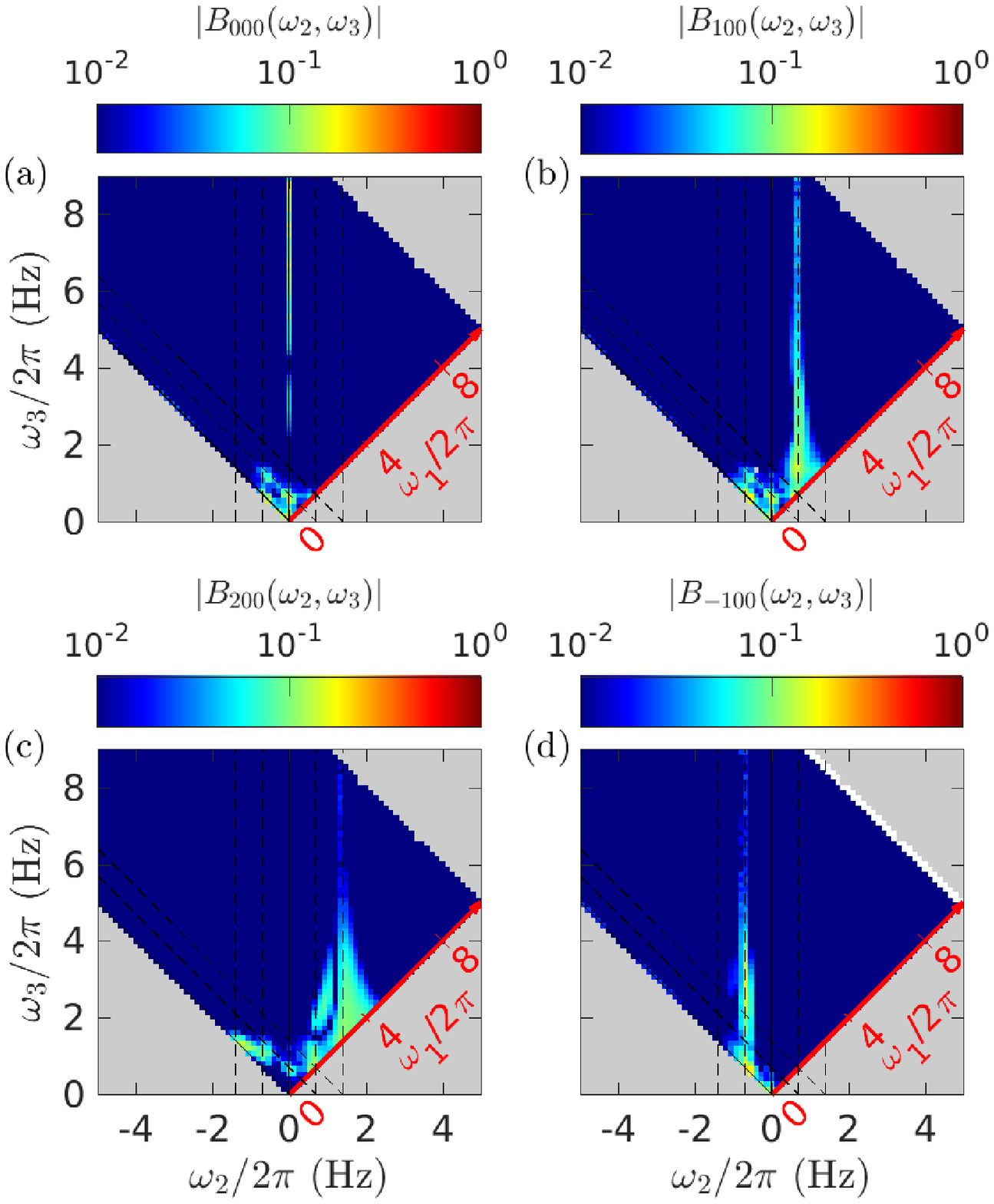}}
\caption{(Top Left) Bicoherence of the filtered temporal Fourier transform of the wave elevation $B_{n_100}(\omega_2, \omega_3)$ with $n_1 = 0,1,2,-1$. This correlation intrinsically have symmetries (see text). For sake of simplicity, the minimalistic part of the correlation is represented here. 
Dashed lines indicate $\omega_1, \omega_2= \pm n \omega_0$ with $n=1, 2$. {No correlation is seen in $B_{000}$ (except at zero frequency, which is not relevant) as expected from the fact that no 3-free wave resonance exists. By contrast, high correlation levels can be observed in $B_{100}$, $B_{200}$ and $B_{-100}$ around the vertical dashed line that corresponds to the relevant bound wave.}}
\label{fig_C3_waves}
\end{figure}
In the previous section we have seen that the main contributions to the third order correlation involve {only resonances involving at least one bound wave}. In addition, as discussed in section~\ref{sec_structure}, it is reasonable to approximate the spatio-temporal signal by $\widetilde{\eta}({\bf k}, \omega)\approx\sum_{n}\widetilde{\eta}_n({\bf k}, \omega)$ as most of the energy lies on these lines (although not all of it). We recall that $\eta_0$ corresponds to free waves and the other contributions to bound waves involving the forcing peak. We then decompose the third order correlation as follows:

%\begin{widetext}
\begin{equation}
c_3(\omega_1, \omega_2, \omega_3) =\sum_{n_1}\sum_{n_2}\sum_{n_3} c_{n_1n_2n_3}(\omega_1, \omega_2, \omega_3)+ MT(\omega_1, \omega_2, \omega_3),
\end{equation}
%\begin{equation}
%\text{with }
with $c_{n_1n_2n_3}(\omega_1, \omega_2, \omega_3) = \langle\widetilde{\eta}_{n_1}^*({\bf x},\omega_1)\widetilde{\eta}_{n_2}({\bf x},\omega_2)\widetilde{\eta}_{n_3}({\bf x},\omega_3) \rangle$
%\end{equation}
%\end{widetext}
being the third order correlation between the components $(n_1,n_2,n_3)$. $MT$ corresponds to missing terms involving other non-identified spatio-temporal contributions. Similarly to the correlation $c_3$, the correlation $c_{n_1n_2n_3}(\omega_1, \omega_2, \omega_3)$ relates to the $n_1\longleftrightarrow(n_2, n_3)$ interactions, i.e. it probes only the spatial resonances such that 
\begin{equation}
k_{n_1}(\omega_2+\omega_3){\bf e}_1=
k_{n_2}(\omega_2){\bf e}_2 +
k_{n_3}(\omega_3){\bf e}_3
\label{eq:3_waves_resonances_or_bound_waves}
\end{equation}
that involve only waves on the specific free or bound waves dispersion relations characterized by the integers $(n_1,n_2,n_3)$.

We note that the specific case $c_{000}$ ($n_1=n_2=n_3=0$), probes the three free waves interactions. In the case of one or more non-zero $n_i$, $c_{n_1n_2n_3}$ probes the correlation involving one or more identified bound waves.

We define accordingly the $(n_1,n_2,n_3)$ bicoherence
\begin{equation}
B_{n_1n_2n_3}(\omega_2, \omega_3) = 
C_{n_1n_2n_3}(\omega_2+\omega_3, \omega_2, \omega_3),
\end{equation}
with
\begin{equation} 
C_{n_1n_2n_3}(\omega_1, \omega_2, \omega_3) = 
\frac{c_{n_1n_2n_3}(\omega_1, \omega_2, \omega_3)}{\langle|\widetilde{\eta}_{n_1}({\bf x},\omega_1)|^2\rangle^{1/2}\langle|\widetilde{\eta}_{n_2}({\bf x},\omega_2)\widetilde{\eta}_{n_3}({\bf x},\omega_3)|^2\rangle^{1/2}},
\end{equation}
defined in the range $[0, 1]$.

These bicoherences actually offer a compact but also a partial way (along the red and green curves of the figure~\ref{fig_C3_kw}) to represent the spatio-temporal bicoherence $\mathcal{B}$. However, we will see in the next section that a generalization of this tool to the fourth order correlation will bring an elegant way to discuss clearly the energy transfers.
Figure~\ref{fig_C3_waves} represents $B_{n_100}$, for $n_1=0,1,2,-1$ in the $(\omega_2, \omega_3)$ plane. {First, $B_{000}$ is zero due to the fact that no 3-wave resonance exist among free waves. In the other cases, significant correlations are concentrated along the bound waves $\omega_2=n_1 \omega_0$ as expected from the way $B_{n_100}$ is constructed.} 

{Following the theroy~\cite{hasselmann_non-linear_1962}}, the three-wave correlations discussed above, even if they {can reach relatively high levels}, are not expected to contribute to the global energy fluxes. Indeed the theory predicts that only resonant or quasi-resonant free waves contribute to the energy cascade. {The overall added contribution of all triads involving bound waves is supposed to have no net contribution to the energy flux. This is why the 3-wave coupling (involving bound waves) can be removed by the canonical change of variables to keep only 4-free-wave coupling. Unfortunately we do not have access experimentally to energy fluxes to check if the bound modes contribute or not to the energy cascade. Nevertheless the core process of the energy cascade is the 4-wave coupling, thus, in the following part, we focus on fourth order wave interactions.}

\section{Fourth order frequency correlation}
%%%%%%%%%%%%%%%%%%%%%
\label{sec_correlation}

\subsection{Definitions}
%%%%%%%%%%%%%%

We now investigate the 4-wave resonant interactions using the 4th-order correlation defined as
\begin{equation}
c_4(\omega_1, \omega_2, \omega_3, \omega_4) =
\langle \widetilde{\eta}^*({\bf x},\omega_1)\widetilde{\eta}^*({\bf x},\omega_2)\widetilde{\eta}({\bf x},\omega_3)\widetilde{\eta}({\bf x},\omega_4) \rangle.
\label{eq:c4}
\end{equation}
For the same reasons that those expressed in the previous section {for equation (\ref{eq_c3_developped})}, the correlation associated to the 4-tuple $(\omega_1, \omega_2, \omega_3, \omega_4)$ is non zero only for $\omega_1+\omega_2=\omega_3+\omega_4$ (due to stationnarity) and it also probes all spatial resonances such that ${\bf k}_1+{\bf k}_2={\bf k}_3+{\bf k}_4$ (due to spatial homogeneity). It should be be noted that the correlation incorporates all resonant 4-tuples $({\bf k}_1,{\bf k}_2,{\bf k}_3,{\bf k}_4)$ including those with $|{\bf k}_i|\neq k_{LDR}(\omega_i)$.  Thus it includes the contribution of bound waves into the correlation $c_4$ but possibly other structures (as well as noise). As discussed above, the concentration of energy observed in fig.~\ref{fig_spectra} on the linear dispersion has a finite width associated to nonlinear spectral widening. Thus the frequency of free waves can be written as $\omega_{LDR}(\mathbf k)+\delta \omega_{NL}(\mathbf k)$ where $\delta \omega_{NL}(\mathbf k)$ is a detuning associated to the nonlinear effects. A consequence is that even though the correlations probes the exact resonances conditions ${\bf k}_1+{\bf k}_2={\bf k}_3+{\bf k}_4$ and $\omega_1+\omega_2=\omega_3+\omega_4$, resonant waves with a slight detuning are actually taken into account. It means that the correlation (\ref{eq:c4}) is also sensitive to approximate resonances of free waves.

We define the normalized fourth order correlation  
\begin{equation}
C_4(\omega_1, \omega_2, \omega_3, \omega_4) =
\frac{|c_4(\omega_1, \omega_2, \omega_3, \omega_4)|}{\langle|\widetilde{\eta}(\omega_1)\widetilde{\eta}(\omega_2)|^2\rangle^{1/2}\langle|\widetilde{\eta}(\omega_3)\widetilde{\eta}(\omega_4)|^2\rangle^{1/2}},
\end{equation}
which is by construction in the range $[0, 1]$. 

\begin{figure}[!htb]
\centerline{
\includegraphics[width=11cm]{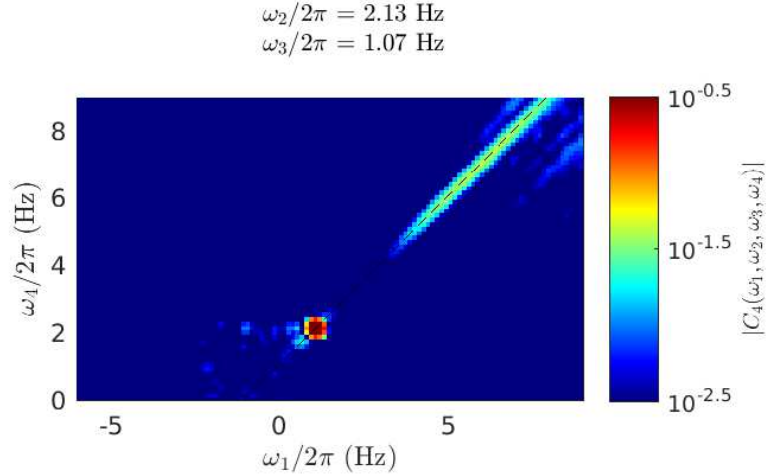}}
\caption{Normalized fourth order correlation of the temporal Fourier transform of the wave elevation $C_4(\omega_1, \omega_2, \omega_3, \omega_4)$ for $\omega_2/2\pi =2.13$~Hz and $\omega_3/2\pi =1.07$~Hz. {The black dashed line is the resonant line $\omega_4 = \omega_2-\omega_3+\omega_1$. A line of correlation emerges on the resonant line as expected for a stationary signal. The red dot is a special case $\omega_1=\omega_3$ and $\omega_2=\omega_4$ for which one expects trivially a correlation level close to 1.}}
\label{fig_C4_uncond_vs_cond}
\end{figure}
Figure~\ref{fig_C4_uncond_vs_cond}~ represents the fourth order correlation $C_4$ in the $(\omega_1, \omega_4)$ plane for $\omega_2/2\pi =2.13$~Hz and $\omega_3/2\pi =1.07$~Hz. We observe that the resonant line $\omega_4 = \omega_2-\omega_3+\omega_1$ clearly emerges from the noise level as expected for a statistically stationary signal. 

In the following we then focus on this resonant condition through the tricoherence 
\begin{equation}
T(\omega_2, \omega_3, \omega_4)=C_4(-\omega_2 +\omega_3+ \omega_4, \omega_2, \omega_3, \omega_4).
\end{equation}

\begin{figure}[!htb]
\centerline{
\includegraphics[width=20cm]{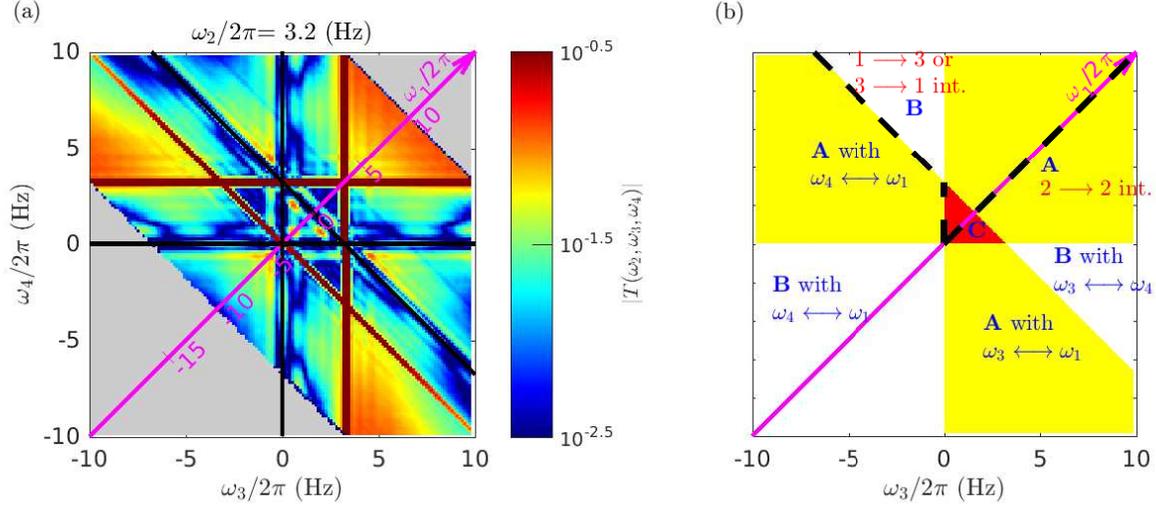}}
\vspace{-1.5cm}\caption{Explanation of symmetries of the tricoherence. (a) Example of a tricoherence estimation of the temporal Fourier transform of the wave elevation $T(\omega_2, \omega_3, \omega_4)$ for $\omega_2/2\pi =3.2$~Hz. (b) sketch of the various zones in the $(\omega_3,\omega_4)$ plane that are equivalent due symmetries under permutations of indices in the definition of the correlation. Regions A (yellow) are equivalent and correspond to $2\leftrightarrow2$ interactions of waves (bound or free). Similarly, regions B are equivalent and correspond to $3\leftrightarrow1$ wave interactions. Region C also correspond to $3\leftrightarrow1$ interactions. The correlation picture is also equivalent by symmetry around the main diagonal. Thus the area surrounded by the thick black dashed line is non redundant and only this subregion will be shown in the next figures.} 
\label{fig_C4_sym}
\end{figure}

The tricoherence shows several symmetries (fig.~\ref{fig_C4_sym}). By construction, obvious symmetries are
\begin{eqnarray}
T(\omega_2, \omega_4, \omega_3)&=&T(\omega_2, \omega_3, \omega_4)\\
T(-\omega_2, -\omega_4, -\omega_3)&=&T(\omega_2, \omega_3, \omega_4).
\end{eqnarray}
A less obvious symmetry is the following: $T(\omega_2, \omega_3, \omega_4)= T(\omega_2, \omega_2-\omega_4-\omega_3, \omega_4)$. This is due to the symmetry of the correlation $c_4(-\omega_1, \omega_2, -\omega_3, \omega_4)=c_4(\omega_3, \omega_2, \omega_1, \omega_4)$. Thus the full map of tricoherence is highly redundant as seen in fig.~\ref{fig_C4_sym}. Due to a combination of all mentioned symetries there is a 3-fold symmetry that can be obtained by permutations over specific pairs of frequency in $c_4$. Consequently the 3 regions labelled A (yellow in fig.~\ref{fig_C4_sym}(b)) are equivalent as well as the three regions labeled B. Region A corresponds to $2\longleftrightarrow 2$ interactions and regions B and C to $3\longleftrightarrow 1$ interactions. In the following only a reduced part of the tricoherence map will be shown that contains the part above the main diagonal of the image made of half of region A located at the top-right, top region B and region C (see fig.~\ref{fig_C4_sym}).

\subsection{Structure of the full correlation map}
%%%%%%
\begin{figure}[!htb]
\centerline{
\includegraphics[width=17cm]{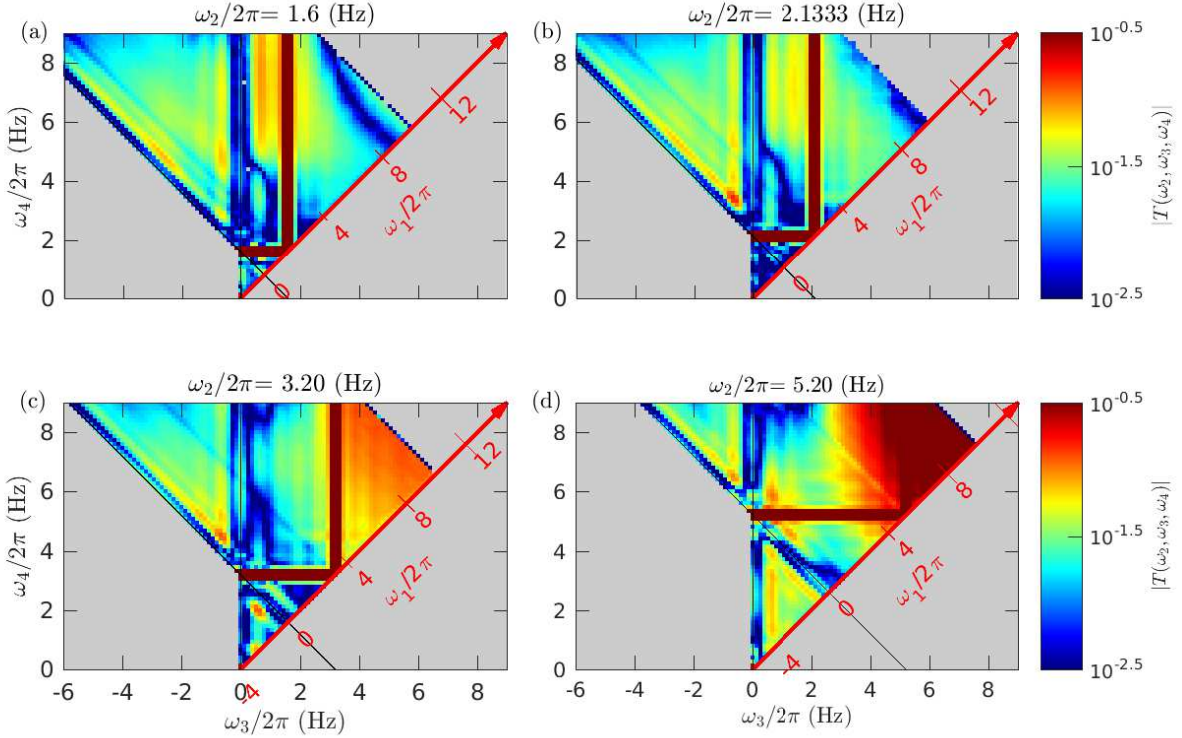}}
\caption{Tricoherence of the temporal Fourier transform of the wave elevation $T(\omega_2, \omega_3, \omega_4)$ for $\omega_2/2\pi =(1.6, 2.13, 3.2, 5.2)$~Hz {((a), (b), (c), (d) respectively).}
This correlation intrinsically have symmetries. For sake of simplicity, the minimalistic part of the correlation is represented here (see fig.~\ref{fig_C4_sym}). 
Black lines indicate $\omega_1, \omega_3=0$. {Distinct regions of significant correlation levels are visible. First, vertical and horizontal lines of correlation level equal to 1 (in red colors) due to trivial cases (corresponding to the red dot in fig.\ref{fig_C4_uncond_vs_cond}) and second, more faint lines in yellow colors and third a continuous variation in the background (see text for descriptions)}.} %Dotted lines indicate $\omega_1, \omega_3, \omega_4=\pm\omega_0$.}
\label{fig_C4}
\end{figure}

Figure~\ref{fig_C4} represents the tricoherence $T(\omega_2, \omega_3, \omega_4)$ in the $(\omega_3, \omega_4)$ plane for 4 distinct values of $\omega_2/2\pi=1.6$, $2.13$, $3.2$ \& $5.2$~Hz. We stress that the level of statistical convergence is about $3.10^{-2}$ so that the range of shown magnitude of correlations corresponds to converged statistics. {It can be seen that the level} of coherence is quite high in some regions of the frequency space (over $0.1$).

The most visible feature is an horizontal line and a vertical line at a level of coherence very close to 1 (saturated red color). They correspond to the cases $\omega_3=\omega_2$ and $\omega_4=\omega_2$ for which the tricoherence is trivially equal to 1. The fact that these lines are well dominating over the background of coherence is a support for the relevance of weak turbulence. Indeed it suggests that the correlation $c_4$ can be written as a development in cumulants
\begin{multline}
c_4(\omega_1, \omega_2, \omega_3, \omega_4)=
\langle|\widetilde{\eta}(\omega_1)|^2\rangle\langle\widetilde{\eta}(\omega_2)|^2\rangle\delta(\omega_1-\omega_3)\delta(\omega_2-\omega_4)+\\
\langle|\widetilde{\eta}(\omega_1)|^2\rangle\langle\widetilde{\eta}(\omega_2)|^2\rangle\delta(\omega_1-\omega_4)\delta(\omega_2-\omega_3)+\\
\langle|\widetilde{\eta}(\omega_1)|^2\rangle\langle\widetilde{\eta}(\omega_3)|^2\rangle\delta(\omega_1+\omega_2)\delta(\omega_3+\omega_4)+Q(\omega_1, \omega_2, \omega_3, \omega_4)
\end{multline}
where $Q$ is a cumulant which is much weaker than the product $\langle|\widetilde{\eta}(\omega_1)|^2\rangle\langle\widetilde{\eta}(\omega_2)|^2\rangle$ (thus a tricoherence level much smaller than 1). This constitutes the first step of the development of the WTT~\cite{zakharov_kolmogorov_1992,newell_wave_2001}. The fact that we see that these lines indeed dominate strongly over the weaker background level of correlation appears consistent with this approximation. {Thus this observation is a strong support of the fact that our system is inherently weakly nonlinear, which is the starting point of the development of the theory. Although the average steepness} of the waves is not that small, the WTT framework may remain relevant. However the interpretation of the correlation picture  is made quite complex due to the contributions of bound waves. For instance in fig.~\ref{fig_C4} the correlation is non zero in the region $\omega_3<0$. As seen in fig.~\ref{fig_C4_sym}, this corresponds to region B associated to $3\leftrightarrow1$ interactions. Such interactions are not possible for free waves due to the curvature of the dispersion relation. Thus it is clear that the observed level of correlation in region B is due only to bound waves. One can see in fig.~\ref{fig_C4} that the level of correlation in region B is of similar order of magnitude than that observed in region A (for $\omega_2/2\pi=1.6$ and $2.13$~Hz in particular), {in which coupling among free waves is possible}.
At this stage, it is unclear whether the observed level of correlation observed in region A is due to free or bound waves (the analysis developed in the appendix \ref{Reson} actually suggests that the whole picture is dominated by bound waves). This ambiguity is due to the fact that our observation variable $\eta$ is not the canonical variable $b$.
The natural next step to go further would be to look at the spatio-temporal fourth order correlation
\begin{equation}
c_4^{k\omega}(\omega_1, \omega_2, \omega_3, \omega_4, {\bf k}_1, {\bf k}_2, {\bf k}_3, {\bf k}_4) =
\langle \widetilde{\eta}^*({\bf k}_1,\omega_1)\widetilde{\eta}^*({\bf k}_2,\omega_2)\widetilde{\eta}({\bf k}_3,\omega_3)\widetilde{\eta}({\bf k}_4,\omega_4) \rangle
\label{eq:c4kw}
\end{equation}
with $\langle \rangle$ corresponding to the use of the usual Welch method.
This estimator has the advantage to separate the correlations between well distinguished spatio-temporal structures {(free waves vs bound waves)} in the plane $(\omega, {\bf k})$. Howerver, its visualisation is made difficult by its huge dimensionality. Moreover, its convergence is strongly reduced compared to $c_4$ by the absence of spatial average and thus it would require a tremendous amount of data. 
To go further in the analysis and circumvent this difficulty, we rather perform the same space-time filtering that we used above for 3-wave correlations. This will enable us to observe correlations of only free wave and to get more insight on the correlation of the full field in terms of bound waves.

\section{Filtered fourth order time correlation}
%%%%%%%%%%%%%%%%%%%%%%%%%
\label{sec_filtC4}

\subsection{Definitions}
%%%%%%

\begin{figure}[!htb]
\centerline{
\includegraphics[width=17cm]{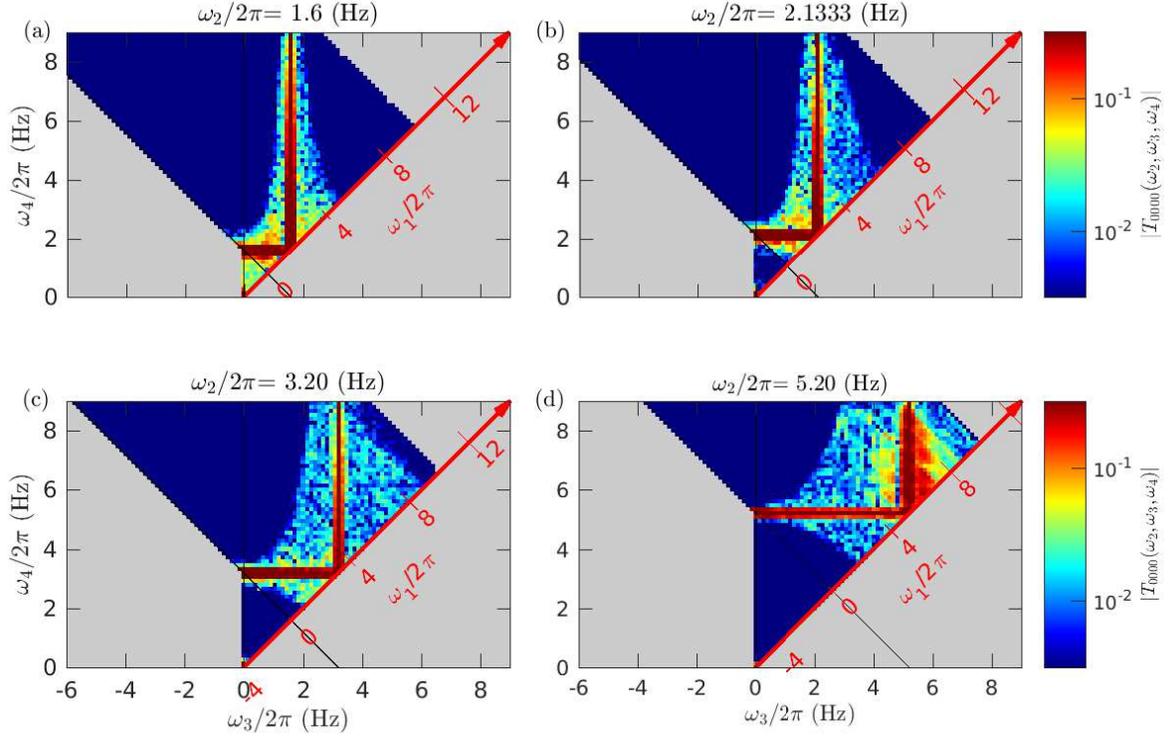}}
\caption{Tricoherence of the filtered temporal Fourier transform of the wave elevation  $T_{0000}(\omega_2, \omega_3, \omega_4)$ 
for $\omega_2/2/\pi =(1.6, 2.13, 3.2, 5.2)$~Hz {((a), (b), (c), (d) respectively). It corresponds to 4-free wave correlations}. This correlation intrinsically have symmetries. For sake of simplicity, 
the minimalistic part of the correlation is represented here. Black lines indicate $\omega_1, \omega_3=0$. {The levels and the map of the correlation are very different than for the unfiltered signal of fig.\ref{fig_C4}. In particular the correlation level is much weaker. See text for more details.}}
%Dotted lines indicate $\omega_1, \omega_3, \omega_4=\pm\omega_0$.}
\label{fig_C4_LDR}
\end{figure}

Justified by the spatio-temporal structure discussed in the section~\ref{sec_structure} and by the dominance of identified bound waves to the third order correlation, we propose to decompose the fourth order correlation into various contributions involving free and bound waves as follows:

\begin{widetext}
\begin{equation}
c_4(\omega_1, \omega_2, \omega_3, \omega_4) =\sum_{n_1}\sum_{n_2}\sum_{n_3}\sum_{n_4} c_{n_1n_2n_3n_4}(\omega_1, \omega_2, \omega_3, \omega_4)+ MT(\omega_1, \omega_2, \omega_3, \omega_4),
\end{equation}
\begin{equation}
\text{with }c_{n_1n_2n_3n_4}(\omega_1, \omega_2, \omega_3, \omega_4) = \langle\widetilde{\eta}_{n_1}^*({\bf x},\omega_1)\widetilde{\eta}_{n_2}^*({\bf x},\omega_2)\widetilde{\eta}_{n_3}({\bf x},\omega_3)\widetilde{\eta}_{n_4}({\bf x},\omega_4) \rangle
\end{equation}
\end{widetext}
being the fourth order correlation between the filtered components $(n_1,n_2,n_3,n_4)$ along $k_{n_i}(\omega)$. $MT$ corresponds to missing terms involving other non-specified bound waves. Similarly to the correlation $c_4$, the correlation $c_{n_1n_2n_3n_4}(\omega_1, \omega_2, \omega_3, \omega_4)$ relates to the $(n_1, n_2)\longleftrightarrow(n_3, n_4)$ interactions, i.e. it probes only the spatial resonances such that 
\begin{equation}
k_{n_1}(-\omega_2+\omega_3+\omega_4){\bf e}_1+
k_{n_2}(\omega_2){\bf e}_2 =
k_{n_3}(\omega_3){\bf e}_3+
k_{n_4}(\omega_4){\bf e}_4,
\label{eq:4_waves_resonances_or_bound_waves}
\end{equation}
that involve only waves on the specific free or bound waves dispersion relations characterized by the integers $(n_1,n_2,n_3,n_4)$.

We note that the specific case $c_{0000}$ ($n_1=n_2=n_3=n_4=0$), for which the latter equation is strictly equivalent to eq.~(\ref{eq:4_waves_resonances}), probes the four free waves interactions. In the case of one or more non-zero $n_i$, $c_{n_1n_2n_3n_4}$ probes the correlation involving one or more identified bound waves.

We define accordingly the $(n_1,n_2,n_3,n_4)$ tricoherence
\begin{equation}
T_{n_1n_2n_3n_4}(\omega_2, \omega_3, \omega_4) = 
C_{n_1n_2n_3n_4}(-\omega_2+\omega_3+\omega_4, \omega_2, \omega_3, \omega_4),
\end{equation}
with
\begin{equation} 
C_{n_1n_2n_3n_4}(\omega_1, \omega_2, \omega_3, \omega_4) = 
\frac{c_{n_1n_2n_3n_4}(\omega_1, \omega_2, \omega_3, \omega_4)}{\langle|\widetilde{\eta}_{n_1}({\bf x},\omega_1)\widetilde{\eta}_{n_2}({\bf x},\omega_2)|^2\rangle^{1/2}\langle|\widetilde{\eta}_{n_3}({\bf x},\omega_3)\widetilde{\eta}_{n_4}({\bf x},\omega_4)|^2\rangle^{1/2}},
\end{equation}
defined in the range $[0, 1]$.

\subsection{Free waves}
%%%%%%

Figure~\ref{fig_C4_LDR} shows $T_{0000}(\omega_2, \omega_3, \omega_4)$ for $\omega_2/2\pi=(1.6,2.13,3.2,5.2)$~Hz. As expected, the tricoherence $T_{0000}$ is exactly equal to zero for $(\omega_3, \omega_4)$ couples for which no solution of the resonant equations exists. We observe a converged level of coherence larger than $10^{-3}$ around couples $(\omega_3, \omega_4)$ for which resonant free waves exists. Except for the trivial lines $\omega_3=\omega_2$ or $\omega_4=\omega_2$ we identify a significative correlation of the order of $10^{-1}$ for $\omega_3, \omega_4\sim \omega_2$ for $\omega_2/2\pi\leq2.13$~Hz. This is actually the range of frequency where the free waves are dominant in the temporal spectrum (see Fig.~\ref{fig_spectra_w}). The dominant wave-wave interactions seem to be local in frequency. This is actually expected in our case since the temporal spectrum is not very wide. We then observe a significative correlation for $\omega_3\sim\omega_2$ for $\omega_2/2\pi=5.2$~Hz. This correlation should be most likely due to noise since free waves have a negligible energy at such high frequency.

{We observe a very weak level of the coherence $T_{0000}$ (0.1 at its maximum and much weaker in most of the pictures). $T_{0000}$ is much weaker than the full coherence $T$. We recall that the coherence is a measurement of the nonlinear activity. The high level of $T$ says that bound modes are very active in terms of nonlinearity. However the theory claims that although bound modes can be active, they do not contribute to the overall energy cascade. The cascade results only from coupling among 4-free waves. By filtering out the bound waves we remove the activity of the bound waves and keep only the activity of the free waves in $T_{0000}$ in the spirit of the canonical change of variable recalled above. The observed weakness of $T_{0000}$ is thus consistent with a weak coupling of free waves with is at the core of the multiscale expansion of the WTT. Thus although the wave field may appear strongly nonlinear with a large steepness ($\epsilon=11\%$), the steepness is largely due to the bound modes. At the level of free waves, the nonlinear activity is much weaker and thus our observation supports the core of the weak turbulence theory.}

\subsection{Resonances with bound waves}
%%%%%%
\begin{figure}[!htb]
\centerline{
\includegraphics[width=17cm]{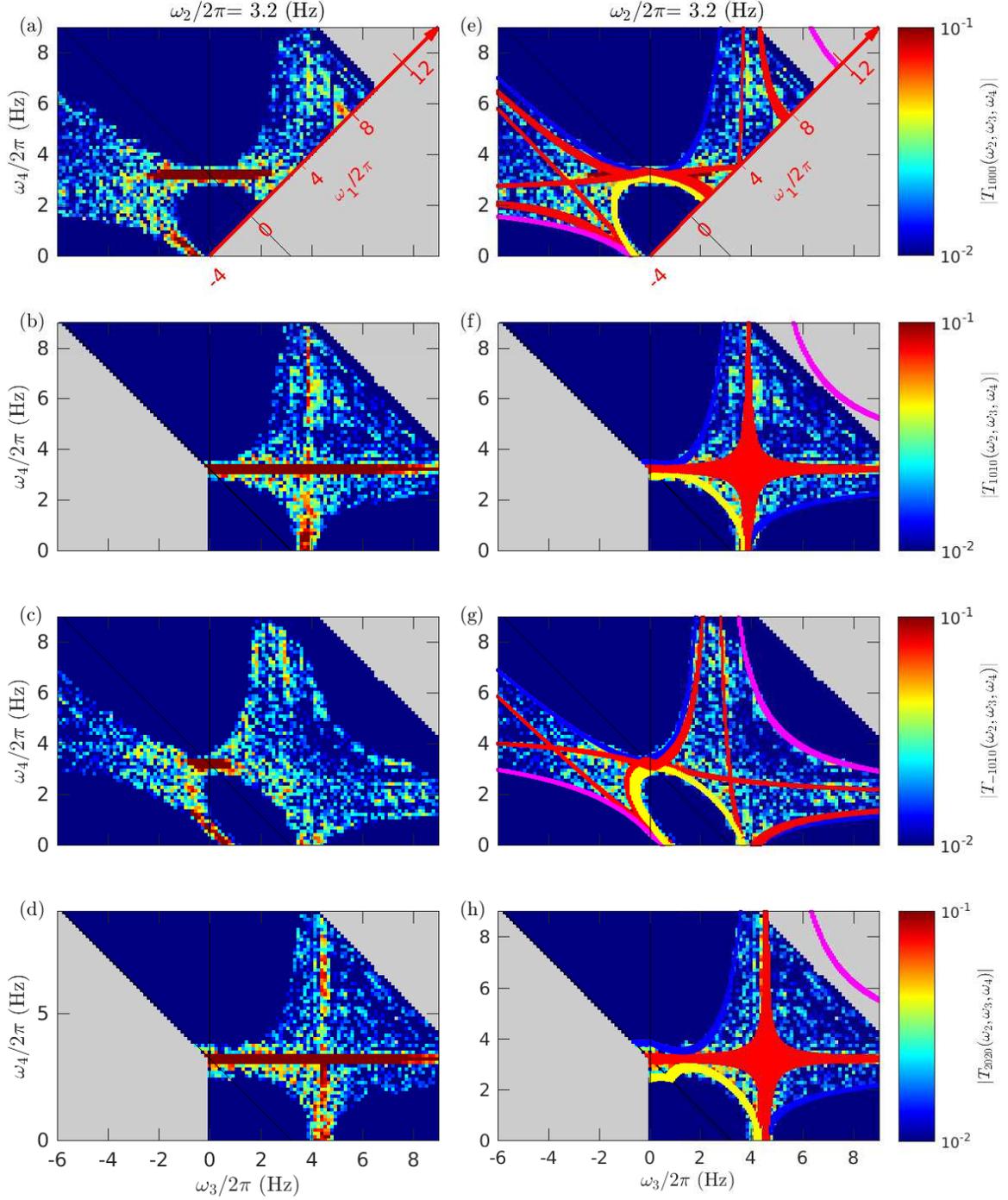}}
\vspace{-1.5cm}\caption{(Left column)Tricoherence of the filtered temporal Fourier transform of the wave elevation  $T_{1000}$, $T_{1010}$ , $T_{-1010}$ and $T_{2020}$ (from top to bottom)
for $\omega_2/2\pi =3.2$~Hz. This correlation intrinsically have symmetries. For sake of simplicity, 
the minimalistic part of the correlations are represented here. We notice that among the represented correlations, by definition of the tricoherence, the horizontal lines defined by $\omega_4=\omega_2$ for $T_{n0n0}$ are the only ones to be trivially equal to 1. Black lines indicate $\omega_1, \omega_3=0$. 
Dotted lines indicate $\omega_1, \omega_3, \omega_4=\pm\omega_0$. (Right column) Same as left column. Colored dots are the 1D exact solutions of the resonant eq~(\ref{eq:4_waves_resonances_or_bound_waves})
(See appendix for details). {One can see that high levels of correlation (in red) are always associated to regions of 1D resonances with bound waves.} }
\label{fig_C4_BW1000}
\end{figure}

We now turn to the fourth order correlation involving at least one bound wave. Figure~\ref{fig_C4_BW1000} represents $T_{1000}$, $T_{1010}$, $T_{-1010}$ and $T_{2020}$ for $\omega_2/2\pi=3.2$~Hz. Solutions of the resonance equations with bound and free waves in the unidirectional case (${\bf e}_1\propto{\bf e}_2\propto{\bf e}_3\propto{\bf e}_4$) are represented by colored dots on the right side of the figure (see appendix for details). 
%We note that, as for quasi resonant conditions~(\ref{eq:4_waves_quasiresonances}) of free waves, the density of 2D solutions of condition~(\ref{eq:4_waves_resonances_or_bound_waves}) is larger in the vicinity of 1D solutions.  
We observe in fig.~\ref{fig_C4_BW1000} correlations of the order of $10^{-1}$ (in dark red color) located around 1D solutions especially for $T_{1010}$ and $T_{2020}$. These correlations due to bound waves are actually visible in the full tricoherence (fig.~\ref{fig_C4}) as faint horizontal and vertical lines. Such correlations do not explicitly appear in the WTT formalism since they do not contribute to the net energy transfers in the limit of vanishing non-linearity (but they are actually hidden in the canonical transformation as mentioned above). 

%We now turn back to the total fourth order tricoherence $T$ (figure~\ref{fig_C4}) and reanalyze the three significative zones of significant levels.
%The correlation $T\simeq10^{-1}$ or more in region Z1 is incompatible with a wave-wave correlation since correlations patterns of $T_{n_1n_2n_3n_4}$ are changing with the value $\omega_2$. This correlation should then be due to singularities or correlated noise. The correlation larger than $10^{-1}$ in region Z3 are observed for 4-tuples at which correlations $T_{n_1n_2n_3n_4}$ are small. It is thus most likely due to non-identified structures. Thus it confirms that a significant part of the full tricoherence $T$ is due to contributions of bound waves or even other structures. 

\section{Comparison with field data}
%%%%%%
\label{field}

%Let us finally consider the relatively high correlation located in region Z2 corresponding roughly to $\omega_3=\omega_2\pm n\omega_0$ with $n=1,2$. Such 4-tuple is a solution of eqs.~(\ref{eq:res_cond4_om}) and~(\ref{eq:res_cond4_k}) whatever $\omega_4$ if two of these frequencies are associated to bound waves $|{\bf k}_1|=k_{\pm n}(\omega_4\pm n\omega_0)$ and $| {\bf k}_3|=k_{\pm n}(\omega_2\pm n\omega_0)$ and the two others to free waves $|{\bf k}_2|=k_0(\omega_2)$ and $|{\bf k}_4|=k_0(\omega_4)$. A non negligeable fourth order correlation for this particular 4-tuples would actually be a reminiscence of the presence of third order correlation due to the presence of bound waves $k_{\pm2}(\omega_3) + k_0(\omega_2)=k_0(\omega_f)=k_{\pm2}(\omega_1) + k_0(\omega_4)$.
%In region Z2, indeed both the correlations $T$ and $T_{n0n0}$ are significant. This is consistent with the dominance of $(n, 0)\longleftrightarrow (n,0)$ or $(n, 0)\longleftrightarrow (0, n)$ interactions involving twice the bound wave associated to $n=1$ or $2$ for those 4-tuples of frequencies.

\begin{figure}[!htb]
\centerline{
\includegraphics[trim=0cm 0cm 0cm 1.2cm, clip=true, totalheight=0.9\textheight, angle=0, width=8cm]{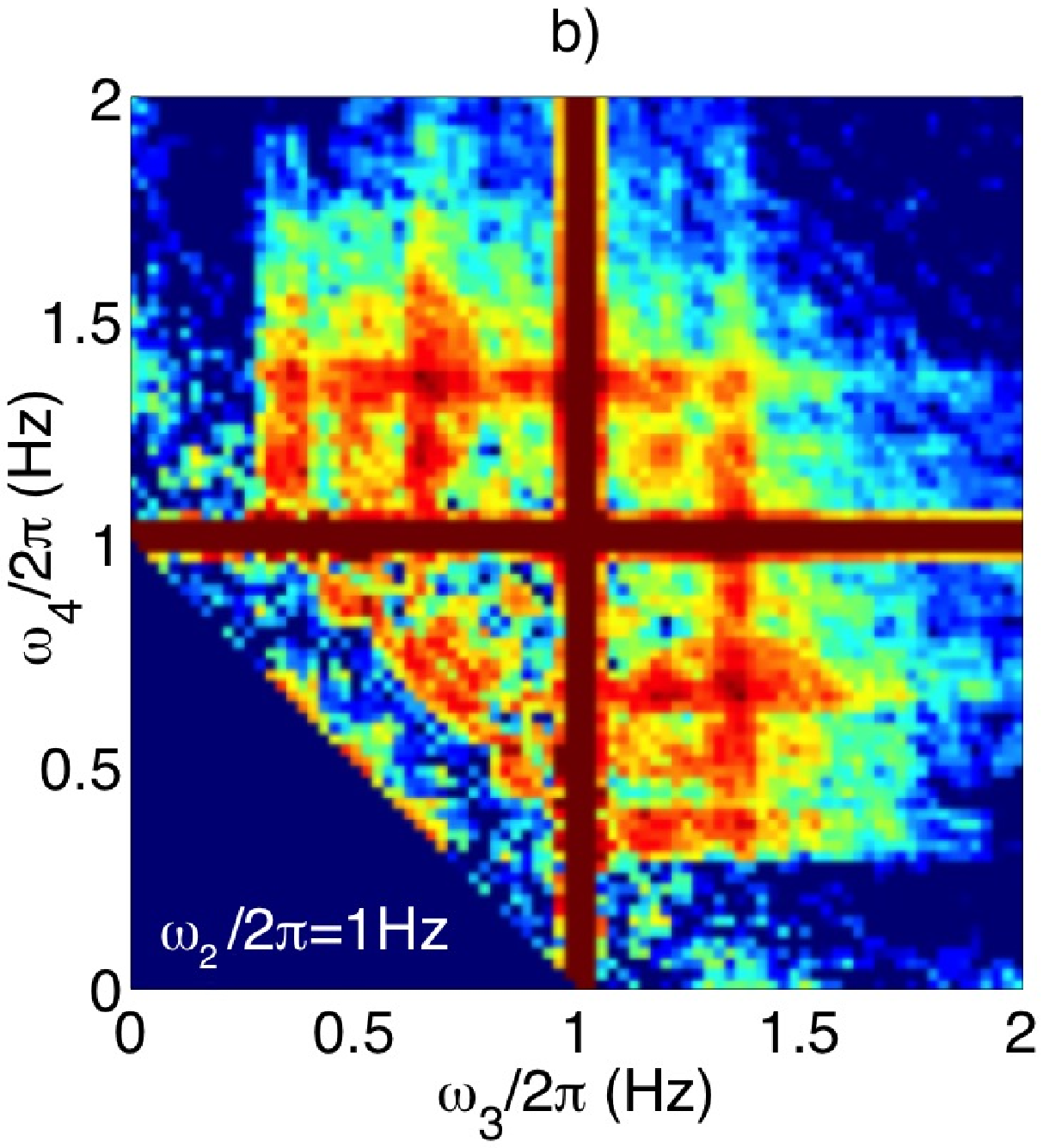}}
\includegraphics[width=7cm]{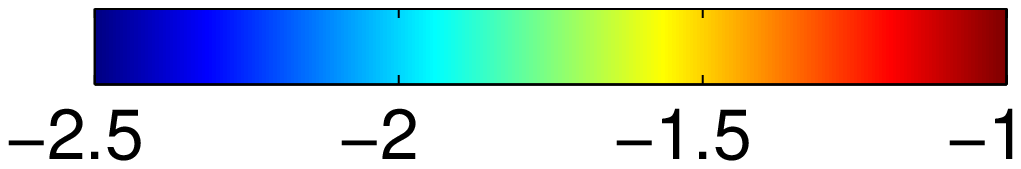}
\caption{Bicoherence computed from field data measured in the Black Sea~\cite{leckler_analysis_2015} and reproduced from ref.~\cite{aubourg_three-wave_2017}. For this dataset the peak of the spectrum occurs at $\omega/2\pi\approx 0.35$~Hz. {Thus the secondary lines are most likely contributions of bound waves as in the experiment. These bound waves have actually been observed in the $(\mathbf k,\omega)$ spectrum in~\cite{leckler_analysis_2015}.}}
\label{fig_field}
\end{figure}

Figure~\ref{fig_field} shows the tricoherence $T$ computed from field data of wave reconstruction from the Black Sea by Leckler {\it et al.}~\cite{leckler_analysis_2015} originally published in~\cite{aubourg_three-wave_2017}. It corresponds to young waves of very low steepness ($\epsilon\approx 2\%$). The image of $T$ is mostly made of horizontal and vertical lines shifted by multiples of $\pm 0.35\times 2\pi$~Hz from $\omega_2$. This frequency corresponds to the peak frequency of the spectrum. These features are reminiscent of the images of $T_{1010}$ or $T_{2020}$ in fig.~\ref{fig_C4_BW1000} and thus can most likely be associated to 1D interactions between bound waves with $n=1$ or $n=2$ in
\begin{eqnarray}
k_n(\omega) = k_{LDR}(|\omega|-n\omega_p) +nk_{LDR}(\omega_p),
\end{eqnarray}
with $\omega_p/2\pi=0.35$~Hz.
The genuine free wave interactions are most likely at a much smaller level of coherence and are hidden below the bound wave contribution as for our laboratory data. Considering the level of non linearity which is much weaker than in our experiment, the fraction of the tricoherence due to free wave interaction is beyond reach considering the tremendous amount of data that would be required to reach the adequate level of statistical convergence. {Our experiment allows us to better understand the observation of the coherence from the field data and highlight that although in the field the nonlinearity is much weaker, the bound waves are nonetheless strongly present in the nonlinear activity measured by the tricoherence.}

%%%%%%%%%%
\section{Conclusion}
%%%%%%%%%%

In summary, we report laboratory investigations of gravity wave turbulence in a relatively strong regime with an average steepness equal to 10\%. 
%As shown in \cite{Socquet} by records in the North Sea, such a steepness corresponds to very tough sea states. 
This strong level of nonlinearity can be seen in the wide tail distribution of the surface gradients as shown in fig.~\ref{fig_slope}. By operating a space-time filtering of the data we could separate free and bound waves. The bound waves are seen to dominate the wave field at frequencies higher than 4~Hz. In particular they are responsible for extreme values of the local slope of the waves.

{We perform an analysis of the nonlinear processes at work in our system by using bicoherence and tricoherence analyses. We could extract the contribution of the free waves by performing a filtering in Fourier space around the dispersion relation. In this way we could extract locally in frequency space the level of nonlinearity of the waves. We observe that,  at the scales present in our experiments (wavelengths comparable of shorter than a meter), the bound waves are very active and actually dominate the 4-wave correlations. They are responsible for very large values of the tricoherence that make the system appear as strongly nonlinear. However we know from the theory that bound waves should not contribute to the energy flux. Unfortunately we cannot measure the energy flux. When filtering out the bound waves, {as a surrogate of the canonical transformation}, we measure the level of nonlinearity of the free waves and observe that it is indeed weakly non linear.}
%With an analysis of the nonlinearity performed through the use of tricoherence estimators, we were able to exhibit for the first time the nonlinear coupling between free waves, which is at the core of the Weak Turbulence Theory. At the scales present in our experiments (wavelengths comparable of shorter than a meter), the nonlinear contribution of bound waves is very strong and dominant in the 4-wave correlations. Nevertheless the use of our space-time filter to extract free waves together with a very large database (required to converge the coherence estimators), enabled us to visualize the nonlinear coupling among them which is actually contributing to the energy flux in scale. We observed that the level of bicoherence is actually very low although the steepness of the full field is relatively high due to a significant contribution of bound waves. 
Such a weak level of correlation supports the fact that the interaction of free waves can nonetheless be described by the WTT even at moderate wave steepness. {This constitutes the main result of this article.}

{The role of bound waves in energy transfer is far from clear and unfortunately we cannot provide a direct answer from our data.}
In the theory, bound waves do not contribute to the energy cascade through the term $\mathcal{H}_3$ of the development of the Hamiltonian. Nonetheless they actually contribute at the 4th order through contributions in $\widetilde{\mathcal{H}}_4$ {in the effective interaction kernel}. The WTT development involving the variables $b({\bf k}, t)$ is supposed to be valid only for vanishing nonlinearity. {In our experiment, the nonlinearity is finite and thus} the contribution of approximate resonances involving bound waves is far from clear. This may have some importance for operational models of sea state predictions at the smallest scales of the gravity wave spectrum, which is actually the one accessible in our experiments. {In \cite{NazLuk}, the authors suggested that the reason for the evolution of the spectral exponent with the steepness could rather be an evolution to a state dominated by singular structures that are strongly non linear. Here we show that although such strongly nonlinear structure can exist due to bound waves, the underlying dynamics may remain weakly non linear nonetheless. The issue may be rather than the observation variables are not the canonical variables. It makes the experimental realisation of weak turbulence difficult in laboratory wave tanks.}

One additional question concerns the impact of the bound waves on dissipation. Assuming a viscous dissipation of a given Fourier mode at $(\mathbf k,\omega)$ with a dissipation rate equal to $2\nu k^2$, one can estimate that in our data, the relative contribution of the bound waves to the total viscous dissipation is of the order of $50\%$. In the WTT, dissipation is assumed to occur at vanishingly small scales. 3-wave correlations are thus expected to circulate energy without a global transfer to small scales. At submeter wavelength, dissipation cannot be neglected and the transfer of energy to bound waves can increase the global dissipation rate. Furthermore, as seen in fig~\ref{fig_slope}, the bound waves cause much higher fluctuations of the local steepness, triggering whitecapping more often than what could be expected from the free wave contribution only. Whitecapping events contribute also to dissipation of the wave energy and this is a very important ingredient of wave forecasting models. 

\begin{acknowledgments}
This project has received funding from the European Research Council (ERC) under the European Union's Horizon 2020 research and innovation programme (grant agreement No 647018-WATU). \end{acknowledgments}

\appendix
%%%%%%%%%%
\section{Resonnant waves}
%%%%%%%%%%
\label{Reson}

\subsection{Free waves resonances}
%%%%%%

\begin{figure}[!htb]
\centerline{\includegraphics[width=10cm]{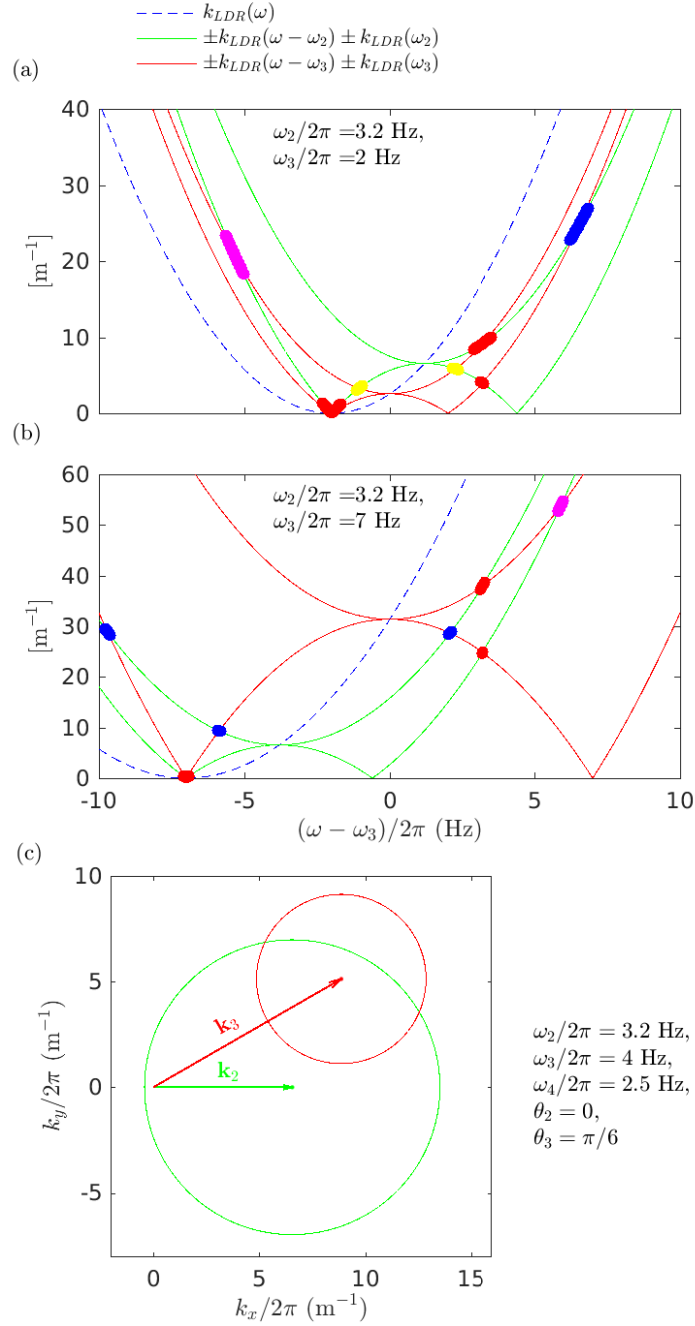}}
\caption{(a,b) Graphical representation of solutions for quasi resonant eq.~(\ref{eq:4_waves_quasiresonances}) in the colinear wave-vectors case, i.e. $\theta_i=\theta_2\pmod\pi$ with $i\in(1,3,4)$, for $\omega_2/2\pi=3.2$~Hz and $\omega_3=(2, 7)$~Hz (in (a) and (b) resp.). Solutions are localized at the crossing between the green and red curves and are highlited by colored dots. (c) Graphical representation of 2D solutions for quasi resonnant equations for $(\omega_2, \omega_3, \omega_4)/2\pi=(3.2,4,2.5)$~Hz and $(\theta_2, \theta_3)=(0,\pi/6)$. Green and red curves are defined by $k_{LDR}(\omega_3+\omega_4-\omega_2)\left(\begin{array}{c} \cos(\theta_1)\\ \sin(\theta_1) \end{array}\right)+k_{LDR}(\omega_2)\left(\begin{array}{c} \cos(\theta_2)\\ \sin(\theta_2) \end{array}\right)$ and  $k_{LDR}(\omega_3)\left(\begin{array}{c} \cos(\theta_3)\\ \sin(\theta_3) \end{array}\right)+k_{LDR}(\omega_4)\left(\begin{array}{c} \cos(\theta_4)\\ \sin(\theta_4) \end{array}\right)$ respectively. The 2D solutions are localized at the crossing between green and red circles.}.
\label{fig_C4_graphical_solutions}
\end{figure}

\begin{figure}[!htb]
\centerline{
\includegraphics[width=17cm]{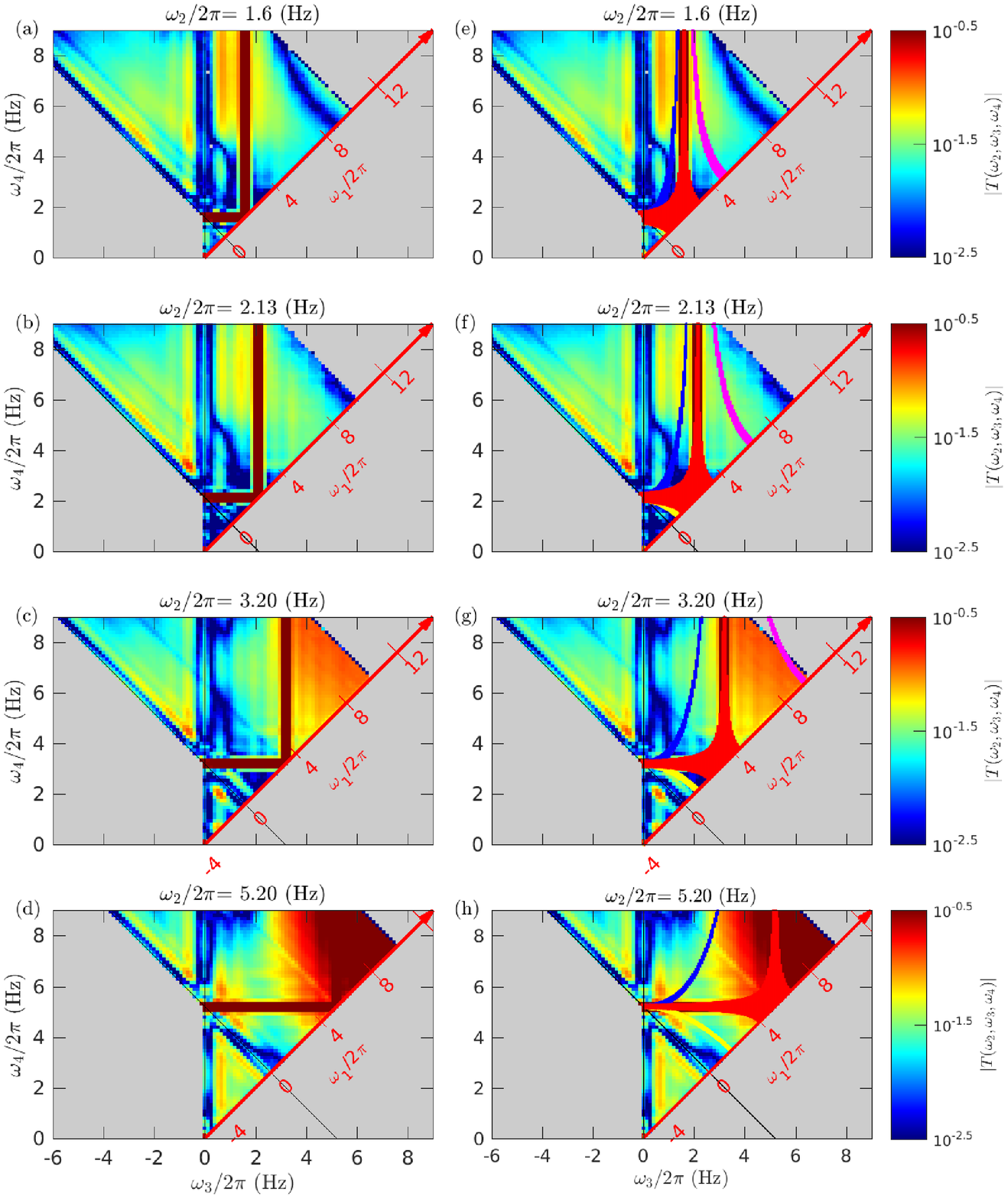}}
\vspace{-1.5cm}\caption{(Left)Tricoherence of the temporal Fourier transform of the wave elevation $T(\omega_2, \omega_3, \omega_4)$ for $\omega_2/2\pi =(1.6, 2.13, 3.2, 5.2)$~Hz.
This correlation intrinsically have symmetries. For sake of simplicity, the minimalistic part of the correlation is represented here. 
Black lines indicate $\omega_1, \omega_3=0$. Dotted lines indicate $\omega_1, \omega_3, \omega_4=\pm\omega_0$.(Right) Same as left figures. Colored dots are the 1D solutions of the resonnant equations~\ref{eq:4_waves_1Dquasiresonances} with the same colors as figure~\ref{fig_C4_graphical_solutions}.}
\label{fig_C4a}
\end{figure}

We consider the generalisation of eq.~(\ref{eq:4_waves_resonances}) for quasi-resonances of free waves, 
\begin{equation}
|k_{LDR}(-\omega_2+\omega_3+\omega_4){\bf e}_1+
k_{LDR}(\omega_2){\bf e}_2 
-(k_{LDR}(\omega_3){\bf e}_3+
k_{LDR}(\omega_4){\bf e}_4)|\leq\Delta_k.	
\label{eq:4_waves_quasiresonances}
\end{equation}
Here, the threshold has been arbitrarily choosen equal to the spatial spectral resolution $\Delta_k$ of our spatial Fourier transform to take into account a nonzero level of detuning of the free waves.
This equation depends on 7 dimensions $(\omega_2, \omega_3, \omega_4, \theta_1, \theta_2, \theta_3, \theta_4)$ which makes uneasy the visualisation of its solutions.
For sake of simplicity, we first look at the 1D case such that $\theta_i=\theta_2\pmod\pi$ with $i\in(1,3,4)$ and set values of 2 frequencies $\omega_2$ and $\omega_3$. Although these resonances are not supposed to contribute to energy transfer (because the nonlinear coupling coefficient is zero~\cite{hasselmann_non-linear_1962}), the explanation of their computation is much easier to explain. The research of 1D solutions is reduced to find the frequency $\omega=\omega_4$ satisfying the equation
\begin{equation}
|\pm k_{LDR}(-\omega_2+\omega_3+\omega)\pm
k_{LDR}(\omega_2) 
-( \pm k_{LDR}(\omega_3)
\pm  k_{LDR}(\omega))|\leq\Delta_k.	
\label{eq:4_waves_1Dquasiresonances}
\end{equation}
Figure~\ref{fig_C4_graphical_solutions} (top) graphically represents examples of these solutions for $\omega_2/2\pi=3.2$~Hz and $\omega_3/2\pi=(2,7)$~Hz. Different solutions are highlighted by coloured dots (red, blue, yellow and magenta). By varying the value of $\omega_3$, all 1D solutions in the plane $(\omega_3, \omega_4)$ are collected at a given $\omega_2$ and plotted on top of the correlations in fig.~\ref{fig_C4a} (right) with the same colours as for fig.~\ref{fig_C4_graphical_solutions}.
% A part of the solutions $(\omega_3, \omega_4)$ is hidden here (gray areas) since they are the redundancy of the white area due to the symmetries of the equation~\ref{eq:4_waves_resonances} regarding the quadruplet $(\omega_1, \omega_2, \omega_3, \omega_4)$.

We observe that the solutions of 1D quasi-resonances are strongly limited in the whole $(\omega_3, \omega_4)$ space. We distinguish 4 different branches:
\begin{itemize}
 \item the red branch with $\omega_3, \omega_4\simeq\omega_2$ and $\omega_4, \omega_3\simeq\omega_1$,   
 \item the magenta branch with $\omega_1, \omega_3, \omega_4>\omega_2$, 
 \item the yellow branch with $\omega_1, \omega_3, \omega_4<\omega_2$,
  \item the blue branch with $\omega_3<\omega_2$ and $\omega_4>\omega_2$.
\end{itemize}

\begin{figure}[!htb]
\centerline{\includegraphics[width=12cm]{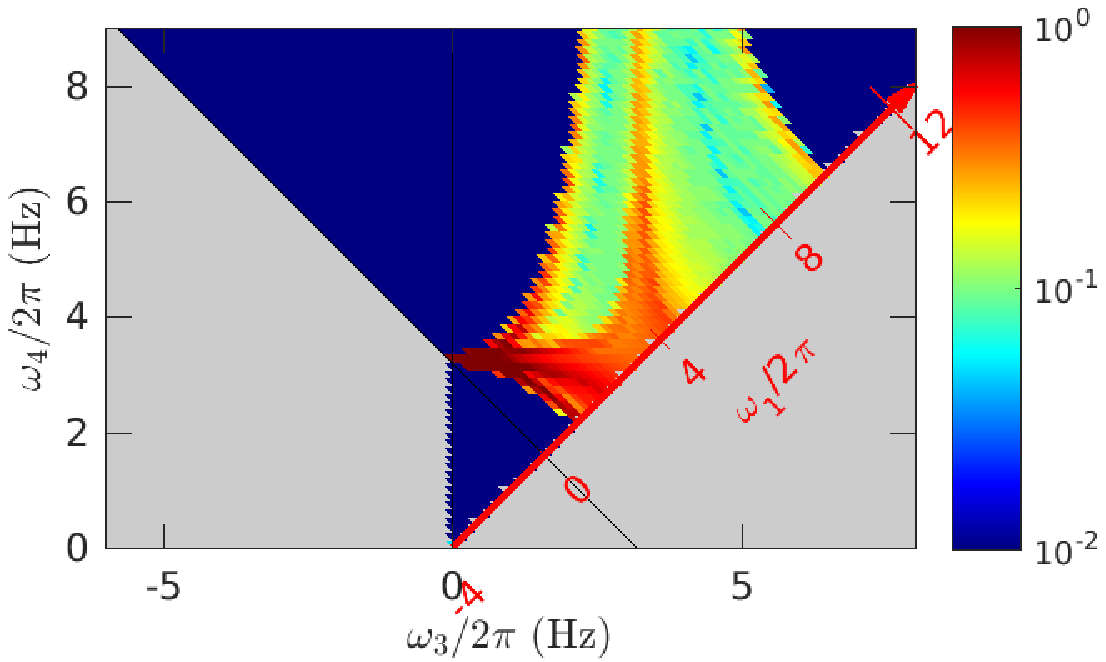}}
\caption{(Bottom) Fraction of angle $\theta_3$ for which solutions 
of quasi resonnant equations~\ref{eq:4_waves_quasiresonances} exist with $\omega_2/2\pi=3.2$ and $\theta_2=0$}.
\label{fig_C4_graphical_solutions_2}
\end{figure}

We now turn to the resolution of full 2D case of quasi resonances. This can be done by setting the values of $(\omega_2, \omega_3, \omega_4, \theta_2, \theta_3)$ and searching for couples $(\theta_1, \theta_4)$ that satisfy eq.~(\ref{eq:4_waves_quasiresonances}). Figure~\ref{fig_C4_graphical_solutions} (bottom) graphically represents an example of these solutions in the ${\bf k}$ space  for $(\omega_2, \omega_3, \omega_4)/2\pi=(3.2,4,2.5)$~Hz and $(\theta_2, \theta_3)=(0,\pi/6)$. Two parts of left hand side of eq.~(\ref{eq:4_waves_quasiresonances}), $k_{LDR}(-\omega_2+\omega_3+\omega_4){\bf e}_1+k_{LDR}(\omega_2){\bf e}_2$ and $k_{LDR}(\omega_3){\bf e}_3+
k_{LDR}(\omega_4){\bf e}_4$ are represented by circles, the solutions being trivially at the crossing of those two circles. By varying the value of $\omega_3$ and $\theta_3$, all 2D solutions are collected at a given $(\omega_2, \theta_2)$ in the plane $(\omega_3, \omega_4, \theta_3)$. Figure~\ref{fig_C4_graphical_solutions_2} (bottom), represents in the $(\omega_2, \omega_3)$ plane the fraction of angle $\theta_3$ for which a 2D solution exists for $\omega_2/2\pi=3.2$~Hz and $\theta_2=0$. We first see that the region for which many solutions (red colors, i.e. a fraction close to 1) exists corresponds also to the regions of 1D approximate solutions. As the fraction of $\theta_3$ is close to one, it means that all angles actually contribute to resonances and not only quasi 1D ones. Note also that the border of the region of existence of 2D solutions corresponds to 1D solutions.

\begin{figure}[!htb]
\centerline{
\includegraphics[width=17cm]{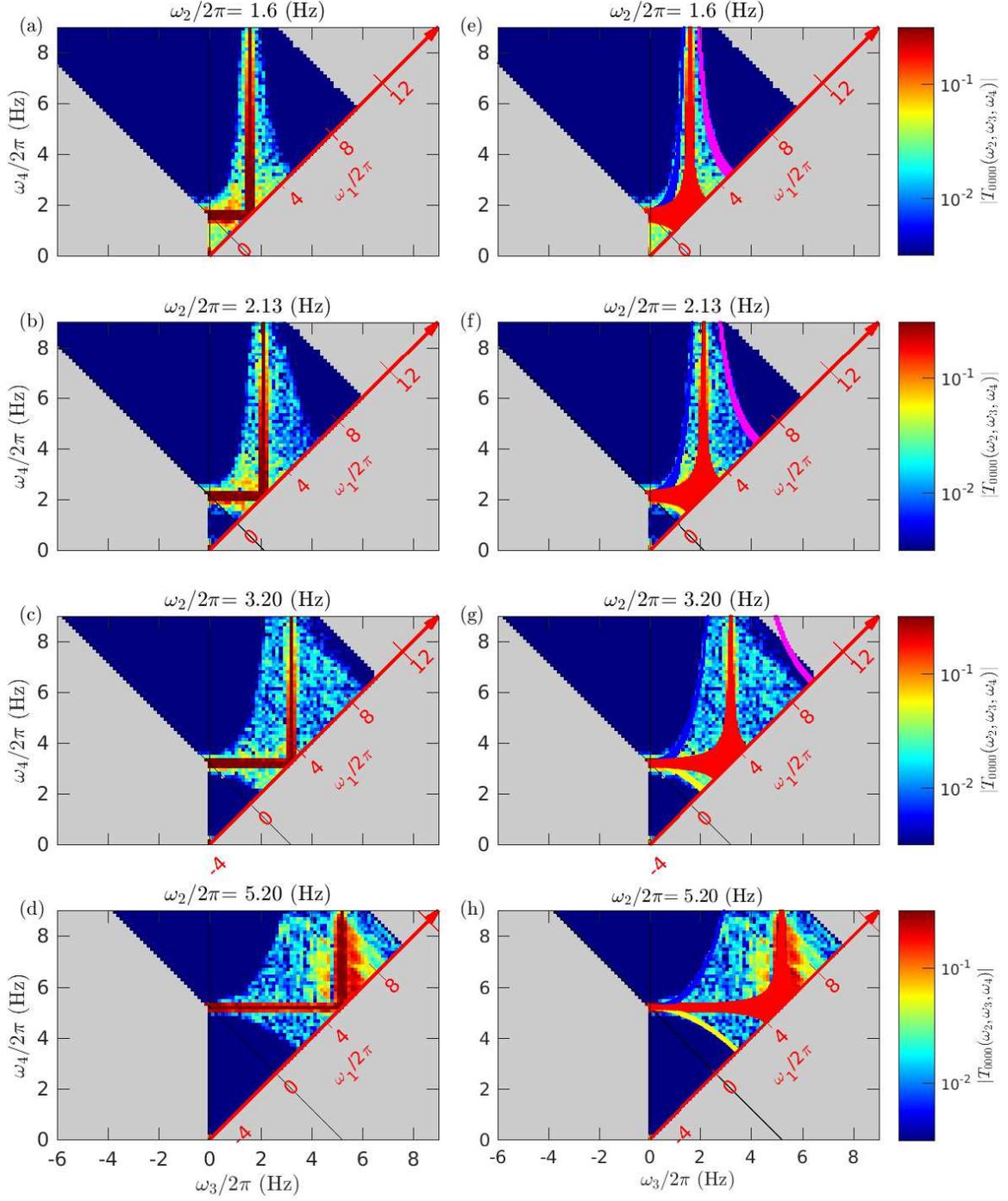}}
\vspace{-1.5cm}\caption{(Left)Tricoherence of the filtered temporal Fourier transform of the wave elevation  $T_{0000}(\omega_2, \omega_3, \omega_4)$ 
for $\omega_2/2/\pi =(1.6, 2.13, 3.2, 5.2)$~Hz. This correlation intrinsically have symmetries. For sake of simplicity, 
the minimalistic part of the correlation is represented here. Black lines indicate $\omega_1, \omega_3=0$. 
Dotted lines indicate $\omega_1, \omega_3, \omega_4=\pm\omega_0$. (Right) Same as left figures. Colored dots are the 1D solutions of the resonnant equations~\ref{eq:4_waves_1Dquasiresonances}.}
\label{fig_C4_LDRa}
\end{figure}

We plot the lines of 1D quasi-resonances on top of tricoherences $T$ and $T_{0000}$ in figs.~\ref{fig_C4a} and \ref{fig_C4_LDRa} as eye guides for convenience. One has to remember that these actually displayed regions correspond to regions with a high number of fully 2D resonances. We see in fig.~\ref{fig_C4_LDRa} that the correlations of free waves are indeed non zero in the regions bounded by the 1D solutions. By contrast, for the correlations of the full field (fig.~\ref{fig_C4a}), even in region A, the level of correlation is quite large in regions where no resonant free waves can exist. Furthermore no visible change in the level of correlation can be seen when crossing the border of the region of allowed resonances among free waves. Thus it supports the fact that the tricoherence of the full field is actually dominated by bound waves which are not relevant in the context of weak turbulence.

\subsection{Free and bound waves resonances}
%%%%%%
We now turn to the resolution of quasi-resonant waves involving both free and bound waves which is described by the following equation 
\begin{equation}
|k_{n_1}(-\omega_2+\omega_3+\omega_4){\bf e}_1+
k_{n_2}(\omega_2){\bf e}_2 
-(k_{n_3}(\omega_3){\bf e}_3+
k_{n_4}(\omega_4){\bf e}_4)|\leq\Delta_k.
\label{eq:4_bound_free_waves_quasiresonances}
\end{equation}
Since the resolution of this equation is done in the same manner as for the previous section, its describtion is not developped here. 
We precise that the 2D solutions are delimited by 1D solutions in the $(\omega_3,\omega_4)$ plane as for free waves quasi-resonances solutions. Furthermore, the areas corresponding to larger number of 2D solutions are located around those 1D solutions.
We plot the lines of 1D quasi-resonances on top of tricoherences $T_{n_1n_2n_3n_4}$ in fig~\ref{fig_C4_BW1000} as eye guides for convenience.

\clearpage

%%%%%%%%%%
\section{Correlation estimators}
%%%%%%%%%%
\label{app_cor}
{Let us consider the correlation $c_3(\mathbf x_1,\omega_1, \mathbf x_2, \omega_2, \mathbf x_3,\omega_3)$ defined as the Fourier transform in time of the correlation
\begin{equation}
c_3(\mathbf x_1,t_1, \mathbf x_2, t_2, \mathbf x_3,t_3) =
\langle \eta(\mathbf x_1,t_1)\eta(\mathbf x_2,t_2)\eta(\mathbf x_3,t_3)\rangle.
\end{equation}
}

{
We can define analogously the correlation $c_3(\mathbf k_1,\omega_1, \mathbf k_2, \omega_2, \mathbf k_3,\omega_3)$ as the Fourier transform in space of $c_3(\mathbf x_1,\omega_1, \mathbf x_2, \omega_2, \mathbf x_3,\omega_3)$ so that
\begin{equation}
c_3(\mathbf x_1,\omega_1, \mathbf x_2, \omega_2, \mathbf x_3,\omega_3)=\frac{1}{(2\pi)^{3/2}}\iiint 
c_3(\mathbf k_1,\omega_1, \mathbf k_2, \omega_2, \mathbf k_3,\omega_3) e^{i(\mathbf k_1\cdot\mathbf x_1+\mathbf k_2\cdot\mathbf x_2+\mathbf k_3\cdot\mathbf x_3)}d\mathbf k_1d\mathbf k_2d\mathbf k_3
\end{equation}
}

{
If the system is homogeneous in space, then $c_3(\mathbf x_1,\omega_1, \mathbf x_2, \omega_2, \mathbf x_3,\omega_3)$ depends only on the position differences $\delta\mathbf x_2=\mathbf x_2-\mathbf x_1$ and $\delta\mathbf x_3=\mathbf x_3-\mathbf x_1$ and not on the three positions independently. In particular $c_3(\mathbf x_1,\omega_1, \mathbf x_1, \omega_2, \mathbf x_1,\omega_3)$ is actually independent of $\mathbf x_1$. Thus one can perform an average over $\mathbf x_1$ so that 
\begin{eqnarray}
c_3(\omega_1, \omega_2, \omega_3)&=&\frac{1}{L^2}\int_{-L/2}^{L/2}\int_{-L/2}^{L/2}c_3(\mathbf x,\omega_1, \mathbf x, \omega_2, \mathbf x,\omega_3)d\mathbf x,\\
&=&\frac{1}{L^2(2\pi)^{3/2}}\int_{-L/2}^{L/2}\int_{-L/2}^{L/2}\iiint 
c_3(\mathbf k_1,\omega_1, \mathbf k_2, \omega_2, \mathbf k_3,\omega_3) e^{i(\mathbf k_1+\mathbf k_2+\mathbf k_3)\cdot\mathbf x}d\mathbf k_1d\mathbf k_2d\mathbf k_3d\mathbf x.
\end{eqnarray}
For $L\rightarrow\infty$, the average of $e^{i(\mathbf k_1+\mathbf k_2+\mathbf k_3)\cdot\mathbf x}$ is zero if  $\mathbf k_1+\mathbf k_2+\mathbf k_3\neq 0$, so that
%\begin{equation}
%c_3(\mathbf x_1,\omega_1, \mathbf x_1+\delta\mathbf x_2, \omega_2, \mathbf x_1+\delta\mathbf x_3,\omega_3)=\frac{1}{(2\pi)^{3/2}}\iiint 
%c_3(\mathbf k_1,\omega_1, \mathbf k_2, \omega_2, \mathbf k_3,\omega_3) e^{i((\mathbf k_1+\mathbf k_2+\mathbf k_3)\cdot\mathbf x_1+\mathbf k_2\cdot\delta\mathbf x_2+\mathbf k_3\cdot\delta\mathbf x_3)}d\mathbf k_1d\mathbf k_2d\mathbf k_3
%\end{equation}
%is only a function of $\delta\mathbf x_2$ and $\delta\mathbf x_3$. Thus one can perform an average over $x_1$ and the value of $c_3$ is unchanged. For an infinite system, as the average of $e^{i((\mathbf k_1+\mathbf k_2+\mathbf k_3)\cdot\mathbf x_1)}$ is zero if  $\mathbf k_1+\mathbf k_2+\mathbf k_3\neq 0$, it gives
%\begin{equation}
%c_3(\mathbf x_1,\omega_1, \mathbf x_1+\delta\mathbf x_2, \omega_2, \mathbf x_1+\delta\mathbf x_3,\omega_3)=\frac{1}{(2\pi)^{3/2}}\iiint 
%c_3(\mathbf k_1,\omega_1, \mathbf k_2, \omega_2, \mathbf k_3,\omega_3) \delta(\mathbf k_1+\mathbf k_2+\mathbf k_3)e^{i(\mathbf k_2\cdot\delta\mathbf x_2+\mathbf k_3\cdot\delta\mathbf x_3)}d\mathbf k_1d\mathbf k_2d\mathbf k_3
%\end{equation}
%In particular for $\delta\mathbf x_2=\delta\mathbf x_3=0$ one obtains
\begin{equation}
c_3(\omega_1, \omega_2, \omega_3)=\frac{1}{(2\pi)^{3/2}}\iiint 
c_3(\mathbf k_1,\omega_1, \mathbf k_2, \omega_2, \mathbf k_3,\omega_3) \delta(\mathbf k_1+\mathbf k_2+\mathbf k_3)d\mathbf k_1d\mathbf k_2d\mathbf k_3
\label{app_c3d}
\end{equation}
}

{
In practice, in (\ref{eq:c3}), our estimator of $c_3(\omega_1, \omega_2, \omega_3)$ is not defined from the space-time correlation as in (B1) but rather as the average of the square modulus of the Fourier transform of the signal which is equivalent for signals with a finite extension as it is the case for experimental signals. Furthermore our signal is discrete. Our field is locally homogeneous and, in (\ref{eq:c3}), we actually define $c_3(\omega_1, \omega_2,\omega_3)$ using the average over $\mathbf x_1$ to enforce the result (\ref{app_c3d}) and also to improve the statistical convergence of the estimator.
}
%----------------------------------------------------------------------------------------
%	BIBLIOGRAPHY
%\bibliographystyle{plain}
\bibliography{mabiblio}

%merlin.mbs apsrev4-1.bst 2010-07-25 4.21a (PWD, AO, DPC) hacked
%Control: key (0)
%Control: author (0) dotless jnrlst
%Control: editor formatted (1) identically to author
%Control: production of article title (0) allowed
%Control: page (1) range
%Control: year (0) verbatim
%Control: production of eprint (0) enabled
\begin{thebibliography}{50}%
\makeatletter
\providecommand \@ifxundefined [1]{%
 \@ifx{#1\undefined}
}%
\providecommand \@ifnum [1]{%
 \ifnum #1\expandafter \@firstoftwo
 \else \expandafter \@secondoftwo
 \fi
}%
\providecommand \@ifx [1]{%
 \ifx #1\expandafter \@firstoftwo
 \else \expandafter \@secondoftwo
 \fi
}%
\providecommand \natexlab [1]{#1}%
\providecommand \enquote  [1]{``#1''}%
\providecommand \bibnamefont  [1]{#1}%
\providecommand \bibfnamefont [1]{#1}%
\providecommand \citenamefont [1]{#1}%
\providecommand \href@noop [0]{\@secondoftwo}%
\providecommand \href [0]{\begingroup \@sanitize@url \@href}%
\providecommand \@href[1]{\@@startlink{#1}\@@href}%
\providecommand \@@href[1]{\endgroup#1\@@endlink}%
\providecommand \@sanitize@url [0]{\catcode `\\12\catcode `\$12\catcode
  `\&12\catcode `\#12\catcode `\^12\catcode `\_12\catcode `\%12\relax}%
\providecommand \@@startlink[1]{}%
\providecommand \@@endlink[0]{}%
\providecommand \url  [0]{\begingroup\@sanitize@url \@url }%
\providecommand \@url [1]{\endgroup\@href {#1}{\urlprefix }}%
\providecommand \urlprefix  [0]{URL }%
\providecommand \Eprint [0]{\href }%
\providecommand \doibase [0]{http://dx.doi.org/}%
\providecommand \selectlanguage [0]{\@gobble}%
\providecommand \bibinfo  [0]{\@secondoftwo}%
\providecommand \bibfield  [0]{\@secondoftwo}%
\providecommand \translation [1]{[#1]}%
\providecommand \BibitemOpen [0]{}%
\providecommand \bibitemStop [0]{}%
\providecommand \bibitemNoStop [0]{.\EOS\space}%
\providecommand \EOS [0]{\spacefactor3000\relax}%
\providecommand \BibitemShut  [1]{\csname bibitem#1\endcsname}%
\let\auto@bib@innerbib\@empty
%</preamble>
\bibitem [{\citenamefont {Hasselmann}(1962)}]{hasselmann_non-linear_1962}%
  \BibitemOpen
  \bibfield  {author} {\bibinfo {author} {\bibfnamefont {K.}~\bibnamefont
  {Hasselmann}},\ }\bibfield  {title} {\enquote {\bibinfo {title} {On the
  non-linear energy transfer in a gravity-wave spectrum {Part} 1. {General}
  theory},}\ }\href {\doibase 10.1017/S0022112062000373} {\bibfield  {journal}
  {\bibinfo  {journal} {Journal of Fluid Mechanics}\ }\textbf {\bibinfo
  {volume} {12}},\ \bibinfo {pages} {481--500} (\bibinfo {year}
  {1962})}\BibitemShut {NoStop}%
\bibitem [{\citenamefont {Zakharov}\ \emph {et~al.}(1992)\citenamefont
  {Zakharov}, \citenamefont {L'vov},\ and\ \citenamefont
  {Falkovich}}]{zakharov_kolmogorov_1992}%
  \BibitemOpen
  \bibfield  {author} {\bibinfo {author} {\bibfnamefont {V.~E.}\ \bibnamefont
  {Zakharov}}, \bibinfo {author} {\bibfnamefont {V.~S.}\ \bibnamefont {L'vov}},
  \ and\ \bibinfo {author} {\bibfnamefont {G.}~\bibnamefont {Falkovich}},\
  }\href {//www.springer.com/gp/book/9783642500541} {\emph {\bibinfo {title}
  {Kolmogorov {Spectra} of {Turbulence} {I}: {Wave} {Turbulence}}}},\ Springer
  {Series} in {Nonlinear} {Dynamics}\ (\bibinfo  {publisher}
  {Springer-Verlag},\ \bibinfo {address} {Berlin Heidelberg},\ \bibinfo {year}
  {1992})\BibitemShut {NoStop}%
\bibitem [{\citenamefont {Tolman}\ \emph {et~al.}()\citenamefont {Tolman},
  \citenamefont {Accensi}, \citenamefont {Alves}, \citenamefont {Ardhuin},
  \citenamefont {Bidlot}, \citenamefont {Booij}, \citenamefont {Bennis},
  \citenamefont {Campbell}, \citenamefont {Chalikov},\ and\ \citenamefont
  {Filipot}}]{tolman18user}%
  \BibitemOpen
  \bibfield  {author} {\bibinfo {author} {\bibfnamefont {HL}~\bibnamefont
  {Tolman}}, \bibinfo {author} {\bibfnamefont {M}~\bibnamefont {Accensi}},
  \bibinfo {author} {\bibfnamefont {H}~\bibnamefont {Alves}}, \bibinfo {author}
  {\bibfnamefont {F}~\bibnamefont {Ardhuin}}, \bibinfo {author} {\bibfnamefont
  {J}~\bibnamefont {Bidlot}}, \bibinfo {author} {\bibfnamefont {N}~\bibnamefont
  {Booij}}, \bibinfo {author} {\bibfnamefont {AC}~\bibnamefont {Bennis}},
  \bibinfo {author} {\bibfnamefont {T}~\bibnamefont {Campbell}}, \bibinfo
  {author} {\bibfnamefont {DV}~\bibnamefont {Chalikov}}, \ and\ \bibinfo
  {author} {\bibfnamefont {JF}~\bibnamefont {Filipot}},\ }\href@noop {}
  {\enquote {\bibinfo {title} {User manual and system documentation of
  wavewatch iii version 4.18. 2014},}\ }\BibitemShut {NoStop}%
\bibitem [{\citenamefont {Galtier}(2003)}]{galtier_weak_2003}%
  \BibitemOpen
  \bibfield  {author} {\bibinfo {author} {\bibfnamefont {S\'ebastien}\
  \bibnamefont {Galtier}},\ }\bibfield  {title} {\enquote {\bibinfo {title}
  {Weak inertial-wave turbulence theory},}\ }\href {\doibase
  10.1103/PhysRevE.68.015301} {\bibfield  {journal} {\bibinfo  {journal}
  {Physical Review E}\ }\textbf {\bibinfo {volume} {68}},\ \bibinfo {pages}
  {015301} (\bibinfo {year} {2003})}\BibitemShut {NoStop}%
\bibitem [{\citenamefont {L’vov}\ \emph {et~al.}(2004)\citenamefont
  {L’vov}, \citenamefont {Nazarenko},\ and\ \citenamefont
  {Volovik}}]{lvov_energy_2004}%
  \BibitemOpen
  \bibfield  {author} {\bibinfo {author} {\bibfnamefont {V.~S.}\ \bibnamefont
  {L’vov}}, \bibinfo {author} {\bibfnamefont {V.}~\bibnamefont {Nazarenko}},
  \ and\ \bibinfo {author} {\bibfnamefont {G.~E.}\ \bibnamefont {Volovik}},\
  }\bibfield  {title} {\enquote {\bibinfo {title} {Energy spectra of developed
  superfluid turbulence},}\ }\href {\doibase 10.1134/1.1839294} {\bibfield
  {journal} {\bibinfo  {journal} {Journal of Experimental and Theoretical
  Physics Letters}\ }\textbf {\bibinfo {volume} {80}},\ \bibinfo {pages}
  {479--483} (\bibinfo {year} {2004})}\BibitemShut {NoStop}%
\bibitem [{\citenamefont {Düring}\ \emph {et~al.}(2006)\citenamefont
  {Düring}, \citenamefont {Josserand},\ and\ \citenamefont
  {Rica}}]{during_weak_2006}%
  \BibitemOpen
  \bibfield  {author} {\bibinfo {author} {\bibfnamefont {Gustavo}\ \bibnamefont
  {Düring}}, \bibinfo {author} {\bibfnamefont {Christophe}\ \bibnamefont
  {Josserand}}, \ and\ \bibinfo {author} {\bibfnamefont {Sergio}\ \bibnamefont
  {Rica}},\ }\bibfield  {title} {\enquote {\bibinfo {title} {Weak {Turbulence}
  for a {Vibrating} {Plate}: {Can} {One} {Hear} a {Kolmogorov} {Spectrum}?}}\
  }\href {\doibase 10.1103/PhysRevLett.97.025503} {\bibfield  {journal}
  {\bibinfo  {journal} {Physical Review Letters}\ }\textbf {\bibinfo {volume}
  {97}},\ \bibinfo {pages} {025503} (\bibinfo {year} {2006})}\BibitemShut
  {NoStop}%
\bibitem [{\citenamefont {Nazarenko}\ and\ \citenamefont
  {Onorato}(2006)}]{nazarenko_wave_2006}%
  \BibitemOpen
  \bibfield  {author} {\bibinfo {author} {\bibfnamefont {Sergey}\ \bibnamefont
  {Nazarenko}}\ and\ \bibinfo {author} {\bibfnamefont {Miguel}\ \bibnamefont
  {Onorato}},\ }\bibfield  {title} {\enquote {\bibinfo {title} {Wave turbulence
  and vortices in {Bose}-{Einstein} condensation},}\ }\href {\doibase
  10.1016/j.physd.2006.05.007} {\bibfield  {journal} {\bibinfo  {journal}
  {Physica D: Nonlinear Phenomena}\ }\textbf {\bibinfo {volume} {219}},\
  \bibinfo {pages} {1--12} (\bibinfo {year} {2006})}\BibitemShut {NoStop}%
\bibitem [{\citenamefont {Zakharov}\ \emph {et~al.}(1985)\citenamefont
  {Zakharov}, \citenamefont {Musher},\ and\ \citenamefont
  {Rubenchik}}]{zakharov_hamiltonian_1985}%
  \BibitemOpen
  \bibfield  {author} {\bibinfo {author} {\bibfnamefont {V.~E.}\ \bibnamefont
  {Zakharov}}, \bibinfo {author} {\bibfnamefont {S.~L.}\ \bibnamefont
  {Musher}}, \ and\ \bibinfo {author} {\bibfnamefont {A.~M.}\ \bibnamefont
  {Rubenchik}},\ }\bibfield  {title} {\enquote {\bibinfo {title} {Hamiltonian
  approach to the description of non-linear plasma phenomena},}\ }\href
  {\doibase 10.1016/0370-1573(85)90040-7} {\bibfield  {journal} {\bibinfo
  {journal} {Physics Reports}\ }\textbf {\bibinfo {volume} {129}},\ \bibinfo
  {pages} {285--366} (\bibinfo {year} {1985})}\BibitemShut {NoStop}%
\bibitem [{\citenamefont {Bortolozzo}\ \emph {et~al.}(2009)\citenamefont
  {Bortolozzo}, \citenamefont {Laurie}, \citenamefont {Nazarenko},\ and\
  \citenamefont {Residori}}]{bortolozzo_optical_2009}%
  \BibitemOpen
  \bibfield  {author} {\bibinfo {author} {\bibfnamefont {Umberto}\ \bibnamefont
  {Bortolozzo}}, \bibinfo {author} {\bibfnamefont {Jason}\ \bibnamefont
  {Laurie}}, \bibinfo {author} {\bibfnamefont {Sergey}\ \bibnamefont
  {Nazarenko}}, \ and\ \bibinfo {author} {\bibfnamefont {Stefania}\
  \bibnamefont {Residori}},\ }\bibfield  {title} {\enquote {\bibinfo {title}
  {Optical wave turbulence and the condensation of light},}\ }\href {\doibase
  10.1364/JOSAB.26.002280} {\bibfield  {journal} {\bibinfo  {journal} {Journal
  of the Optical Society of America B}\ }\textbf {\bibinfo {volume} {26}},\
  \bibinfo {pages} {2280} (\bibinfo {year} {2009})}\BibitemShut {NoStop}%
\bibitem [{\citenamefont {Phillips}(1960)}]{phillips_dynamics_1960}%
  \BibitemOpen
  \bibfield  {author} {\bibinfo {author} {\bibfnamefont {O.~M.}\ \bibnamefont
  {Phillips}},\ }\bibfield  {title} {\enquote {\bibinfo {title} {On the
  dynamics of unsteady gravity waves of finite amplitude {Part} 1. {The}
  elementary interactions},}\ }\href {\doibase 10.1017/S0022112060001043}
  {\bibfield  {journal} {\bibinfo  {journal} {Journal of Fluid Mechanics}\
  }\textbf {\bibinfo {volume} {9}},\ \bibinfo {pages} {193--217} (\bibinfo
  {year} {1960})}\BibitemShut {NoStop}%
\bibitem [{\citenamefont {Nazarenko}(2011)}]{nazarenko_wave_2011}%
  \BibitemOpen
  \bibfield  {author} {\bibinfo {author} {\bibfnamefont {Sergey}\ \bibnamefont
  {Nazarenko}},\ }\href {//www.springer.com/us/book/9783642159411} {\emph
  {\bibinfo {title} {Wave {Turbulence}}}},\ Lecture {Notes} in {Physics}\
  (\bibinfo  {publisher} {Springer-Verlag},\ \bibinfo {address} {Berlin
  Heidelberg},\ \bibinfo {year} {2011})\BibitemShut {NoStop}%
\bibitem [{\citenamefont {Hammack}\ and\ \citenamefont
  {Henderson}(1993)}]{Hammack:1993vz}%
  \BibitemOpen
  \bibfield  {author} {\bibinfo {author} {\bibfnamefont {Joseph~L}\
  \bibnamefont {Hammack}}\ and\ \bibinfo {author} {\bibfnamefont {D~M}\
  \bibnamefont {Henderson}},\ }\bibfield  {title} {\enquote {\bibinfo {title}
  {{RESONANT INTERACTIONS AMONG SURFACE WATER WAVES}},}\ }\href@noop {}
  {\bibfield  {journal} {\bibinfo  {journal} {Annual Review Of Fluid
  Mechanics}\ }\textbf {\bibinfo {volume} {25}},\ \bibinfo {pages} {55--97}
  (\bibinfo {year} {1993})}\BibitemShut {NoStop}%
\bibitem [{\citenamefont {Annenkov}\ and\ \citenamefont
  {Shrira}(2006)}]{annenkov_role_2006}%
  \BibitemOpen
  \bibfield  {author} {\bibinfo {author} {\bibfnamefont {Sergei~Yu}\
  \bibnamefont {Annenkov}}\ and\ \bibinfo {author} {\bibfnamefont {Victor~I.}\
  \bibnamefont {Shrira}},\ }\bibfield  {title} {\enquote {\bibinfo {title}
  {Role of non-resonant interactions in the evolution of nonlinear random water
  wave fields},}\ }\href {\doibase 10.1017/S0022112006000632} {\bibfield
  {journal} {\bibinfo  {journal} {Journal of Fluid Mechanics}\ }\textbf
  {\bibinfo {volume} {561}},\ \bibinfo {pages} {181--207} (\bibinfo {year}
  {2006})}\BibitemShut {NoStop}%
\bibitem [{\citenamefont {Shemer}\ \emph {et~al.}(2007)\citenamefont {Shemer},
  \citenamefont {Goulitski},\ and\ \citenamefont
  {Kit}}]{shemer_evolution_2007}%
  \BibitemOpen
  \bibfield  {author} {\bibinfo {author} {\bibfnamefont {L.}~\bibnamefont
  {Shemer}}, \bibinfo {author} {\bibfnamefont {K.}~\bibnamefont {Goulitski}}, \
  and\ \bibinfo {author} {\bibfnamefont {E.}~\bibnamefont {Kit}},\ }\bibfield
  {title} {\enquote {\bibinfo {title} {Evolution of wide-spectrum
  unidirectional wave groups in a tank: an experimental and numerical study},}\
  }\href {\doibase 10.1016/j.euromechflu.2006.06.004} {\bibfield  {journal}
  {\bibinfo  {journal} {European Journal of Mechanics - B/Fluids}\ }\textbf
  {\bibinfo {volume} {26}},\ \bibinfo {pages} {193--219} (\bibinfo {year}
  {2007})}\BibitemShut {NoStop}%
\bibitem [{\citenamefont {Kartashova}\ \emph {et~al.}(2008)\citenamefont
  {Kartashova}, \citenamefont {Nazarenko},\ and\ \citenamefont
  {Rudenko}}]{Kartashova:2008bv}%
  \BibitemOpen
  \bibfield  {author} {\bibinfo {author} {\bibfnamefont {Elena}\ \bibnamefont
  {Kartashova}}, \bibinfo {author} {\bibfnamefont {Sergei~V}\ \bibnamefont
  {Nazarenko}}, \ and\ \bibinfo {author} {\bibfnamefont {Oleksii}\ \bibnamefont
  {Rudenko}},\ }\bibfield  {title} {\enquote {\bibinfo {title} {{Resonant
  interactions of nonlinear water waves in a finite basin}},}\ }\href@noop {}
  {\bibfield  {journal} {\bibinfo  {journal} {Physical Review E}\ }\textbf
  {\bibinfo {volume} {78}},\ \bibinfo {pages} {520} (\bibinfo {year}
  {2008})}\BibitemShut {NoStop}%
\bibitem [{\citenamefont {Hwang}\ \emph {et~al.}(2000)\citenamefont {Hwang},
  \citenamefont {Wang}, \citenamefont {Walsh}, \citenamefont {Krabill},\ and\
  \citenamefont {Swift}}]{hwang_airborne_2000}%
  \BibitemOpen
  \bibfield  {author} {\bibinfo {author} {\bibfnamefont {Paul~A.}\ \bibnamefont
  {Hwang}}, \bibinfo {author} {\bibfnamefont {David~W.}\ \bibnamefont {Wang}},
  \bibinfo {author} {\bibfnamefont {Edward~J.}\ \bibnamefont {Walsh}}, \bibinfo
  {author} {\bibfnamefont {William~B.}\ \bibnamefont {Krabill}}, \ and\
  \bibinfo {author} {\bibfnamefont {Robert~N.}\ \bibnamefont {Swift}},\
  }\bibfield  {title} {\enquote {\bibinfo {title} {Airborne {Measurements} of
  the {Wavenumber} {Spectra} of {Ocean} {Surface} {Waves}. {Part} {I}:
  {Spectral} {Slope} and {Dimensionless} {Spectral} {Coefficient}},}\ }\href
  {\doibase 10.1175/1520-0485(2001)031<2753:AMOTWS>2.0.CO;2} {\bibfield
  {journal} {\bibinfo  {journal} {Journal of Physical Oceanography}\ }\textbf
  {\bibinfo {volume} {30}},\ \bibinfo {pages} {2753--2767} (\bibinfo {year}
  {2000})}\BibitemShut {NoStop}%
\bibitem [{\citenamefont {Romero}\ and\ \citenamefont
  {Melville}(2010)}]{romero_airborne_2010}%
  \BibitemOpen
  \bibfield  {author} {\bibinfo {author} {\bibfnamefont {Leonel}\ \bibnamefont
  {Romero}}\ and\ \bibinfo {author} {\bibfnamefont {W.~Kendall}\ \bibnamefont
  {Melville}},\ }\bibfield  {title} {\enquote {\bibinfo {title} {Airborne
  {Observations} of {Fetch}-{Limited} {Waves} in the {Gulf} of
  {Tehuantepec}},}\ }\href {\doibase 10.1175/2009JPO4127.1} {\bibfield
  {journal} {\bibinfo  {journal} {Journal of Physical Oceanography}\ }\textbf
  {\bibinfo {volume} {40}},\ \bibinfo {pages} {441--465} (\bibinfo {year}
  {2010})}\BibitemShut {NoStop}%
\bibitem [{\citenamefont {Leckler}\ \emph {et~al.}(2015)\citenamefont
  {Leckler}, \citenamefont {Ardhuin}, \citenamefont {Peureux}, \citenamefont
  {Benetazzo}, \citenamefont {Bergamasco},\ and\ \citenamefont
  {Dulov}}]{leckler_analysis_2015}%
  \BibitemOpen
  \bibfield  {author} {\bibinfo {author} {\bibfnamefont {Fabien}\ \bibnamefont
  {Leckler}}, \bibinfo {author} {\bibfnamefont {Fabrice}\ \bibnamefont
  {Ardhuin}}, \bibinfo {author} {\bibfnamefont {Charles}\ \bibnamefont
  {Peureux}}, \bibinfo {author} {\bibfnamefont {Alvise}\ \bibnamefont
  {Benetazzo}}, \bibinfo {author} {\bibfnamefont {Filippo}\ \bibnamefont
  {Bergamasco}}, \ and\ \bibinfo {author} {\bibfnamefont {Vladimir}\
  \bibnamefont {Dulov}},\ }\bibfield  {title} {\enquote {\bibinfo {title}
  {Analysis and {Interpretation} of {Frequency}{Wavenumber} {Spectra} of
  {Young} {Wind} {Waves}},}\ }\href {\doibase 10.1175/JPO-D-14-0237.1}
  {\bibfield  {journal} {\bibinfo  {journal} {Journal of Physical
  Oceanography}\ }\textbf {\bibinfo {volume} {45}},\ \bibinfo {pages}
  {2484--2496} (\bibinfo {year} {2015})}\BibitemShut {NoStop}%
\bibitem [{\citenamefont {Lvov}\ \emph {et~al.}(2006)\citenamefont {Lvov},
  \citenamefont {Nazarenko},\ and\ \citenamefont
  {Pokorni}}]{lvov_discreteness_2006}%
  \BibitemOpen
  \bibfield  {author} {\bibinfo {author} {\bibfnamefont {Yuri~V.}\ \bibnamefont
  {Lvov}}, \bibinfo {author} {\bibfnamefont {Sergey}\ \bibnamefont
  {Nazarenko}}, \ and\ \bibinfo {author} {\bibfnamefont {Boris}\ \bibnamefont
  {Pokorni}},\ }\bibfield  {title} {\enquote {\bibinfo {title} {Discreteness
  and its effect on water-wave turbulence},}\ }\href {\doibase
  10.1016/j.physd.2006.04.003} {\bibfield  {journal} {\bibinfo  {journal}
  {Physica D: Nonlinear Phenomena}\ }\textbf {\bibinfo {volume} {218}},\
  \bibinfo {pages} {24--35} (\bibinfo {year} {2006})}\BibitemShut {NoStop}%
\bibitem [{\citenamefont {Yokoyama}(2004)}]{yokoyama_statistics_2004}%
  \BibitemOpen
  \bibfield  {author} {\bibinfo {author} {\bibfnamefont {Naoto}\ \bibnamefont
  {Yokoyama}},\ }\bibfield  {title} {\enquote {\bibinfo {title} {Statistics of
  gravity waves obtained by direct numerical simulation},}\ }\href {\doibase
  10.1017/S0022112003007444} {\bibfield  {journal} {\bibinfo  {journal}
  {Journal of Fluid Mechanics}\ }\textbf {\bibinfo {volume} {501}},\ \bibinfo
  {pages} {169--178} (\bibinfo {year} {2004})}\BibitemShut {NoStop}%
\bibitem [{\citenamefont {Onorato}\ \emph {et~al.}(2002)\citenamefont
  {Onorato}, \citenamefont {Osborne}, \citenamefont {Serio}, \citenamefont
  {Resio}, \citenamefont {Pushkarev}, \citenamefont {Zakharov},\ and\
  \citenamefont {Brandini}}]{onorato_freely_2002}%
  \BibitemOpen
  \bibfield  {author} {\bibinfo {author} {\bibfnamefont {M.}~\bibnamefont
  {Onorato}}, \bibinfo {author} {\bibfnamefont {A.~R.}\ \bibnamefont
  {Osborne}}, \bibinfo {author} {\bibfnamefont {M.}~\bibnamefont {Serio}},
  \bibinfo {author} {\bibfnamefont {D.}~\bibnamefont {Resio}}, \bibinfo
  {author} {\bibfnamefont {A.}~\bibnamefont {Pushkarev}}, \bibinfo {author}
  {\bibfnamefont {V.~E.}\ \bibnamefont {Zakharov}}, \ and\ \bibinfo {author}
  {\bibfnamefont {C.}~\bibnamefont {Brandini}},\ }\bibfield  {title} {\enquote
  {\bibinfo {title} {Freely {Decaying} {Weak} {Turbulence} for {Sea} {Surface}
  {Gravity} {Waves}},}\ }\href {\doibase 10.1103/PhysRevLett.89.144501}
  {\bibfield  {journal} {\bibinfo  {journal} {Physical Review Letters}\
  }\textbf {\bibinfo {volume} {89}},\ \bibinfo {pages} {144501} (\bibinfo
  {year} {2002})}\BibitemShut {NoStop}%
\bibitem [{\citenamefont {Zakharov}\ \emph {et~al.}(2005)\citenamefont
  {Zakharov}, \citenamefont {Korotkevich}, \citenamefont {Pushkarev},\ and\
  \citenamefont {Dyachenko}}]{zakharov_mesoscopic_2005}%
  \BibitemOpen
  \bibfield  {author} {\bibinfo {author} {\bibfnamefont {V.~E.}\ \bibnamefont
  {Zakharov}}, \bibinfo {author} {\bibfnamefont {A.~O.}\ \bibnamefont
  {Korotkevich}}, \bibinfo {author} {\bibfnamefont {A.~N.}\ \bibnamefont
  {Pushkarev}}, \ and\ \bibinfo {author} {\bibfnamefont {A.~I.}\ \bibnamefont
  {Dyachenko}},\ }\bibfield  {title} {\enquote {\bibinfo {title} {Mesoscopic
  wave turbulence},}\ }\href {\doibase 10.1134/1.2150867} {\bibfield  {journal}
  {\bibinfo  {journal} {Journal of Experimental and Theoretical Physics
  Letters}\ }\textbf {\bibinfo {volume} {82}},\ \bibinfo {pages} {487--491}
  (\bibinfo {year} {2005})}\BibitemShut {NoStop}%
\bibitem [{\citenamefont {Campagne}\ \emph {et~al.}(2018)\citenamefont
  {Campagne}, \citenamefont {Hassaini}, \citenamefont {Redor}, \citenamefont
  {Sommeria}, \citenamefont {Valran}, \citenamefont {Viboud},\ and\
  \citenamefont {Mordant}}]{campagne_impact_2018}%
  \BibitemOpen
  \bibfield  {author} {\bibinfo {author} {\bibfnamefont {Antoine}\ \bibnamefont
  {Campagne}}, \bibinfo {author} {\bibfnamefont {Roumaissa}\ \bibnamefont
  {Hassaini}}, \bibinfo {author} {\bibfnamefont {Ivan}\ \bibnamefont {Redor}},
  \bibinfo {author} {\bibfnamefont {Jo\"el}\ \bibnamefont {Sommeria}}, \bibinfo
  {author} {\bibfnamefont {Thomas}\ \bibnamefont {Valran}}, \bibinfo {author}
  {\bibfnamefont {Samuel}\ \bibnamefont {Viboud}}, \ and\ \bibinfo {author}
  {\bibfnamefont {Nicolas}\ \bibnamefont {Mordant}},\ }\bibfield  {title}
  {\enquote {\bibinfo {title} {Impact of dissipation on the energy spectrum of
  experimental turbulence of gravity surface waves},}\ }\href {\doibase
  10.1103/PhysRevFluids.3.044801} {\bibfield  {journal} {\bibinfo  {journal}
  {Physical Review Fluids}\ }\textbf {\bibinfo {volume} {3}},\ \bibinfo {pages}
  {044801} (\bibinfo {year} {2018})}\BibitemShut {NoStop}%
\bibitem [{\citenamefont {Aubourg}\ \emph {et~al.}(2017)\citenamefont
  {Aubourg}, \citenamefont {Campagne}, \citenamefont {Peureux}, \citenamefont
  {Ardhuin}, \citenamefont {Sommeria}, \citenamefont {Viboud},\ and\
  \citenamefont {Mordant}}]{aubourg_three-wave_2017}%
  \BibitemOpen
  \bibfield  {author} {\bibinfo {author} {\bibfnamefont {Quentin}\ \bibnamefont
  {Aubourg}}, \bibinfo {author} {\bibfnamefont {Antoine}\ \bibnamefont
  {Campagne}}, \bibinfo {author} {\bibfnamefont {Charles}\ \bibnamefont
  {Peureux}}, \bibinfo {author} {\bibfnamefont {Fabrice}\ \bibnamefont
  {Ardhuin}}, \bibinfo {author} {\bibfnamefont {Joel}\ \bibnamefont
  {Sommeria}}, \bibinfo {author} {\bibfnamefont {Samuel}\ \bibnamefont
  {Viboud}}, \ and\ \bibinfo {author} {\bibfnamefont {Nicolas}\ \bibnamefont
  {Mordant}},\ }\bibfield  {title} {\enquote {\bibinfo {title} {Three-wave and
  four-wave interactions in gravity wave turbulence},}\ }\href {\doibase
  10.1103/PhysRevFluids.2.114802} {\bibfield  {journal} {\bibinfo  {journal}
  {Physical Review Fluids}\ }\textbf {\bibinfo {volume} {2}},\ \bibinfo {pages}
  {114802} (\bibinfo {year} {2017})}\BibitemShut {NoStop}%
\bibitem [{\citenamefont {Deike}\ \emph {et~al.}(2015)\citenamefont {Deike},
  \citenamefont {Miquel}, \citenamefont {Gutiérrez}, \citenamefont {Jamin},
  \citenamefont {Semin}, \citenamefont {Berhanu}, \citenamefont {Falcon},\ and\
  \citenamefont {Bonnefoy}}]{deike_role_2015}%
  \BibitemOpen
  \bibfield  {author} {\bibinfo {author} {\bibfnamefont {L.}~\bibnamefont
  {Deike}}, \bibinfo {author} {\bibfnamefont {B.}~\bibnamefont {Miquel}},
  \bibinfo {author} {\bibfnamefont {P.}~\bibnamefont {Gutiérrez}}, \bibinfo
  {author} {\bibfnamefont {T.}~\bibnamefont {Jamin}}, \bibinfo {author}
  {\bibfnamefont {B.}~\bibnamefont {Semin}}, \bibinfo {author} {\bibfnamefont
  {M.}~\bibnamefont {Berhanu}}, \bibinfo {author} {\bibfnamefont
  {E.}~\bibnamefont {Falcon}}, \ and\ \bibinfo {author} {\bibfnamefont
  {F.}~\bibnamefont {Bonnefoy}},\ }\bibfield  {title} {\enquote {\bibinfo
  {title} {Role of the basin boundary conditions in gravity wave turbulence},}\
  }\href {\doibase 10.1017/jfm.2015.494} {\bibfield  {journal} {\bibinfo
  {journal} {Journal of Fluid Mechanics}\ }\textbf {\bibinfo {volume} {781}},\
  \bibinfo {pages} {196--225} (\bibinfo {year} {2015})}\BibitemShut {NoStop}%
\bibitem [{\citenamefont {Nazarenko}\ and\ \citenamefont
  {Lukaschuk}(2016)}]{nazarenko_wave_2016}%
  \BibitemOpen
  \bibfield  {author} {\bibinfo {author} {\bibfnamefont {Sergey}\ \bibnamefont
  {Nazarenko}}\ and\ \bibinfo {author} {\bibfnamefont {Sergei}\ \bibnamefont
  {Lukaschuk}},\ }\bibfield  {title} {\enquote {\bibinfo {title} {Wave
  {Turbulence} on {Water} {Surface}},}\ }\href {\doibase
  10.1146/annurev-conmatphys-071715-102737} {\bibfield  {journal} {\bibinfo
  {journal} {Annual Review of Condensed Matter Physics}\ }\textbf {\bibinfo
  {volume} {7}},\ \bibinfo {pages} {61--88} (\bibinfo {year}
  {2016})}\BibitemShut {NoStop}%
\bibitem [{\citenamefont {Denissenko}\ \emph {et~al.}(2007)\citenamefont
  {Denissenko}, \citenamefont {Lukaschuk},\ and\ \citenamefont
  {Nazarenko}}]{denissenko_gravity_2007}%
  \BibitemOpen
  \bibfield  {author} {\bibinfo {author} {\bibfnamefont {Petr}\ \bibnamefont
  {Denissenko}}, \bibinfo {author} {\bibfnamefont {Sergei}\ \bibnamefont
  {Lukaschuk}}, \ and\ \bibinfo {author} {\bibfnamefont {Sergey}\ \bibnamefont
  {Nazarenko}},\ }\bibfield  {title} {\enquote {\bibinfo {title} {Gravity
  {Wave} {Turbulence} in a {Laboratory} {Flume}},}\ }\href {\doibase
  10.1103/PhysRevLett.99.014501} {\bibfield  {journal} {\bibinfo  {journal}
  {Physical Review Letters}\ }\textbf {\bibinfo {volume} {99}},\ \bibinfo
  {pages} {014501} (\bibinfo {year} {2007})}\BibitemShut {NoStop}%
\bibitem [{\citenamefont {Lukaschuk}\ \emph {et~al.}(2009)\citenamefont
  {Lukaschuk}, \citenamefont {Nazarenko}, \citenamefont {McLelland},\ and\
  \citenamefont {Denissenko}}]{lukaschuk_gravity_2009}%
  \BibitemOpen
  \bibfield  {author} {\bibinfo {author} {\bibfnamefont {Sergei}\ \bibnamefont
  {Lukaschuk}}, \bibinfo {author} {\bibfnamefont {Sergey}\ \bibnamefont
  {Nazarenko}}, \bibinfo {author} {\bibfnamefont {Stuart}\ \bibnamefont
  {McLelland}}, \ and\ \bibinfo {author} {\bibfnamefont {Petr}\ \bibnamefont
  {Denissenko}},\ }\bibfield  {title} {\enquote {\bibinfo {title} {Gravity
  {Wave} {Turbulence} in {Wave} {Tanks}: {Space} and {Time} {Statistics}},}\
  }\href {\doibase 10.1103/PhysRevLett.103.044501} {\bibfield  {journal}
  {\bibinfo  {journal} {Physical Review Letters}\ }\textbf {\bibinfo {volume}
  {103}},\ \bibinfo {pages} {044501} (\bibinfo {year} {2009})}\BibitemShut
  {NoStop}%
\bibitem [{\citenamefont {Miles}(1967)}]{miles_surface-wave_1967}%
  \BibitemOpen
  \bibfield  {author} {\bibinfo {author} {\bibfnamefont {J.~W.}\ \bibnamefont
  {Miles}},\ }\bibfield  {title} {\enquote {\bibinfo {title} {Surface-wave
  damping in closed basins},}\ }\href {\doibase 10.1098/rspa.1967.0081}
  {\bibfield  {journal} {\bibinfo  {journal} {Proc. R. Soc. Lond. A}\ }\textbf
  {\bibinfo {volume} {297}},\ \bibinfo {pages} {459--475} (\bibinfo {year}
  {1967})}\BibitemShut {NoStop}%
\bibitem [{\citenamefont {Dorn}(1966)}]{dorn_boundary_1966}%
  \BibitemOpen
  \bibfield  {author} {\bibinfo {author} {\bibfnamefont {W.~G.~Van}\
  \bibnamefont {Dorn}},\ }\bibfield  {title} {\enquote {\bibinfo {title}
  {Boundary dissipation of oscillatory waves},}\ }\href {\doibase
  10.1017/S0022112066000995} {\bibfield  {journal} {\bibinfo  {journal}
  {Journal of Fluid Mechanics}\ }\textbf {\bibinfo {volume} {24}},\ \bibinfo
  {pages} {769--779} (\bibinfo {year} {1966})}\BibitemShut {NoStop}%
\bibitem [{\citenamefont {Henderson}\ and\ \citenamefont
  {Miles}(1990)}]{henderson_single-mode_1990}%
  \BibitemOpen
  \bibfield  {author} {\bibinfo {author} {\bibfnamefont {Diane~M.}\
  \bibnamefont {Henderson}}\ and\ \bibinfo {author} {\bibfnamefont {John~W.}\
  \bibnamefont {Miles}},\ }\bibfield  {title} {\enquote {\bibinfo {title}
  {Single-mode {Faraday} waves in small cylinders},}\ }\href {\doibase
  10.1017/S0022112090002233} {\bibfield  {journal} {\bibinfo  {journal}
  {Journal of Fluid Mechanics}\ }\textbf {\bibinfo {volume} {213}},\ \bibinfo
  {pages} {95--109} (\bibinfo {year} {1990})}\BibitemShut {NoStop}%
\bibitem [{\citenamefont {Humbert}\ \emph {et~al.}(2013)\citenamefont
  {Humbert}, \citenamefont {Cadot}, \citenamefont {D\"uring}, \citenamefont
  {Josserand}, \citenamefont {Rica},\ and\ \citenamefont {Touz\'e}}]{Humbert}%
  \BibitemOpen
  \bibfield  {author} {\bibinfo {author} {\bibfnamefont {T.}~\bibnamefont
  {Humbert}}, \bibinfo {author} {\bibfnamefont {O.}~\bibnamefont {Cadot}},
  \bibinfo {author} {\bibfnamefont {G.}~\bibnamefont {D\"uring}}, \bibinfo
  {author} {\bibfnamefont {C.}~\bibnamefont {Josserand}}, \bibinfo {author}
  {\bibfnamefont {S.}~\bibnamefont {Rica}}, \ and\ \bibinfo {author}
  {\bibfnamefont {C.}~\bibnamefont {Touz\'e}},\ }\bibfield  {title} {\enquote
  {\bibinfo {title} {Wave turbulence in vibrating plates : the effect of
  damping},}\ }\href@noop {} {\bibfield  {journal} {\bibinfo  {journal} {EPL}\
  }\textbf {\bibinfo {volume} {102}},\ \bibinfo {pages} {30002} (\bibinfo
  {year} {2013})}\BibitemShut {NoStop}%
\bibitem [{\citenamefont {Miquel}\ \emph {et~al.}(2014)\citenamefont {Miquel},
  \citenamefont {Alexakis},\ and\ \citenamefont {Mordant}}]{R23}%
  \BibitemOpen
  \bibfield  {author} {\bibinfo {author} {\bibfnamefont {B.}~\bibnamefont
  {Miquel}}, \bibinfo {author} {\bibfnamefont {A.}~\bibnamefont {Alexakis}}, \
  and\ \bibinfo {author} {\bibfnamefont {N.}~\bibnamefont {Mordant}},\
  }\bibfield  {title} {\enquote {\bibinfo {title} {Role of dissipation in
  flexural wave turbulence: from experimental spectrum to kolmogorov-zakharov
  spectrum},}\ }\href@noop {} {\bibfield  {journal} {\bibinfo  {journal} {Phys.
  Rev. E}\ }\textbf {\bibinfo {volume} {89}},\ \bibinfo {pages} {062925}
  (\bibinfo {year} {2014})}\BibitemShut {NoStop}%
\bibitem [{\citenamefont {Deike}\ \emph {et~al.}(2014)\citenamefont {Deike},
  \citenamefont {Berhanu},\ and\ \citenamefont {Falcon}}]{Deikedis2}%
  \BibitemOpen
  \bibfield  {author} {\bibinfo {author} {\bibfnamefont {L.}~\bibnamefont
  {Deike}}, \bibinfo {author} {\bibfnamefont {M.}~\bibnamefont {Berhanu}}, \
  and\ \bibinfo {author} {\bibfnamefont {E.}~\bibnamefont {Falcon}},\
  }\bibfield  {title} {\enquote {\bibinfo {title} {Energy flux measurement from
  the dissipated energy in capillary wave turbulence},}\ }\href@noop {}
  {\bibfield  {journal} {\bibinfo  {journal} {Phys. Rev. E}\ }\textbf {\bibinfo
  {volume} {89}},\ \bibinfo {pages} {023003} (\bibinfo {year}
  {2014})}\BibitemShut {NoStop}%
\bibitem [{\citenamefont {Socquet-Juglard}\ \emph {et~al.}(2005)\citenamefont
  {Socquet-Juglard}, \citenamefont {Dysthe}, \citenamefont {Trulsen},
  \citenamefont {Krogstad},\ and\ \citenamefont {Liu}}]{Socquet}%
  \BibitemOpen
  \bibfield  {author} {\bibinfo {author} {\bibfnamefont {H.}~\bibnamefont
  {Socquet-Juglard}}, \bibinfo {author} {\bibfnamefont {K.}~\bibnamefont
  {Dysthe}}, \bibinfo {author} {\bibfnamefont {K.}~\bibnamefont {Trulsen}},
  \bibinfo {author} {\bibfnamefont {H.~E.}\ \bibnamefont {Krogstad}}, \ and\
  \bibinfo {author} {\bibfnamefont {J.}~\bibnamefont {Liu}},\ }\bibfield
  {title} {\enquote {\bibinfo {title} {{Probability distributions of surface
  gravity waves during spectral changes}},}\ }\href@noop {} {\bibfield
  {journal} {\bibinfo  {journal} {Journal Of Fluid Mechanics}\ }\textbf
  {\bibinfo {volume} {542}},\ \bibinfo {pages} {195} (\bibinfo {year}
  {2005})}\BibitemShut {NoStop}%
\bibitem [{\citenamefont {Nazarenko}\ \emph {et~al.}(2009)\citenamefont
  {Nazarenko}, \citenamefont {Lukaschuk}, \citenamefont {McLelland},\ and\
  \citenamefont {Denissenko}}]{NazLuk}%
  \BibitemOpen
  \bibfield  {author} {\bibinfo {author} {\bibfnamefont {S.}~\bibnamefont
  {Nazarenko}}, \bibinfo {author} {\bibfnamefont {S.}~\bibnamefont
  {Lukaschuk}}, \bibinfo {author} {\bibfnamefont {S.}~\bibnamefont
  {McLelland}}, \ and\ \bibinfo {author} {\bibfnamefont {P.}~\bibnamefont
  {Denissenko}},\ }\bibfield  {title} {\enquote {\bibinfo {title} {Statistics
  of surface gravity wave turbulence in the space and time domains},}\
  }\href@noop {} {\bibfield  {journal} {\bibinfo  {journal} {J. Fluid Mech.}\
  }\textbf {\bibinfo {volume} {642}},\ \bibinfo {pages} {395} (\bibinfo {year}
  {2009})}\BibitemShut {NoStop}%
\bibitem [{\citenamefont {Phillips}(2006)}]{Phillips:2006jt}%
  \BibitemOpen
  \bibfield  {author} {\bibinfo {author} {\bibfnamefont {O~M}\ \bibnamefont
  {Phillips}},\ }\bibfield  {title} {\enquote {\bibinfo {title} {{Spectral and
  statistical properties of the equilibrium range in wind-generated gravity
  waves}},}\ }\href@noop {} {\bibfield  {journal} {\bibinfo  {journal} {Journal
  Of Fluid Mechanics}\ }\textbf {\bibinfo {volume} {156}},\ \bibinfo {pages}
  {505} (\bibinfo {year} {2006})}\BibitemShut {NoStop}%
\bibitem [{\citenamefont {Newell}\ and\ \citenamefont {Rumpf}(2011)}]{NewellR}%
  \BibitemOpen
  \bibfield  {author} {\bibinfo {author} {\bibfnamefont {A.~C.}\ \bibnamefont
  {Newell}}\ and\ \bibinfo {author} {\bibfnamefont {B.}~\bibnamefont {Rumpf}},\
  }\bibfield  {title} {\enquote {\bibinfo {title} {Wave turbulence},}\
  }\href@noop {} {\bibfield  {journal} {\bibinfo  {journal} {Ann. Rev. Fluid
  Mech.}\ }\textbf {\bibinfo {volume} {43}} (\bibinfo {year}
  {2011})}\BibitemShut {NoStop}%
\bibitem [{\citenamefont {Aubourg}(2016)}]{aubourg_etude_2016}%
  \BibitemOpen
  \bibfield  {author} {\bibinfo {author} {\bibfnamefont {Quentin}\ \bibnamefont
  {Aubourg}},\ }\emph {\bibinfo {title} {\'Etude exp\'erimentale de la
  turbulence d'ondes \`a la surface d'un fluide. {La} th\'eorie de la
  turbulence faible \`a l'\'epreuve de la r\'ealit\'e pour les ondes de
  capillarit\'e et gravit\'e}},\ \href
  {https://tel.archives-ouvertes.fr/tel-01597576/document} {\bibinfo {type}
  {phdthesis}},\ \bibinfo  {school} {Universit\'e Grenoble Alpes} (\bibinfo
  {year} {2016})\BibitemShut {NoStop}%
\bibitem [{\citenamefont {{Aubourg}}\ \emph {et~al.}(2019)\citenamefont
  {{Aubourg}}, \citenamefont {{Sommeria}}, \citenamefont {{Viboud}},\ and\
  \citenamefont {{Mordant}}}]{2019arXiv190205819A}%
  \BibitemOpen
  \bibfield  {author} {\bibinfo {author} {\bibfnamefont {Quentin}\ \bibnamefont
  {{Aubourg}}}, \bibinfo {author} {\bibfnamefont {Joel}\ \bibnamefont
  {{Sommeria}}}, \bibinfo {author} {\bibfnamefont {Samuel}\ \bibnamefont
  {{Viboud}}}, \ and\ \bibinfo {author} {\bibfnamefont {Nicolas}\ \bibnamefont
  {{Mordant}}},\ }\bibfield  {title} {\enquote {\bibinfo {title} {{Combined
  stereoscopic wave mapping and particle image velocimetry}},}\ }\href@noop {}
  {\bibfield  {journal} {\bibinfo  {journal} {arXiv e-prints}\ ,\ \bibinfo
  {eid} {arXiv:1902.05819}} (\bibinfo {year} {2019})},\ \Eprint
  {http://arxiv.org/abs/1902.05819} {arXiv:1902.05819 [physics.flu-dyn]}
  \BibitemShut {NoStop}%
\bibitem [{\citenamefont {Benetazzo}(2006)}]{Benetazzo:2006dp}%
  \BibitemOpen
  \bibfield  {author} {\bibinfo {author} {\bibfnamefont {Alvise}\ \bibnamefont
  {Benetazzo}},\ }\bibfield  {title} {\enquote {\bibinfo {title} {{Measurements
  of short water waves using stereo matched image sequences}},}\ }\href@noop {}
  {\bibfield  {journal} {\bibinfo  {journal} {Coastal Engineering}\ }\textbf
  {\bibinfo {volume} {53}},\ \bibinfo {pages} {1013--1032} (\bibinfo {year}
  {2006})}\BibitemShut {NoStop}%
\bibitem [{\citenamefont {Zavadsky}\ \emph {et~al.}(2017)\citenamefont
  {Zavadsky}, \citenamefont {Benetazzo},\ and\ \citenamefont
  {Shemer}}]{Zavadsky:2017fm}%
  \BibitemOpen
  \bibfield  {author} {\bibinfo {author} {\bibfnamefont {Andrey}\ \bibnamefont
  {Zavadsky}}, \bibinfo {author} {\bibfnamefont {Alvise}\ \bibnamefont
  {Benetazzo}}, \ and\ \bibinfo {author} {\bibfnamefont {Lev}\ \bibnamefont
  {Shemer}},\ }\bibfield  {title} {\enquote {\bibinfo {title} {{On the
  two-dimensional structure of short gravity waves in a wind wave tank}},}\
  }\href@noop {} {\bibfield  {journal} {\bibinfo  {journal} {Physics Of
  Fluids}\ }\textbf {\bibinfo {volume} {29}},\ \bibinfo {pages} {016601}
  (\bibinfo {year} {2017})}\BibitemShut {NoStop}%
\bibitem [{\citenamefont {Stokes}(1880)}]{stokes1880theory}%
  \BibitemOpen
  \bibfield  {author} {\bibinfo {author} {\bibfnamefont {George~G}\
  \bibnamefont {Stokes}},\ }\bibfield  {title} {\enquote {\bibinfo {title} {On
  the theory of oscillatory waves},}\ }\href@noop {} {\bibfield  {journal}
  {\bibinfo  {journal} {Transactions of the Cambridge Philosophical Society}\ }
  (\bibinfo {year} {1880})}\BibitemShut {NoStop}%
\bibitem [{\citenamefont {Tayfun}(1980)}]{Tayfun}%
  \BibitemOpen
  \bibfield  {author} {\bibinfo {author} {\bibfnamefont {M.~A.}\ \bibnamefont
  {Tayfun}},\ }\bibfield  {title} {\enquote {\bibinfo {title} {Narrow-band
  nonlinear sea waves},}\ }\href@noop {} {\bibfield  {journal} {\bibinfo
  {journal} {J. Geophys. Res.}\ }\textbf {\bibinfo {volume} {85}} (\bibinfo
  {year} {1980})}\BibitemShut {NoStop}%
\bibitem [{\citenamefont {Peureux}(2017)}]{peureux_observation_2017}%
  \BibitemOpen
  \bibfield  {author} {\bibinfo {author} {\bibfnamefont {Charles}\ \bibnamefont
  {Peureux}},\ }\emph {\bibinfo {title} {Observation et mod\'elisation des
  propri\'et\'es directionnelles des ondes de gravit\'e courtes}},\ \href
  {https://tel.archives-ouvertes.fr/tel-01738611/document} {\bibinfo {type}
  {phdthesis}},\ \bibinfo  {school} {Universit\'e de Bretagne occidentale -
  Brest} (\bibinfo {year} {2017})\BibitemShut {NoStop}%
\bibitem [{\citenamefont {Zakharov}(1999)}]{zakharov_statistical_1999}%
  \BibitemOpen
  \bibfield  {author} {\bibinfo {author} {\bibfnamefont {V}~\bibnamefont
  {Zakharov}},\ }\bibfield  {title} {\enquote {\bibinfo {title} {Statistical
  theory of gravity and capillary waves on the surface of a finite-depth
  fluid},}\ }\href {\doibase 10.1016/S0997-7546(99)80031-4} {\bibfield
  {journal} {\bibinfo  {journal} {European Journal of Mechanics - B/Fluids}\
  }\bibinfo {series} {Three-{Dimensional} {Aspects} of {Air}-{Sea}
  {Interaction}},\ \textbf {\bibinfo {volume} {18}},\ \bibinfo {pages}
  {327--344} (\bibinfo {year} {1999})}\BibitemShut {NoStop}%
\bibitem [{\citenamefont {Krasitskii}(1994)}]{krasitskii_reduced_1994}%
  \BibitemOpen
  \bibfield  {author} {\bibinfo {author} {\bibfnamefont {Vladimir~P.}\
  \bibnamefont {Krasitskii}},\ }\bibfield  {title} {\enquote {\bibinfo {title}
  {On reduced equations in the {Hamiltonian} theory of weakly nonlinear surface
  waves},}\ }\href {\doibase 10.1017/S0022112094004350} {\bibfield  {journal}
  {\bibinfo  {journal} {Journal of Fluid Mechanics}\ }\textbf {\bibinfo
  {volume} {272}},\ \bibinfo {pages} {1--20} (\bibinfo {year}
  {1994})}\BibitemShut {NoStop}%
\bibitem [{\citenamefont {Webb}(1978)}]{webb_non-linear_1978}%
  \BibitemOpen
  \bibfield  {author} {\bibinfo {author} {\bibfnamefont {D.~J.}\ \bibnamefont
  {Webb}},\ }\bibfield  {title} {\enquote {\bibinfo {title} {Non-linear
  transfers between sea waves},}\ }\href {\doibase
  10.1016/0146-6291(78)90593-3} {\bibfield  {journal} {\bibinfo  {journal}
  {Deep Sea Research}\ }\textbf {\bibinfo {volume} {25}},\ \bibinfo {pages}
  {279--298} (\bibinfo {year} {1978})}\BibitemShut {NoStop}%
\bibitem [{\citenamefont {elgar}\ and\ \citenamefont
  {Guza}(2006)}]{elgar:2006hj}%
  \BibitemOpen
  \bibfield  {author} {\bibinfo {author} {\bibfnamefont {STEVE}\ \bibnamefont
  {elgar}}\ and\ \bibinfo {author} {\bibfnamefont {R~T}\ \bibnamefont {Guza}},\
  }\bibfield  {title} {\enquote {\bibinfo {title} {{Observations of bispectra
  of shoaling surface gravity waves}},}\ }\href@noop {} {\bibfield  {journal}
  {\bibinfo  {journal} {Journal Of Fluid Mechanics}\ }\textbf {\bibinfo
  {volume} {161}},\ \bibinfo {pages} {425} (\bibinfo {year}
  {2006})}\BibitemShut {NoStop}%
\bibitem [{\citenamefont {Newell}\ \emph {et~al.}(2001)\citenamefont {Newell},
  \citenamefont {Nazarenko},\ and\ \citenamefont {Biven}}]{newell_wave_2001}%
  \BibitemOpen
  \bibfield  {author} {\bibinfo {author} {\bibfnamefont {Alan~C.}\ \bibnamefont
  {Newell}}, \bibinfo {author} {\bibfnamefont {Sergey}\ \bibnamefont
  {Nazarenko}}, \ and\ \bibinfo {author} {\bibfnamefont {Laura}\ \bibnamefont
  {Biven}},\ }\bibfield  {title} {\enquote {\bibinfo {title} {Wave turbulence
  and intermittency},}\ }\href {\doibase 10.1016/S0167-2789(01)00192-0}
  {\bibfield  {journal} {\bibinfo  {journal} {Physica D: Nonlinear Phenomena}\
  }\bibinfo {series} {Advances in {Nonlinear} {Mathematics} and {Science}: {A}
  {Special} {Issue} to {Honor} {Vladimir} {Zakharov}},\ \textbf {\bibinfo
  {volume} {152-153}},\ \bibinfo {pages} {520--550} (\bibinfo {year}
  {2001})}\BibitemShut {NoStop}%
\end{thebibliography}%
%---------------------------------------------------------------------------------------
\end{document}